\documentclass{SciPost}

\hypersetup{
    colorlinks,
    linkcolor={red!50!black},
    citecolor={blue!50!black},
    urlcolor={blue!80!black}
}

\usepackage[bitstream-charter]{mathdesign}
\usepackage{amsmath}
\usepackage{tikz}
\usetikzlibrary{arrows.meta}
\usepackage{subcaption}
\usepackage{babel}
\usepackage[normalem]{ulem}
\DeclareSymbolFont{usualmathcal}{OMS}{cmsy}{m}{n}
\DeclareSymbolFontAlphabet{\mathcal}{usualmathcal}

\fancypagestyle{SPstyle}{
\fancyhf{}
\lhead{\colorbox{scipostblue}{\bf \color{white} ~SciPost Physics }}
\rhead{{\bf \color{scipostdeepblue} ~Submission }}

\fancyfoot[C]{\textbf{\thepage}}
}

\begin{document}
\pagestyle{SPstyle}

\begin{center}{\Large \textbf{\color{scipostdeepblue}{
New class of exact flat bands - compact localised states protected by local graph topology
}}}\end{center}

\begin{center}\textbf{
Tamaghna Hazra\textsuperscript{$\star$}
}\end{center}

\begin{center}
Institute for Theory of Condensed Matter, Karlsruhe Institute of Technology, Kaiserstrasse 12, Karlsruhe 76131, Germany
\\[\baselineskip]
$\star$ \href{mailto:tamaghnahazra@gmail.com}{\small tamaghnahazra@gmail.com}
\end{center}

 \section*{\color{scipostdeepblue}{Abstract}}
 \textbf{\boldmath{%
Strongly correlated quantum matter is fundamentally defined by the tension between non-commuting quantum operators. Hamiltonians exhibiting macroscopic degeneracies are of general interest in this field because they imply an infinite susceptibility to any non-commuting perturbation. In moir\'e heterostructures, engineering such extensive degeneracies in the kinetic Hamiltonian creates a fertile garden for exotic strongly correlated phases of matter to emerge from the resulting flat bands. Here, we introduce a prescription to construct an infinite family of exact flat-band Hamiltonians on arbitrary hypergraphs and cell complexes - containing line-graphs as its simplest case and extending to faces, polyhedra, and higher-order cells - whenever orbitals on the cells couple exclusively through a smaller number of shared vertices. As the minimal extension beyond line-graphs, we demonstrate this construction on the face-graphs of four three-dimensional Bravais lattices. For all flat band Hamiltonians thus defined, we prove that the extensive degeneracies are protected by the Euler characteristic of the dual hypergraph - the topological index of the discrete Atiyah-Singer index theorem. The degeneracy of the flat bands is therefore insensitive to the detailed weights of local coupling, as long as the local topology of the cell-vertex connectivity is maintained. On amenable graphs, the resulting macroscopic nullspaces yield compact localised states that remain localised over time due to destructive interference across shared vertices. We discuss the broad implications of such non-dispersing quantum modes in diverse settings, from arrested dynamics in quantum networks and quantum machine learning algorithms to Majorana-free topological quantum computation.
 }}

\vspace{\baselineskip}

\noindent\textcolor{white!90!black}{%
\fbox{\parbox{0.975\linewidth}{%
\textcolor{white!40!black}{\begin{tabular}{lr}%
  \begin{minipage}{0.6\textwidth}%
    {\small Copyright attribution to authors. \newline
    This work is a submission to SciPost Physics. \newline
    License information to appear upon publication. \newline
    Publication information to appear upon publication.}
  \end{minipage} & \begin{minipage}{0.4\textwidth}
    {\small Received Date \newline Accepted Date \newline Published Date}%
  \end{minipage}
\end{tabular}}
}}
}

\vspace{10pt}
\noindent\rule{\textwidth}{1pt}
\tableofcontents
\noindent\rule{\textwidth}{1pt}
\vspace{10pt}

\section{Introduction}

Positive semi-definite operators of the form $BB^{T}$ can host an extensive degeneracy of their lowest eigenvalue, guaranteed by dimension counting alone. For any linear map $B:\mathbb{R}^{n}\to\mathbb{R}^{m}$, the operator $BB^{T}$ has at least $m-n$ zero eigenvalues, and because its spectrum is non-negative, these zero modes span the lowest eigenspace. The degeneracy is therefore extensive whenever $m-n$ grows linearly with system size, irrespective of the matrix elements of $B$. This mechanism underlies flat bands in condensed matter systems, zero-energy `floppy modes' of mechanical networks, lattice models for supersymmetry and has direct implications across multiple scientific and computational disciplines. 

In condensed matter physics, such macroscopic degeneracies of the Hamiltonian operator that describes the kinetic energy of electrons result in flat electronic bands; prominent examples include Landau levels and flat bands in medial lattices like the Kagome lattice and subdivided graphs like the Lieb lattice\cite{landau1930,lieb1989,vonklitzing1986,mielkeFerromagnetismHubbardmodel1991,tasakiFerromagnetismHubbardModels1992,tasakiPhysicsMathematicsQuantum2020}. The complete quenching of kinetic energy in these bands causes electron-electron interactions to dominate the low energy physics, providing a natural platform for various strongly correlated phases of matter, such as unconventional superconductivity, Wigner crystals, strange metals, and fractionalised topological phases. In recent years, there has been a resurgence in such flat bands following the discovery of moir\'e heterostructures\cite{li2010,caoCorrelatedInsulatorBehaviour2018,caoUnconventionalSuperconductivityMagicangle2018} where different layers of two-dimensional materials are placed in contact with a relative twist. The resulting moir\'e superlattice has a large spatial periodicity and consequently narrow bands that can be tuned by the relative twist angle. Many novel strongly correlated phases of matter have been discovered in these platforms\cite{andrei2020,andrei2021,mak2022a,nuckolls2024a,mellado2025}, including a fractional quantum Hall state at zero magnetic field\cite{cai2023,zeng2023,xu2023,lu2024a}. This work introduces a prescription to engineer an infinite family of such flat band Hamiltonians of the form $B B^T$, supported on arbitrary lattices, by interpreting the hitherto abstract operator $B$ as the incidence matrix of the faces of the lattice on its vertices.

The topology of the positive semi-definite form $B B^T$ and its macroscopic null-space have profound consequences beyond standard band theory. For instance, this incidence matrix structure governs the dynamical matrices of isostatic mechanical networks, where the null space identifies the number of topologically protected, zero-frequency floppy modes\cite{kane2014}. The same incidence matrix structure has been exploited\cite{roychowdhury2024} to provide fresh perspectives on lattice models for supersymmetry\cite{rana1993,fendley2003}. The connection of the macroscopic nullspace to the local topology of graph connectivity was first pointed out by Sutherland\cite{sutherland1986} and later interpreted\cite{roychowdhury2024} as a discrete lattice analogue of the Witten index~\cite{witten1982constraints}. Quite generally, frustration-free Hamiltonians of the form $B B^\dagger$ have a long history as exactly solvable points of a wide variety of strongly correlated systems, from frustrated magnetism to Fractional Quantum Hall phases~\cite{arovas1992,ardonneTopologicalOrderConformal2004}. More recently~\cite{biswas2022,tan2025}, extensive degeneracies in the many-body eigenspectrum of kinetically constrained models of various origins have been discussed as a novel form of eigenstate order\cite{ben-amiManybodyCagesDisorderfree2025}, variously dubbed as glasses, scars, cages and collective bound states by different groups\cite{ben-amiManybodyCagesDisorderfree2025,tan2025,jonay2026,mohapatra2026,nicolau2026,hazra2026}. These extensive many-body degeneracies have been demonstrated to originate from compact localised states (CLSs) in the many-body transition graph, mapping directly to the single-particle flat bands of a fictitious particle hopping on the transition graph\cite{tan2025}.

Beyond line-graphs and subdivided graphs of an arbitrary root graph\cite{mielkeFerromagnetismHubbardmodel1991,leykam2018,kollarLineGraphLatticesEuclidean2020}, positive semi-definite flat-band Hamiltonians of the form $t B B^\dagger$ also appear in Tasaki's cell construction\cite{tasakiFerromagnetismHubbardModels1992,mielkeTasaki1993,tasaki1998,tasakiPhysicsMathematicsQuantum2020,tanaka2020}, molecular-orbital representations\cite{hatsugai2011,mizoguchi2019,mizoguchi2023} and Gram-matrix constructions\cite{xu2020}. Meanwhile, other systematic prescriptions for engineering compact localised states have relied on local permutation symmetries~\cite{roentgen2018} or hidden ``latent'' symmetries revealed upon isospectral reduction\cite{roentgen2021,morfonios2021}. By contrast, the incidence-matrix mechanism developed here requires neither geometric nor latent symmetry - its extensive degeneracy follows solely from the rank deficiency of $B$ whenever cells outnumber vertices.

The known flat-band results on line-graphs - such as those in the Kagome and pyrochlore lattices - are understood here as the $k=1$ case of an infinite family of flat-band models. This framework extends straightforwardly from the edges of a graph to its faces, polyhedra, and arbitrary $k$-cells in a CW complex defined on the root graph or any of its derivative graphs. Section~\ref{sec:summary} summarises this general hypergraph framework in self-contained terms: the extensive nullspace of $BB^T$, its restriction to compactly supported zero modes on finite subgraphs, and a sufficient condition for the flat-band manifold to host compact localised states. Section~\ref{sec:motivation} then presents the physical motivation for line-graphs and their higher-dimensional counterparts such as face-graphs, showing how vertex-mediated hybridisation naturally realises the $BB^T$ connectivity. In the bulk of this paper, we demonstrate this prescription on various face-graphs - built on the faces of an arbitrary graph as defined in Section~\ref{sec:faces} - which host exactly flat bands and an extensive degeneracy of compact localised states whenever the number of faces $|F|$ exceeds the number of vertices $|V|$. In Section~\ref{sec:cubic}, we apply this construction to the simplest case of the nearest-neighbour (NN) graph of a cubic lattice. In Section \ref{sec:homo}, we generalise to an arbitrary graph whose faces have the same coordination (triangles or rectangles). In Section \ref{sec:hex}, we demonstrate the application to the NN graph of the simple hexagonal lattice, which has both triangular and rectangular faces. In Section \ref{sec:hetero}, we generalise to an arbitrary graph with such a heterogeneous face set, and discuss the compact localised states. In Section \ref{sec:topo}, we connect the extensive degeneracy of face-graph adjacency matrices to a topological index, which is a discrete version of the Atiyah-Singer index, rigorously establishing that the compact localised states in the bulk of the face-graph are protected by the local topology of the graph connectivity. Finally, in Section \ref{sec:disc}, we discuss the broad-ranging implications of such compact localised states for quantum algorithms, disordered quantum networks, many-body quantum dynamics and a potentially new route to topologically protected quantum information without Majorana fermions or Majorana bound states.

\section{Main Results}\label{sec:summary}

Here, we show an infinite family of flat band models and Hamiltonians defined on arbitrary graphs, of which the line-graph construction of Mielke~\cite{mielkeFerromagnetismHubbardmodel1991} is the first member. 
The explicit demonstrations in the following sections focus on the faces of crystal lattices: face-graphs are the next member of this infinite family. 
In this section, we present the general mechanism in terms of an arbitrary hypergraph in self-contained terms so that every subsequent case we demonstrate: from line graphs to homogeneous and heterogeneous face-graphs, graphs of polyhedra, and higher cells, is an instance of this generic mechanism for obtaining exact extensive degeneracies and CLS from the flat band eigenspace. 

\subsection{Extensive nullspace of the dual signless Laplacian $BB^T$}

Given any set of vertices $V$, one can define a set of hyper-edges $X$ and specify how the set $X$ covers the set $V$ by an incidence matrix $B$ with dimensions $|X|\times|V|$ such that $\ensuremath{B_{xv}=1}$ if $v\in x$ and 0 otherwise. 
This gives us a hypergraph $G = (V, X)$ where each cell or hyper-edge $x \subset V$ covers $|x| \ge 2$ vertices.
Then the key idea is that one can use the square matrix $BB^{T}$ to define an adjacency condition between the hyper-edges, similar to the adjacency
condition of the line-graph, that guarantees an exact flat band for hopping between adjacent hyper-edges when $|X|-|V|\sim\mathcal{O}(|V|)$.

Consider the tight-binding Hamiltonian $H = t B B^T$ where $t \neq 0$ is an arbitrary non-zero scalar. 
Now $BB^T$ inherits the rank of the rectangular matrix $B$ but has dimension $|X|\times |X|$, so by the rank-nullity theorem, must have a nullspace of dimension at least $|X|-|V|$, and $H$ has as many exact zero modes. 
In essence, $BB^T$ is the \emph{signless Laplacian of the dual hypergraph $\widetilde{G}(X,V)$}~\footnote{cf. $B^T B$ is the signless Laplacian of the original hypergraph $G(V,X)$}: cells and vertices interchange roles, so cells are connected through every vertex they share. 
We can decompose $BB^{T}=\sum_{k\ge1}^{p}kA^{(k)}$ where $A^{(k)}$ is the adjacency matrix connecting hyper-edges that share exactly $k$ vertices
\footnote{For a graph $G(V,E)$, this reduces to $BB^{T}=2I_{|E|}+A[\mathcal{L}(G)]$ where $A$ is the adjacency matrix of the linegraph $\mathcal{L}(G)$.}, so that the zero modes of hopping between adjacent cells span the nullspace of $BB^T$. 
For instance, if all hyper-edges cover $p$ vertices, then $A^{(p)}=I_{|X|}$ and the hopping Hamiltonian $H_{{\rm hop}}=t\left(BB^{T}-pI_{|X|}\right)$ realises a flat band at energy $E=-pt$ if $V$ is imbued with lattice translation symmetry. This is the reason all line-graphs have flat-bands, and it generalises smoothly from edges to hyperedges with degree $|x|>2$. 

Every model analysed in the remainder of this paper fits inside this single framework. Taking cells to be edges ($|x| = 2$) recovers line-graph flat bands, such as the kagome lattice derived from the honeycomb lattice ($|X|=3N, |V| = 2N$), and the checkerboard lattice derived from the square lattice ($|X|=2N, |V| = N$). Defining faces as irreducible cycles (Section~\ref{sec:faces}) yields the  face-graph flat bands of the simple cubic, FCC, and BCC lattices (Sections~\ref{sec:cubic} and~\ref{sec:homo}), as well as the face-graph of the simple hexagonal lattice with a heterogeneous face-set, where destructive interference between triangular and rectangular faces produces an additional mixed flat band (Sections~\ref{sec:hex} and~\ref{sec:hetero}). Iterating the same construction to polyhedra and higher-dimensional cells generates an entire hierarchy of flat-band models on any root graph, and Section~\ref{sec:topo} recasts the nullity of the dual signless Laplacian as a topological index protected by the geometry of local graph connectivity.

\subsection{Zero modes on subgraphs and compact support}

The expectation value $\langle \psi | H | \psi \rangle = t \|B^T \psi\|^2 = 0$ shows that the zero modes of the dual signless Laplacian necessarily satisfy 
\begin{equation}
(B^T \psi)_v = \sum_{x \ni v} \psi_x = 0 \qquad \text{for every } v \in V.
\label{eq:summary_interference}
\end{equation}
They must have zero incidence at \emph{every vertex}. This allows us a direct route to address the question of when the large degeneracy of zero modes corresponds to compact localised states on the set $X$. 

By definition, a zero mode of $B^T$ with compact support inside a finite subset of cells $S \subset X$, is also a zero mode of the submatrix $B_S^T$, which is derived from the transpose of the incidence matrix $B$ by restricting the columns to $S$ and the rows to $V(S)= \bigcup_{x \in S} x$ the set of all vertices covered by the cells in $S$ (including boundary vertices).
\begin{equation}
(B_S^T \psi)_v = \sum_{x\in S, x \ni v} \psi_x = 0 \qquad \text{for every } v \in V(S).
\label{eq:summary_interference_sub}
\end{equation}
The converse is also true, every null vector of $B_S^T$ extended by zero outside $S$, is an exact zero mode of the full $B^T$. 
This means that we can count the zero modes supported inside $S$ by the nullity of $B_S^T$
\begin{equation}
\dim \ker B_S^T = |S| - \operatorname{rank}(B_S) \ge |S| - |V(S)|.
\label{eq:summary_subgraph}
\end{equation}
Whenever the number of cells in a subset exceeds the number of vertices covered by it, this is a sufficient condition for there to be a zero mode with compact support within that subset. This suggests a local quantity that adds up to the lower bound $|X|-|V|$ on the nullity of $BB^T$ and diagnoses when finite subgraphs must host compact localised states.

\subsection{When does an extensive nullspace yield compact localised states?}\label{subsec:dual_curvature}

Recall the familiar Euler characteristic of a polyhedron, $\chi = |V| - |E| + |F|$, which generalises to the alternating sum $\chi = \sum_k (-1)^k |X_k|$ over $k$-cells of a $k$-dimensional cell complex (illustrated for the surface of a cube in Fig.~\ref{fig:cell_complex})~\cite{levitt1992,knill2012}. 

\begin{figure}[htbp]
    \centering
    \definecolor{cellred}{RGB}{235,55,45}
    \begin{tikzpicture}[scale=1.0, line cap=round, line join=round]
        \node[font=\small] at (4.96,2.08) {$\begin{array}{r@{\;}c@{\;}c@{\;}c@{\;}c@{\;}c@{\;}c}
          {\rm Euler\, char\,}  \chi & = & |V|   & - & |E|   & + & |F|   = 2 \\
                 & = & |X_0| & - & |X_1| & + & |X_2|
        \end{array}$};
        \begin{scope}[xshift=0cm]
            \draw[line width=1.05pt] (0,0) -- (1.1,0) -- (1.1,1.1) -- (0,1.1) -- cycle;
            \draw[line width=1.05pt] (0,1.1) -- (0.42,1.46) -- (1.52,1.46) -- (1.52,0.36) -- (1.1,0);
            \draw[line width=1.05pt] (1.1,1.1) -- (1.52,1.46);
            \fill[cellred, draw=white, line width=1.6pt] (1.1,0) circle (3.6pt);
            \node[align=center, font=\small] at (0.76,-0.42) {$X_0 = \{8$ $0$-cells/vertices\}};
        \end{scope}
        \begin{scope}[xshift=4.2cm]
            \draw[line width=1.05pt] (0,0) -- (1.1,0) -- (1.1,1.1) -- (0,1.1) -- cycle;
            \draw[line width=1.05pt] (0,1.1) -- (0.42,1.46) -- (1.52,1.46) -- (1.52,0.36) -- (1.1,0);
            \draw[line width=1.05pt] (1.1,1.1) -- (1.52,1.46);
            \draw[white, line width=4.6pt] (1.1,0) -- (1.1,1.1);
            \draw[cellred, line width=2.8pt] (1.1,0) -- (1.1,1.1);
            \node[align=center, font=\small] at (0.76,-0.42) {$X_1 = \{ 12$ $1$-cells/edges\}};
        \end{scope}
        \begin{scope}[xshift=8.4cm]
            \draw[line width=1.05pt] (0,1.1) -- (0.42,1.46) -- (1.52,1.46) -- (1.52,0.36) -- (1.1,0);
            \draw[line width=1.05pt] (1.1,1.1) -- (1.52,1.46);
            \fill[cellred, draw=white, line width=1.6pt] (0,0) rectangle (1.1,1.1);
            \node[align=center, font=\small] at (0.76,-0.42) {$X_2 = \{ 6$ $2$-cells/faces\}};
        \end{scope}
    \end{tikzpicture}
    \caption{Cells of a cube's surface ($0$-, $1$-, and $2$-cells) and its Euler characteristic. Adapted from \href{https://commons.wikimedia.org/wiki/File:Vertex_edge_face.svg}{``Vertex edge face''} by C.~Lawson-Perfect.}
    \label{fig:cell_complex}
\end{figure}
For a $1$-dimensional cell complex (such as the graph of vertices and edges in Fig.~\ref{fig:cell_complex}) or a hypergraph $G(V,X)$ (with $X_0 = V$ and $X_1 = X$), this reduces to $\chi(G) = |V| - |X|$. For the dual hypergraph $\widetilde{G}(X,V)$, the Euler characteristic is
\begin{equation}
\chi(\widetilde{G}) = |X| - |V|,
\end{equation}
precisely the lower bound on the zero-mode degeneracy of $BB^T$.

To convert this global count into a sum over local quantities, we define $d(v)$ as the degree of vertex $v$, i.e., the number of cells in $X$ covering $v$ and rewrite the vertex count as \\
$|V| = \sum_{v \in V} \sum_{x \ni v} 1/d(v) = \sum_{x \in X} \sum_{v \in x} 1/d(v)$, so the dual Euler characteristic decomposes as a local sum over cells:
\begin{equation}
\chi(\widetilde{G}) = |X| - |V| = \sum_{x \in X} K_x, \qquad K_x = 1 - \sum_{v \in x} \frac{1}{d(v)}.
\label{eq:summary_dual_curvature}
\end{equation}
We call $K_x$ the \emph{dual discrete curvature} of cell $x$. It is positive whenever the vertices covered by $x$ are shared by sufficiently many cells - for example, $K_x = 1 - 2/3 = 1/3$ for the edges of the honeycomb lattice and $K_x = 1 - 4/12 = 2/3$ for the square faces of the simple cubic lattice.

Now sum $K_x$ over any finite subset of cells $S \subset X$:
\begin{equation}
\sum_{x \in S} K_x = \sum_{x \in S} \left(1 - \sum_{v \in x} \frac{1}{d(v)}\right) = |S| - \sum_{v \in V(S)} \frac{d_S(v)}{d(v)},
\end{equation}
where $d_S(v)$ is the number of cells in $S$ covering vertex $v$. In $|S| - |V(S)|$, by contrast, every vertex $v \in V(S)$ is subtracted with unit weight. Writing $|V(S)| = \sum_{v \in V(S)} 1$ therefore gives
\begin{equation}
|S| - |V(S)| = \sum_{x \in S} K_x - \sum_{v \in V(S)} \left(1 - \frac{d_S(v)}{d(v)}\right).
\end{equation}
For every interior vertex of $S$, all cells covering $v$ lie in $S$, so $d_S(v) = d(v)$ and its term vanishes identically; only boundary vertices remain, that are also covered by cells outside $S$: $\partial V(S) \subset V(S)$,
\begin{equation}
|S| - |V(S)| = \sum_{x \in S} K_x - \sum_{v \in \partial V(S)} \left(1 - \frac{d_S(v)}{d(v)}\right).
\label{eq:summary_bulk_boundary}
\end{equation}
Every boundary vertex satisfies $1 \le d_S(v) < d(v)$, so $0 < 1 - d_S(v)/d(v) < 1$ and the boundary sum is bounded above by $|\partial V(S)|$. Introducing the average curvature density $\bar{K} = |S|^{-1} \sum_{x \in S} K_x$ on $S$ and combining with Eq.~\eqref{eq:summary_subgraph} yields
\begin{equation}
\dim \ker B_S^T \ge |S| - |V(S)| \ge \bar{K}\,|S| - |\partial V(S)|,
\label{eq:bst}
\end{equation}
i.e., $(\text{curvature density}) \times (\text{volume}) - (\text{boundary area})$. For any positive curvature density $\bar{K} > 0$, this bound becomes positive on a sufficiently large subset $S$ whenever the hypergraph admits large finite subsets whose boundary-to-volume ratio vanishes:
\begin{equation}
\frac{|\partial V(S)|}{|S|} \to 0.
\label{eq:summary_amenability}
\end{equation}
Graphs with this property are called \emph{amenable graphs}, and if the average discrete curvature is positive, these graphs must have CLS as their flat band eigenstates: the zero modes of the dual signless Laplacian $BB^T$. The average density of the dual discrete curvature $\bar{K}$ also prescribes the minimal size of the CLS through \eqref{eq:bst}. Note that we use the ideas of size, area and volume colloquially for intuition: the quantities we use are defined on an arbitrary hyper-graph without any reference to a metric space in which the hyper-graph may be embedded.

Every $d$-dimensional Euclidean crystal (including all lattices studied in this paper) and every quasicrystal is amenable: for a block of linear size $\ell$, $|\partial V(S)|/|S| = \mathcal{O}(1/\ell)$, so the bulk curvature $\bar{K}|S|$ inevitably overtakes the boundary area and guarantees an extensive manifold of CLS. 
By contrast, on infinite trees (Bethe lattices) and hyperbolic lattices, the boundary area $|\partial V(S)|$ always remains a finite fraction of the volume $|S|$, so the bulk curvature need not outgrow the boundary penalty on such non-amenable graphs, although they have flat bands whenever $|X|>|V|$. Amenability is therefore the geometric condition separating graphs whose extensive nullspace of $BB^T$ is guaranteed to host CLS from non-amenable networks whose zero modes may remain non-compact. Flat-band zero modes on such non-amenable tree and hyperbolic line-graph networks are explicitly shown, for instance, in Fig.~2 of Koll\'ar \textit{et al.}~\cite{kollarLineGraphLatticesEuclidean2020}, where hyperbolic graphs do have CLS on even cycles, since amenability is sufficient but not necessary to admit CLS.

\section{Physical motivation for line-graphs and face-graphs in quantum condensed matter}\label{sec:motivation}

In general, a tight-binding model proportional to the dual signless Laplacian $H = t B B^T$ is fine-tuned if one considers particles hopping directly between cell centres, because adjacent cells determined by the dual Laplacian $BB^T$ are not necessarily nearest neighbours in Euclidean space~\footnote{The Kagome lattice in 2D and the pyrochlore lattice in 3D are prominent exceptions}. On the checkerboard line-graph of the square lattice, collinear edges at separation $a$ must hop with the same amplitude as orthogonal nearest neighbours at $a/\sqrt{2}$ while opposite parallel edges of a square at separation $a$ have zero hopping. When are the adjacency relations specified by the dual Laplacian natural for any physical system? 

Here we show that they arise automatically whenever the cell orbitals couple exclusively through states localised at the shared vertices $V$. Consider a microscopic bipartite Hamiltonian on $\mathcal{H}_X \oplus \mathcal{H}_V$ that places orbitals $|x\rangle$ at reference energy $0$ on each cell $x \in X$ hybridising with coupling $g$ to off-resonant atomic levels $|v\rangle$ at site energy $\Delta$ ($|\Delta| \gg |g|$) on each vertex $v \in V$:
\begin{equation}
H_{\text{micro}} = \begin{pmatrix} 0 & g B \\ g B^T & \Delta I_{|V|} \end{pmatrix}.
\label{eq:mot_bipartite}
\end{equation}
Eliminating the vertex levels to second order in $g/\Delta$ via Schrieffer-Wolff yields:
\begin{equation}
H_{\text{eff}} = t B B^T, \qquad t = -\frac{g^2}{\Delta},
\label{eq:mot_downfold}
\end{equation}
Each virtual excursion $x \xrightarrow{g} v \xrightarrow{g} x'$ through a shared vertex $v \in x \cap x'$ contributes hopping amplitude $t = -g^2/\Delta$: on a line-graph ($|x|=2$), two edges sharing a vertex couple with amplitude $t$, while on a face-graph two faces sharing an edge (two vertices) couple with amplitude $2t$, two faces sharing a single corner couple with amplitude $t$, a $p$-vertex face acquires an on-site shift $pt$, and opposite faces sharing no vertices remain uncoupled.

Note that the presence of a flat band in $H_{\text{micro}}$ is a non-perturbative statement, independent of $|\Delta| \gg |g|$. A zero mode $|\Psi_0\rangle = (\psi_F, 0)^T$ supported only on the faces with vanishing incidence at every vertex remains an exact zero-energy eigenstate of $H_{\text{micro}}$, because $B^T \psi_F =0$.

\begin{figure}[htbp]
    \centering
    \includegraphics[width=\linewidth]{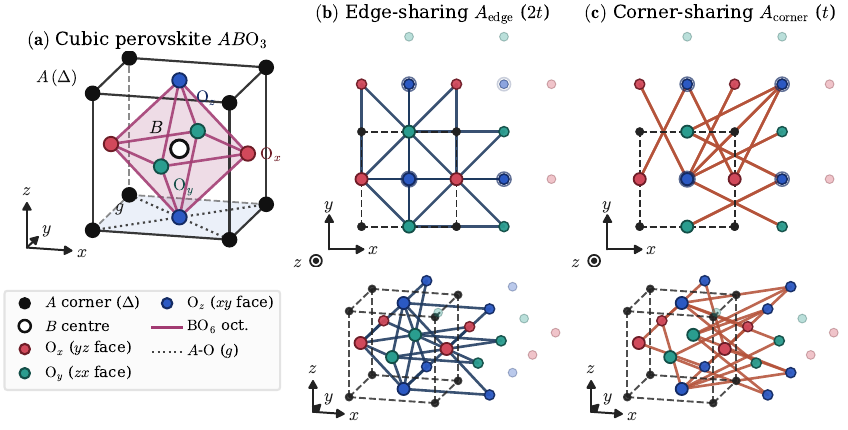}
    \caption{Realisation of face-graphs on the $\rm{O}$ sites of a perovskite crystal. (a) Cubic perovskite $AB\mathrm{O}_3$: $A$ cations occupy the corners which realise the vertex set $V$ (filled black circles), $B$ cations occupy the body centres (open black circles), and oxygen anions forming the face set $F$ $\mathrm{O}_x, \mathrm{O}_y, \mathrm{O}_z$ occupy the $yz$, $zx$, and $xy$ face centres (red, teal, and blue circles). Dotted lines on the bottom $xy$ plaquette illustrate the coupling $g$ between a face-centred $\mathrm{O}_z$ orbital and its four surrounding $A$-site corners. (b)~Edge-sharing face-graph $A_{\text{edge}}$ and (c)~corner-sharing face-graph $A_{\text{corner}}$, shown in top-down ($xy$) and orthographic views. For clarity, only edges incident on the six faces of the highlighted reference unit cell (dashed cube) are drawn.}
    \label{fig:perovskite}
\end{figure}

In solid-state crystals, the cubic perovskite structure $AB\mathrm{O}_3$ (Fig.~\ref{fig:perovskite}(a)) provides an immediate structural realisation of the simple cubic face-graph. Its $A$ cations sit at the cubic corners ($|V| = N$), its $B$ cations sit at the body centres, and its oxygen anions $\mathrm{O}_x, \mathrm{O}_y, \mathrm{O}_z$ sit at the $yz$, $zx$, and $xy$ face centres ($|F| = 3N$), forming a three-dimensional network of corner-sharing $\mathrm{BO}_6$ octahedra. This geometry admits two distinct cation-mediated hopping regimes for orbitals supported on the oxygen sublattice. When hopping proceeds predominantly through the body-centred $B$ cations, each $\mathrm{O}$ bridges exactly two $B$ sites, realising the $|x| = 2$ line-graph of the $B$-site simple cubic lattice. When instead the $\mathrm{O}$ orbitals hybridise predominantly through $A$-site levels, eliminating the $A$ sublattice via Eq.~\eqref{eq:mot_downfold} generates $H_{\text{eff}} = t(4I + 2A_{\text{edge}} + A_{\text{corner}})$, precisely the face-graph hopping Hamiltonian we derive later in Sec.~\ref{sec:cubic}. Figure~\ref{fig:perovskite}(b) isolates the resulting edge-sharing connectivity $A_{\text{edge}}$ (amplitude $2t$) and Fig.~\ref{fig:perovskite}(c) isolates the corner-sharing connectivity $A_{\text{corner}}$ (amplitude $t$). In real perovskites and antiperovskites, direct $\mathrm{O\text{-}O}$ overlap and presence of finite $B\text{-}\mathrm{O}$ and $A\text{-}\mathrm{O}$ hybridisation act as dispersive perturbations around these idealised flat-band parent limits.

Beyond atomic crystals, the same vertex-mediated mechanism of face-graphs with flat bands arises naturally in two broad arenas. First, in systems whose low-energy Wannier or molecular orbitals are centred on elementary plaquettes with weight concentrated away from the plaquette centre towards the periphery or corners - prominently the three-lobed ``fidget-spinner'' Wannier orbitals of magic-angle twisted bilayer graphene~\cite{koshinoMaximallyLocalizedWannier2018,po-prx-2018,kangSymmetryMaximallyLocalized2018}, whose AB/BA-centred charge density peaks on the three corners of elementary AA triangles, and open-shell triangulene flakes~\cite{clar1953,pavlicek2017,mishra2020,mishra2021} assembled into corner-linked molecular networks (Section~\ref{sec:disc}) - overlap between neighbouring cells is concentrated at shared vertices. Second, in synthetic metamaterials and quantum simulators, including superconducting circuit-QED resonator lattices~\cite{kollarLineGraphLatticesEuclidean2020} as well as photonic, acoustic, and topolectrical networks~\cite{leykam2018,roentgen2024}, resonators representing faces can be coupled exclusively through off-resonant vertex-localised modes. Because the hopping matrix is then fixed by incidence topology rather than Euclidean distance, the exact $B B^T$ connectivity and its macroscopic flat-band manifold are realised without fine-tuning.

\section{Defining the Faces of an Arbitrary Graph}\label{sec:faces}

Before exploring the consequences of hopping models on face-graphs, which are the minimal extensions of the infinite family of flat band models that go beyond line-graphs, we need an unambiguous definition of what constitutes a ``face'' if we are given an abstract graph $G(V,E)$ with a set of vertices $V$ and a set of edges $E$. In regular lattices, faces are intuitively identified as the elementary plaquettes that geometrically tile 3D voids. For an arbitrary graph, we define the faces $F$ of the graph as the set of its relevant cycles, which is identical to the set of its irreducible cycles~\cite{vismara1997}; taking the hyperedges of Section~\ref{sec:summary} to be these irreducible cycles ($X = F$) maps any root graph $G(V,E)$ to its face-graph $G(V,F)$ (just as $X = E$ gives the line-graph). Physically, an irreducible cycle is simply a contiguous sequence of edges that cannot be expressed as a modulo-2 sum of shorter cycles. Thus specified, the face-set of a graph can be enumerated in polynomial time \cite{horton1987,vismara1997,kavitha2009}, and individual faces of an arbitrary graph can be identified from the local edge-connectivity. To verify if a candidate cycle is a face, one only needs to search its immediate structural neighbourhood to verify that it cannot be decomposed into a modulo-2 sum of shorter cycles. 
For example, in the Face-Centred Cubic (FCC) lattice in Fig.~\ref{fig:FCCFaces}, an equatorial square of the octahedral void is \textit{reducible} because it can be decomposed into four shorter triangles that share the apex of the octahedron; therefore, it is not a face. 
In Appendix~\ref{app:face_definition_physics}, we use physical intuition to motivate our chosen definition of a face of a graph.
Note that this definition of a face is not unique, but this is the one that we will use for demonstration in the next four sections. Other definitions yield distinct face-graphs, which also have extensive degeneracies. 

\section{Tight binding model on the faces of a cubic lattice}\label{sec:cubic}

We first apply the general hypergraph construction of Section~\ref{sec:summary} to the simplest three-dimensional example of face-graphs: the nearest-neighbour graph $G(V,E)$ of the simple cubic lattice. Its irreducible cycles (Section~\ref{sec:faces}) are the elementary unit squares normal to the three coordinate axes, giving $|F|=3N$ uniform square faces ($p=4$) for $|V|=N$ vertices and a $3N\times N$ face-vertex incidence matrix $B$.

Because two distinct square faces on the cubic lattice share at most one edge ($k=2$ vertices) or one corner ($k=1$ vertex), the decomposition $BB^T = \sum_{k\ge 1}^{p} k A^{(k)}$ of Section~\ref{sec:summary} reduces to
\begin{equation}
BB^T = 4I + 2A_{\text{edge}} + A_{\text{corner}},
\end{equation}
where $A_{\text{edge}}$ and $A_{\text{corner}}$ are the adjacency matrices of the edge-sharing and corner-sharing face-graphs illustrated in Fig.~\ref{fig:perovskite}(b,c). Subtracting the uniform on-site shift for any non-zero hopping scale $t$ defines the pure hopping Hamiltonian
\begin{equation}
H = 2tA_{\text{edge}} + tA_{\text{corner}} = t(BB^T - 4I),
\end{equation}
describing bosonic or fermionic orbitals on the $yz$, $zx$, and $xy$ faces that hop to edge-sharing neighbours with twice the amplitude of corner-sharing neighbours (as motivated microscopically in Section~\ref{sec:motivation}). Since $\dim\ker(BB^T) \ge |F|-|V| = 2N$, $H$ is guaranteed an at least $2N$-fold degeneracy, corresponding to two exact flat bands at $E=-4t$.

\subsection{Basis Elements and Face-Graph Adjacency Matrices}

To construct the momentum-space Hamiltonian $H=\sum_{\mathbf{k}} H(\mathbf{k})$, we first define the basis elements for the vector space of faces and vertices. Let the vertices of the simple cubic lattice be located at integer coordinates $\mathbf{R}\in\mathbb{Z}^{3}$. The basis for the vertex vector space is given by the localised states $|\mathbf{R}\rangle$. The faces $f_{\mu}$ ($\mu\in\{x,y,z\}$) are centred at half-integer coordinates. For instance, the $x$-faces are centred at $\mathbf{R}+\frac{1}{2}\hat{y}+\frac{1}{2}\hat{z}$, and we denote their basis states as $|f_{x,\mathbf{R}}\rangle$.

We define the corresponding Bloch wavefunctions using a Fourier transform:
\begin{align}
|v_{\mathbf{k}}\rangle & =\frac{1}{\sqrt{N}}\sum_{\mathbf{R}}e^{i\mathbf{k}\cdot\mathbf{R}}|\mathbf{R}\rangle\\
|f_{\mu,\mathbf{k}}\rangle & =\frac{1}{\sqrt{N}}\sum_{\mathbf{R}}e^{i\mathbf{k}\cdot(\mathbf{R}+\mathbf{r}_{\mu})}|f_{\mu,\mathbf{R}}\rangle
\end{align}
where $\mathbf{r}_{x}=\frac{1}{2}(\hat{y}+\hat{z})$, $\mathbf{r}_{y}=\frac{1}{2}(\hat{x}+\hat{z})$, and $\mathbf{r}_{z}=\frac{1}{2}(\hat{x}+\hat{y})$ are the relative centres of the faces within the unit cell.

\begin{figure}[htbp]
    \centering
    \begin{subfigure}[b]{0.46\textwidth}
        \centering
        \begin{tikzpicture}[scale=1.0]
            \draw[thick, fill=green!10] (0,0) rectangle (1,1);
            \draw[thick, fill=green!10] (1,0) rectangle (2,1);
            \draw[thick, fill=green!10] (0,1) rectangle (1,2);
            \draw[thick, fill=green!10] (1,1) rectangle (2,2);
            
            \draw[ultra thick, blue!60!black] (1,0) -- (1,1);
            \draw[ultra thick, blue!60!black] (0,1) -- (1,1);
            \fill[green!60!black] (1,1) circle (2pt);
            
            \draw[->, thick, blue!60!black, shorten >=4pt, shorten <=4pt] (0.5, 0.5) -- (1.5, 0.5);
            \draw[->, thick, blue!60!black, shorten >=4pt, shorten <=4pt] (0.5, 0.5) -- (0.5, 1.5);
            \draw[->, thick, green!60!black, dashed, shorten >=4pt, shorten <=4pt] (0.5, 0.5) -- (1.5, 1.5);
            
            \draw[->, thick] (-0.3, -0.3) -- (-0.05, -0.3) node[right] {$x$};
            \draw[->, thick] (-0.3, -0.3) -- (-0.3, -0.05) node[above] {$y$};
            
        \end{tikzpicture}
        \caption{Intra-orbital hopping\\ ($z$-face to $z$-face)}
        \label{fig:intra}
    \end{subfigure}
    \hfill
    \begin{subfigure}[b]{0.46\textwidth}
        \centering
        \begin{tikzpicture}[x={(-0.5cm,-0.4cm)}, y={(1cm,0cm)}, z={(0cm,1cm)}, scale=1.0]
            \draw[thick, fill=blue!10] (0,0,0) -- (0,1,0) -- (0,1,1) -- (0,0,1) -- cycle;
            
            \draw[thick, fill=blue!10] (0,1,0) -- (0,2,0) -- (0,2,1) -- (0,1,1) -- cycle;
            
            \draw[thick, fill=green!10] (0,0,0) -- (1,0,0) -- (1,1,0) -- (0,1,0) -- cycle;
            
            \draw[ultra thick, blue!60!black] (0,0,0) -- (0,1,0);
            
            \fill[green!60!black] (0,1,0) circle (2pt);
            
            \draw[->, thick, blue!60!black, shorten >=3pt, shorten <=3pt] (0.5, 0.5, 0) -- (0, 0.5, 0.5);
            \draw[->, thick, green!60!black, dashed, shorten >=3pt, shorten <=3pt] (0.5, 0.5, 0) -- (0, 1.5, 0.5);
            
            \draw[->, thick] (0, 2.5, -0.5) -- (0.25, 2.5, -0.5) node[below left] {$x$};
            \draw[->, thick] (0, 2.5, -0.5) -- (0, 2.75, -0.5) node[right] {$y$};
            \draw[->, thick] (0, 2.5, -0.5) -- (0, 2.5, -0.25) node[above] {$z$};
        \end{tikzpicture}
        \caption{Inter-orbital hopping\\ ($z$-face to $x$-face)}
        \label{fig:inter}
    \end{subfigure}
    \caption{Illustration of hopping processes on the cubic lattice. (a) Intra-orbital hoppings between coplanar $z$-faces. Edge-sharing hoppings (solid dark blue arrows) have amplitude $2t$, while corner-sharing hoppings (dashed dark green arrows) have amplitude $t$. (b) Inter-orbital hoppings between an orthogonal $z$-face and $x$-faces, exhibiting analogous edge- and corner-sharing amplitudes.}
    \label{fig:hoppings}
\end{figure}

With this symmetric gauge choice in the Bloch basis, we can block-diagonalise the edge-sharing adjacency matrix $A_{\text{edge}}=\sum_{\mathbf{k}}A_{\text{edge}}(\mathbf{k})$ and the corner-sharing adjacency matrix $A_{\text{corner}}=\sum_{\mathbf{k}} A_{\text{corner}}(\mathbf{k})$. For the diagonal elements (intra-orbital hopping), coplanar faces of the same orientation share edges along two axes and corners along the face diagonals (see Fig.~\ref{fig:hoppings}(a)), leading to
\begin{align}
[A_{\text{edge}}(\mathbf{k})]_{\mu\mu} & =2(\cos k_{\nu}+\cos k_{\lambda})\\
[A_{\text{corner}}(\mathbf{k})]_{\mu\mu} & =4\cos k_{\nu}\cos k_{\lambda}
\end{align}
where $(\mu,\nu,\lambda)$ is a permutation of $(x,y,z)$.

For the off-diagonal elements (inter-orbital hopping), orthogonal faces share a single edge along their intersecting axis, while corner-sharing orthogonal faces are displaced along the remaining spatial dimension (see Fig.~\ref{fig:hoppings}(b)), leading to
\begin{align}
[A_{\text{edge}}(\mathbf{k})]_{\mu\nu} & =4\cos\frac{k_{\mu}}{2}\cos\frac{k_{\nu}}{2}\\
[A_{\text{corner}}(\mathbf{k})]_{\mu\nu} & =8\cos k_{\lambda}\cos\frac{k_{\mu}}{2}\cos\frac{k_{\nu}}{2}
\end{align}
with $\mu\neq\nu\neq\lambda$.

\subsection{Total Hamiltonian and Incidence Matrix}\label{sec:Bmat}

Summing these components according to $H(\mathbf{k})=t(2 A_{\text{edge}}(\mathbf{k})+A_{\text{corner}}(\mathbf{k}))$ provides the full tight-binding Hamiltonian. For instance, the diagonal elements evaluate to $4t(\cos k_{y}+\cos k_{z}+\cos k_{y}\cos k_{z})$, which can be rewritten as:
\begin{align}
4t(\cos k_{y}+\cos k_{z}+\cos k_{y}\cos k_{z}) & =4t\left((1+\cos k_{y})(1+\cos k_{z})-1\right)\nonumber \\
 & =16t\left(\cos^{2}\frac{k_{y}}{2}\cos^{2}\frac{k_{z}}{2}-\frac{1}{4}\right)
\end{align}
By applying this to all diagonal elements, we assemble the full tight-binding Hamiltonian:
\begin{equation}
H(\mathbf{k})=16t\begin{pmatrix}\cos^{2}\frac{k_{y}}{2}\cos^{2}\frac{k_{z}}{2}-\frac{1}{4} & \cos\frac{k_{x}}{2}\cos\frac{k_{y}}{2}\cos^{2}\frac{k_{z}}{2} & \cos\frac{k_{x}}{2}\cos^{2}\frac{k_{y}}{2}\cos\frac{k_{z}}{2}\\
\cos\frac{k_{x}}{2}\cos\frac{k_{y}}{2}\cos^{2}\frac{k_{z}}{2} & \cos^{2}\frac{k_{x}}{2}\cos^{2}\frac{k_{z}}{2}-\frac{1}{4} & \cos^{2}\frac{k_{x}}{2}\cos\frac{k_{y}}{2}\cos\frac{k_{z}}{2}\\
\cos\frac{k_{x}}{2}\cos^{2}\frac{k_{y}}{2}\cos\frac{k_{z}}{2} & \cos^{2}\frac{k_{x}}{2}\cos\frac{k_{y}}{2}\cos\frac{k_{z}}{2} & \cos^{2}\frac{k_{x}}{2}\cos^{2}\frac{k_{y}}{2}-\frac{1}{4}
\end{pmatrix}
\end{equation}

To write the face-vertex incidence matrix $B(\mathbf{k})$ in the Bloch basis, we maintain the same symmetric gauge choice defined in the previous subsection. The incidence matrix elements $B_{\mu}(\mathbf{k})=\langle f_{\mu,\mathbf{k}}|\hat{B}|v_{\mathbf{k}}\rangle$ connect a face to its four bounding vertices. For the $x$-faces, the four vertices are displaced from the face centre by $\pm\frac{1}{2}\hat{y}\pm\frac{1}{2}\hat{z}$. This yields the momentum-space incidence vector 
\begin{align}
B(\mathbf{k})&=[\gamma_{x}(\mathbf{k}),\gamma_{y}(\mathbf{k}),\gamma_{z}(\mathbf{k})]^T, {\rm \, with}\\
\gamma_{x}(\mathbf{k}) & =4\cos\frac{k_{y}}{2}\cos\frac{k_{z}}{2}\\
\gamma_{y}(\mathbf{k}) & =4\cos\frac{k_{x}}{2}\cos\frac{k_{z}}{2}\\
\gamma_{z}(\mathbf{k}) & =4\cos\frac{k_{x}}{2}\cos\frac{k_{y}}{2}
\end{align}
from which it is easy to identify the Hamiltonian 
\begin{equation}
H(\mathbf{k})=t\begin{pmatrix}
\gamma_x^2  & \gamma_x \gamma_y   & \gamma_x \gamma_z\\
\gamma_y \gamma_x & \gamma_y^2 & \gamma_y\gamma_z\\
\gamma_z \gamma_x & \gamma_z\gamma_y & \gamma_z^2
\end{pmatrix}(\mathbf{k}) -4t I = 
t(B(\mathbf{k})B^{\dagger}(\mathbf{k})-4I)
\end{equation}
as expected from Section~\ref{sec:summary}.

\subsection{Band Structure, Band Touchings, and $\Gamma$-Point Eigenvectors}\label{sec:CubicBS}

\begin{figure}[htbp]
    \centering
    \includegraphics[width=\linewidth]{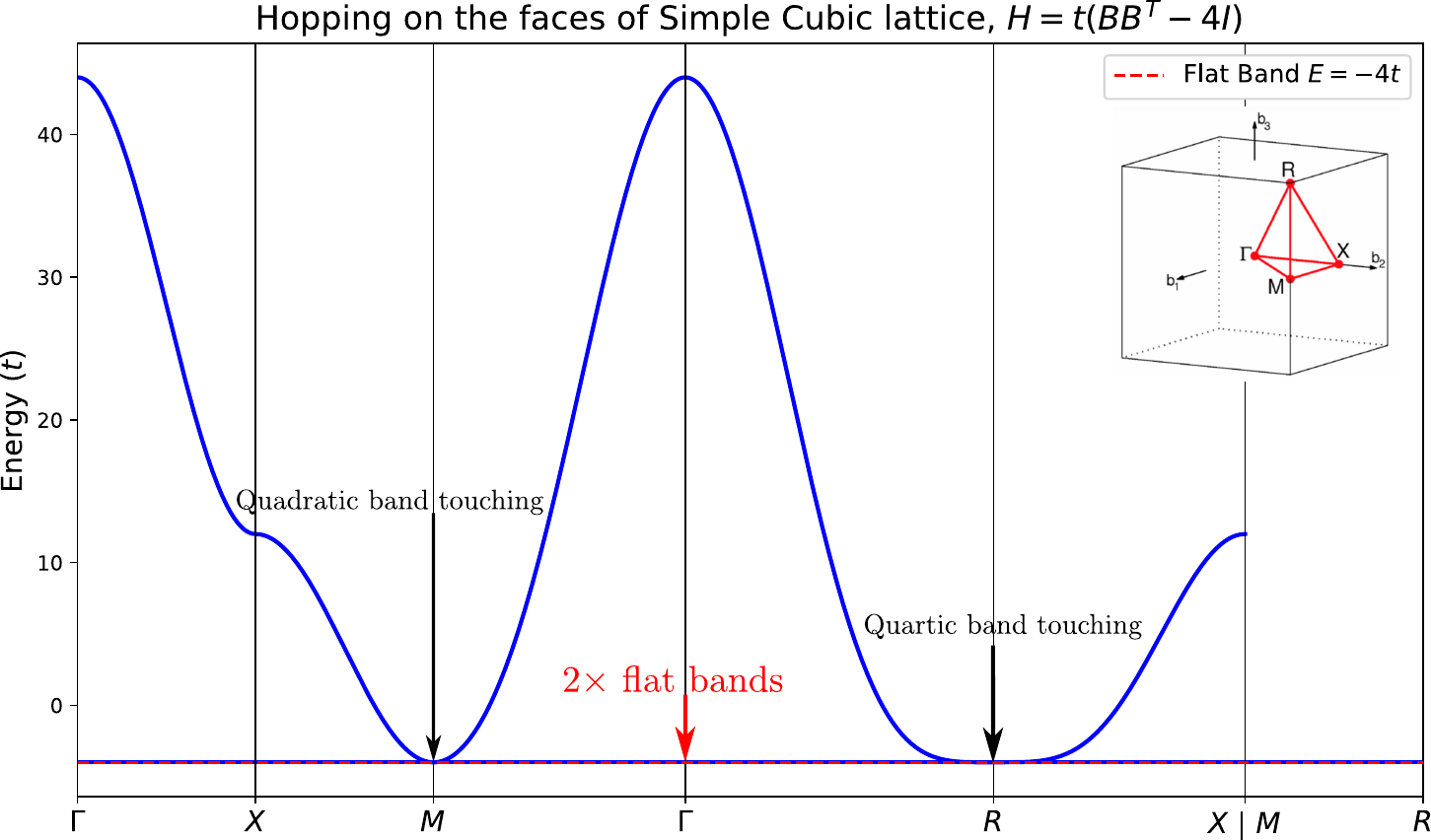}
    \caption{Band structure of the simple cubic face-graph hopping Hamiltonian $H = t(BB^T - 4I)$ along the high-symmetry path $\Gamma$-$X$-$M$-$\Gamma$-$R$-$X\mid M$-$R$, showing two exact flat bands at $E = -4t$ and one dispersive band that touches the flat manifold along the $M$-$R$ Brillouin-zone edges.}
    \label{fig:cubicBS}
\end{figure}

Because $B(\mathbf{k})B^{\dagger}(\mathbf{k})$ is the outer product of a single vector with itself, it has rank at most 1 at every momentum $\mathbf{k}$, with non-zero eigenvalue $|B(\mathbf{k})|^{2}=|\gamma_{x}|^{2}+|\gamma_{y}|^{2}+|\gamma_{z}|^{2}$. Consequently, $H(\mathbf{k})$ hosts two exact flat bands pinned at $E=-4t$ and one dispersive band (Fig.~\ref{fig:cubicBS}),
\begin{align}
E(\mathbf{k}) &= t\left(|\gamma_{x}(\mathbf{k})|^{2}+|\gamma_{y}(\mathbf{k})|^{2}+|\gamma_{z}(\mathbf{k})|^{2}-4\right)\\
&= 16t\left(\cos^{2}\frac{k_{x}}{2}\cos^{2}\frac{k_{y}}{2}+\cos^{2}\frac{k_{y}}{2}\cos^{2}\frac{k_{z}}{2}+\cos^{2}\frac{k_{z}}{2}\cos^{2}\frac{k_{x}}{2}\right)-4t.
\end{align}
Whenever at least two momentum components equal $\pi$, all three components of $B(\mathbf{k})$ vanish and the dispersive band touches the flat-band manifold at $E=-4t$, realising the general molecular-orbital criterion that flat and dispersive bands touch precisely where the columns of $B(\mathbf{k})$ become linearly dependent~\cite{mizoguchi2019} (here $B(\mathbf{k}) = 0$ for a single vertex per primitive cell). Expanding around $M=(\pi,\pi,0)$ with $\mathbf{k}=\mathbf{k}_{M}+\mathbf{q}$ gives a quadratic touching $E(\mathbf{k}_{M}+\mathbf{q}) \approx -4t + 4t(q_{x}^{2}+q_{y}^{2})$ transverse to the Brillouin-zone edge, while along the entire $M$-$R$ line ($\mathbf{k}=(\pi,\pi,k_{z})$) the dispersive band flattens completely onto $E=-4t$. Around $R=(\pi,\pi,\pi)$, each $\cos((\pi+q_{i})/2) \approx -q_{i}/2$, producing a purely quartic band touching
\begin{equation}
E(\mathbf{k}_{R}+\mathbf{q}) \approx -4t + t\left(q_{x}^{2}q_{y}^{2}+q_{y}^{2}q_{z}^{2}+q_{z}^{2}q_{x}^{2}\right).
\end{equation}

At the $\Gamma$-point ($\mathbf{k}=\mathbf{0}$), $\gamma_{x}=\gamma_{y}=\gamma_{z}=4$ so  $B(\mathbf{0})=[4,4,4]^{T}$. The flat-band eigenvectors $|\psi_{\text{flat}}\rangle=(u_{x},u_{y},u_{z})^{T}$ satisfy $B^{\dagger}(\mathbf{0})|\psi_{\text{flat}}\rangle=0$, yielding
\begin{equation}
u_{x}+u_{y}+u_{z}=0,
\end{equation}
which defines the two-dimensional nullspace at the $\Gamma$-point.

\subsection{Compact Localised States}

The degenerate manifold of flat-band eigenstates is spanned by CLSs, augmented by localised states that wind around the 3-torus, as in flat bands on line-graphs~\cite{bergman2008,maimaiti2017,maimaiti2021,kollarLineGraphLatticesEuclidean2020}. Unlike line-graphs, where the minimal-support null vectors of $B^T$ are simple even cycles or bowties on the root graph~\cite{cvetkovic2004spectral}, face-graphs can have even cycles of faces that do not support a CLS, as in Fig.~\ref{fig:cubicCLS_fig}(a) (cf.\ Section~\ref{subsec:dual_curvature}). On the cubic lattice, minimal CLSs are formed on belts chosen from the six faces $|f_{x,\mathbf{0}}\rangle, |f_{x,\hat{\mathbf{x}}}\rangle, |f_{y,\mathbf{0}}\rangle, |f_{y,\hat{\mathbf{y}}}\rangle, |f_{z,\mathbf{0}}\rangle, |f_{z,\hat{\mathbf{z}}}\rangle$ bounding each cubic unit cell (see Fig.~\ref{fig:cubicCLS_fig}(b)), with alternating signs around the belt:
\begin{equation}
|\psi_{\text{CLS},xy}\rangle = |f_{x,\mathbf{0}}\rangle - |f_{y,\mathbf{0}}\rangle + |f_{x,\hat{\mathbf{x}}}\rangle - |f_{y,\hat{\mathbf{y}}}\rangle .
\end{equation}
There are two others related to it by symmetry: $|\psi_{\text{CLS},yz}\rangle$ and $|\psi_{\text{CLS},zx}\rangle$. Of these three symmetric states, only two are linearly independent (since $|\psi_{\text{CLS},xy}\rangle + |\psi_{\text{CLS},yz}\rangle + |\psi_{\text{CLS},zx}\rangle = 0$), consistent with the two-fold degeneracy of the flat bands.

\begin{figure}[htbp]
    \centering
    \definecolor{clsplus}{RGB}{255,198,204}
    \definecolor{clsminus}{RGB}{198,228,252}
    \definecolor{clscancel}{RGB}{20,130,60}
    \definecolor{clsuncancel}{RGB}{215,65,30}
    \begin{subfigure}[b]{0.47\textwidth}
        \centering
        \begin{tikzpicture}[scale=1.15,
            canceled/.style={circle, fill=clscancel, draw=black, line width=0.6pt, inner sep=0pt, minimum size=5.4pt},
            uncanceled/.style={circle, draw=clsuncancel, line width=1.5pt, fill=white, inner sep=0pt, minimum size=7.6pt}
        ]
            \draw[gray!35, line width=0.5pt] (-0.45,-0.45) grid[step=1] (2.45,2.45);

            \fill[clsplus, fill opacity=0.85] (0,1) rectangle (1,2);
            \fill[clsminus, fill opacity=0.85] (1,1) rectangle (2,2);
            \fill[clsminus, fill opacity=0.85] (0,0) rectangle (1,1);
            \fill[clsplus, fill opacity=0.85] (1,0) rectangle (2,1);

            \draw[thick] (0,0) grid (2,2);

            \node at (0.5,1.5) {\small $+1$};
            \node at (1.5,1.5) {\small $-1$};
            \node at (0.5,0.5) {\small $-1$};
            \node at (1.5,0.5) {\small $+1$};

            \node[canceled] at (1,0) {};
            \node[canceled] at (0,1) {};
            \node[canceled] at (1,1) {};
            \node[canceled] at (2,1) {};
            \node[canceled] at (1,2) {};

            \node[uncanceled] (c02) at (0,2) {};
            \node[uncanceled] (c22) at (2,2) {};
            \node[uncanceled] at (0,0) {};
            \node[uncanceled] at (2,0) {};

            \node[align=center, font=\scriptsize, anchor=east] (lblL) at (-0.35, 2.02) {degree-1\\corner};
            \draw[-{Stealth[length=4pt]}, line width=0.6pt] (lblL.east) -- (c02.west);

            \node[align=center, font=\scriptsize, anchor=west] (lblR) at (2.35, 1.92) {degree-1\\corner\\(uncancelled)};
            \draw[-{Stealth[length=4pt]}, line width=0.6pt] (lblR.west) -- (c22.east);

            \draw[-{Stealth[length=4pt]}, thick] (-0.7, -0.35) -- (-0.3, -0.35) node[right, font=\scriptsize] {$x$};
            \draw[-{Stealth[length=4pt]}, thick] (-0.7, -0.35) -- (-0.7, 0.05) node[above, font=\scriptsize] {$y$};
        \end{tikzpicture}
        \caption{Planar $2\times 2$ square face loop (no CLS)}
        \label{fig:notcls}
    \end{subfigure}
    \hfill
    \begin{subfigure}[b]{0.50\textwidth}
        \centering
        \begin{tikzpicture}[x={(-0.5cm,-0.4cm)}, y={(1cm,0cm)}, z={(0cm,1cm)}, scale=1.75,
            canceled/.style={circle, fill=clscancel, draw=black, line width=0.6pt, inner sep=0pt, minimum size=5.4pt}
        ]
            \fill[clsplus, fill opacity=0.38] (0,0,0) -- (0,1,0) -- (0,1,1) -- (0,0,1) -- cycle;
            \fill[clsminus, fill opacity=0.38] (0,0,0) -- (1,0,0) -- (1,0,1) -- (0,0,1) -- cycle;

            \draw[dashed, gray!70!black] (0,0,0) -- (0,1,0);
            \draw[dashed, gray!70!black] (0,0,0) -- (1,0,0);
            \draw[dashed, gray!70!black] (0,0,0) -- (0,0,1);

            \fill[clsminus, fill opacity=0.75] (0,1,0) -- (1,1,0) -- (1,1,1) -- (0,1,1) -- cycle;
            \fill[clsplus, fill opacity=0.75] (1,0,0) -- (1,1,0) -- (1,1,1) -- (1,0,1) -- cycle;

            \draw[thick] (1,0,0) -- (1,1,0) -- (1,1,1) -- (1,0,1) -- cycle;
            \draw[thick] (1,1,0) -- (0,1,0) -- (0,1,1) -- (1,1,1);
            \draw[thick] (1,0,1) -- (0,0,1) -- (0,1,1);

            \node[canceled, fill opacity=0.55, draw opacity=0.55] at (0,0,0) {};
            \foreach \p in {(1,0,0),(0,1,0),(1,1,0),(0,0,1),(1,0,1),(0,1,1),(1,1,1)}
                \node[canceled] at \p {};

            \node[align=center, inner sep=1pt] at (1, 0.5, 0.62) {\scriptsize $|f_{x,\hat{\mathbf{x}}}\rangle$ \\[-2pt] \scriptsize $+1$};
            \node[align=center, inner sep=1pt] at (0.45, 1, 0.52) {\scriptsize $|f_{y,\hat{\mathbf{y}}}\rangle$ \\[-2pt] \scriptsize $-1$};

            \node[font=\scriptsize, anchor=south west] (lblBack) at (0, 0.85, 1.18) {$|f_{x,\mathbf{0}}\rangle\ (+1)$};
            \draw[-{Stealth[length=4pt]}, line width=0.6pt] (lblBack.south west) -- (0.05, 0.55, 0.82);

            \node[font=\scriptsize, anchor=east] (lblLeft) at (0.35, -0.35, 0.65) {$|f_{y,\mathbf{0}}\rangle\ (-1)$};
            \draw[-{Stealth[length=4pt]}, line width=0.6pt] (lblLeft.east) -- (0.45, 0.02, 0.55);

            \draw[-{Stealth[length=4pt]}, thick] (0.2, 1.45, -0.25) -- (0.6, 1.45, -0.25) node[below left, font=\scriptsize] {$x$};
            \draw[-{Stealth[length=4pt]}, thick] (0.2, 1.45, -0.25) -- (0.2, 1.85, -0.25) node[right, font=\scriptsize] {$y$};
            \draw[-{Stealth[length=4pt]}, thick] (0.2, 1.45, -0.25) -- (0.2, 1.45, 0.15) node[above, font=\scriptsize] {$z$};
        \end{tikzpicture}
        \caption{Minimal CLS $|\psi_{\text{CLS},xy}\rangle$ ($x$- and $y$-face belt)}
        \label{fig:cubicCLS}
    \end{subfigure}
    \caption{Contrast between an even face loop that fails to support a compact localised state (CLS) and a minimal CLS on the simple cubic lattice. Faces with amplitude $+1$ ($-1$) are shaded light red (light blue); filled green circles mark vertices where the incident amplitudes cancel ($+1-1=0$), while hollow red rings mark uncancelled degree-1 corners. (a) A planar $2\times 2$ alternating loop of $z$-faces cancels at the five interior and edge-midpoint vertices, but leaves four degree-1 corners uncancelled, so it is not a null vector of $B^T$. (b) The minimal CLS $|\psi_{\text{CLS},xy}\rangle = |f_{x,\mathbf{0}}\rangle - |f_{y,\mathbf{0}}\rangle + |f_{x,\hat{\mathbf{x}}}\rangle - |f_{y,\hat{\mathbf{y}}}\rangle$ supported on a four-face belt around a cubic unit cell: every one of the eight corners lies on one $+1$ $x$-face and one $-1$ $y$-face, yielding exact cancellation on all vertices. }
    \label{fig:cubicCLS_fig}
\end{figure}

\section{Generalisation to arbitrary homogeneous face-graphs}\label{sec:homo}

This derivation extends naturally to any cell complex defined by the vertices $V$, edges $E$ (specified as pairs of vertices) and faces $F$ (specified by irreducible cycles on the graph $G(V,E)$),
assuming every face has $p$ vertices. The face-vertex incidence matrix $B$ is a $|F| \times |V|$ matrix and therefore $B B^T$ is an $|F| \times |F|$ matrix with rank at most $|V|$ and nullity at least $|F| - |V|$.

Thus the face-graph hopping Hamiltonian defined as $H = t(B B^T - pI)$ ($t \neq 0$) by removing the diagonal entries is guaranteed to have an at least $(|F|-|V|)$-fold degeneracy. Note that this extensive degeneracy does not require a regular crystal lattice. When the cell complex has discrete translation symmetry, $H$ is guaranteed to have exact flat bands at $E=-pt$ where $p$ is the degree of the faces. For the simple cubic lattice, $|F| = 3N$ and $|V| = N$, guaranteeing at least 2 flat bands as we found in Sec.~\ref{sec:CubicBS}. For demonstration, we show the flat bands in the cubic (Fig.~\ref{fig:cubicBS}), FCC (Fig.~\ref{fig:FCCBS}), and BCC (Fig.~\ref{fig:BCCBS}) lattices below. Consistent with the linear-dependence criterion for $B(\mathbf{k})$~\cite{mizoguchi2019}, the dispersive band touches the flat-band manifold where $B(\mathbf{k}) = 0$ on the simple cubic and BCC lattices (along $M$-$R$ and at $H$, respectively), whereas on the FCC lattice $B(\mathbf{k})$ remains non-zero throughout the Brillouin zone, separating the seven flat bands from the dispersive band by a gap of $8t$. Other 3D Bravais lattices can also be similarly shown to generate flat bands on the face-graph hopping Hamiltonian defined herein, except the hexagonal lattice which we discuss next. 

\begin{figure}[p]
    \centering
    \includegraphics[width=0.90\linewidth]{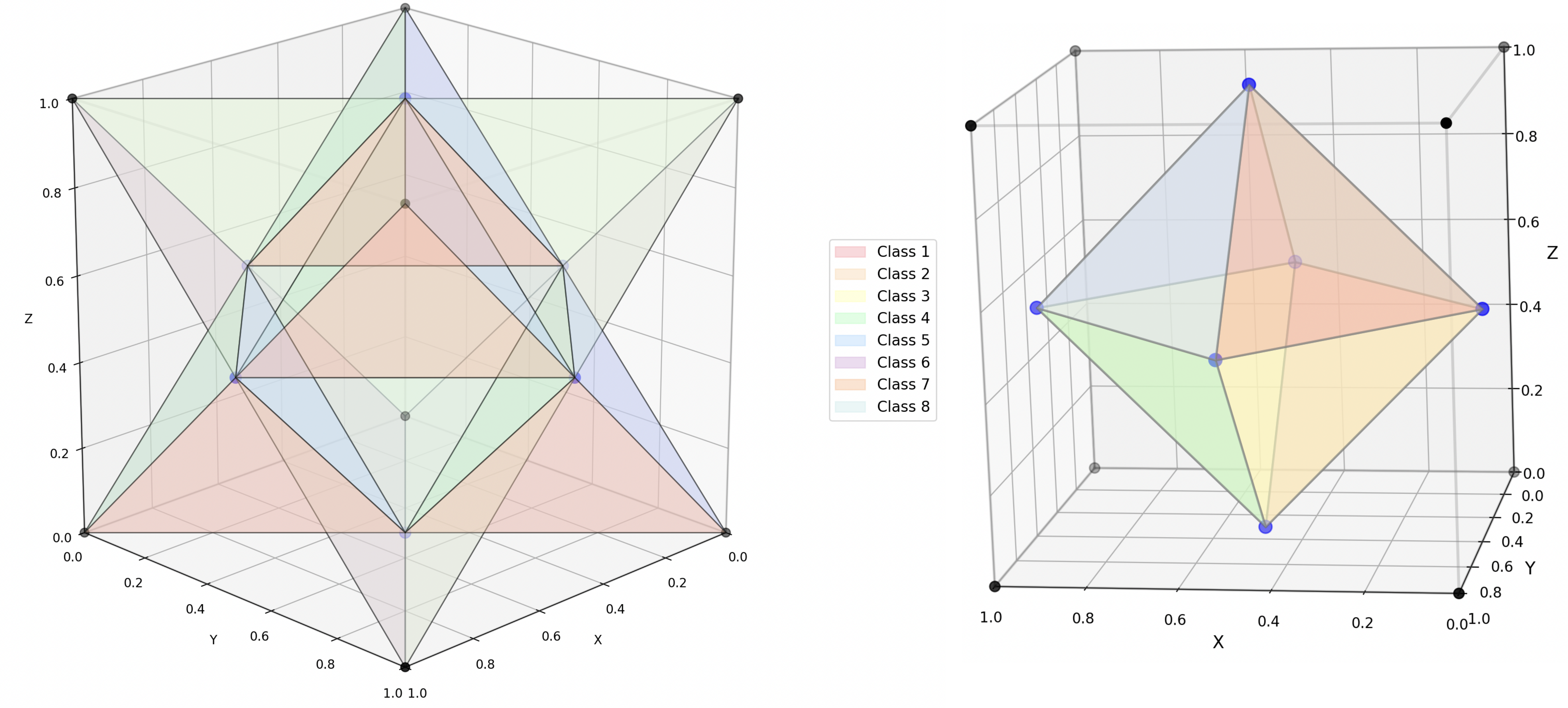}
    \caption{8 distinct classes of faces of an FCC lattice, shown in the conventional unit cell (left) and primitive unit cell (right).}
    \label{fig:FCCFaces}
\end{figure}

\begin{figure}[p]
    \centering
    \includegraphics[width=0.90\linewidth]{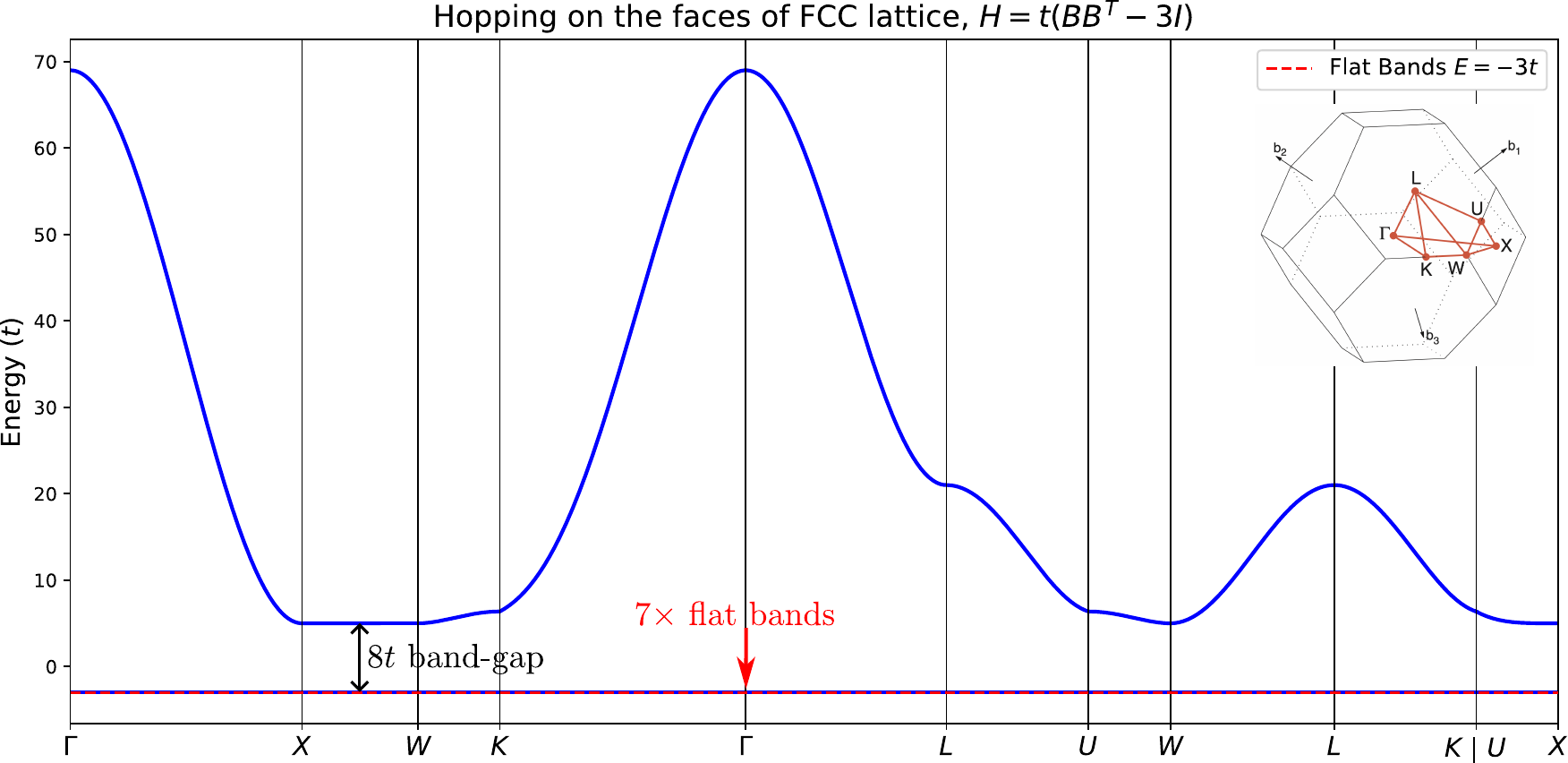}
    \caption{Band structure of the FCC face-graph hopping Hamiltonian $H = t(BB^T - 3I)$ along the path in the inset, showing seven exact flat bands at $E = -3t$ separated from the single dispersive band by a gap of $8t$.}
    \label{fig:FCCBS}
\end{figure}

\begin{figure}[p]
    \centering
    \begin{subfigure}[b]{0.42\linewidth}
        \centering
        \includegraphics[width=\linewidth]{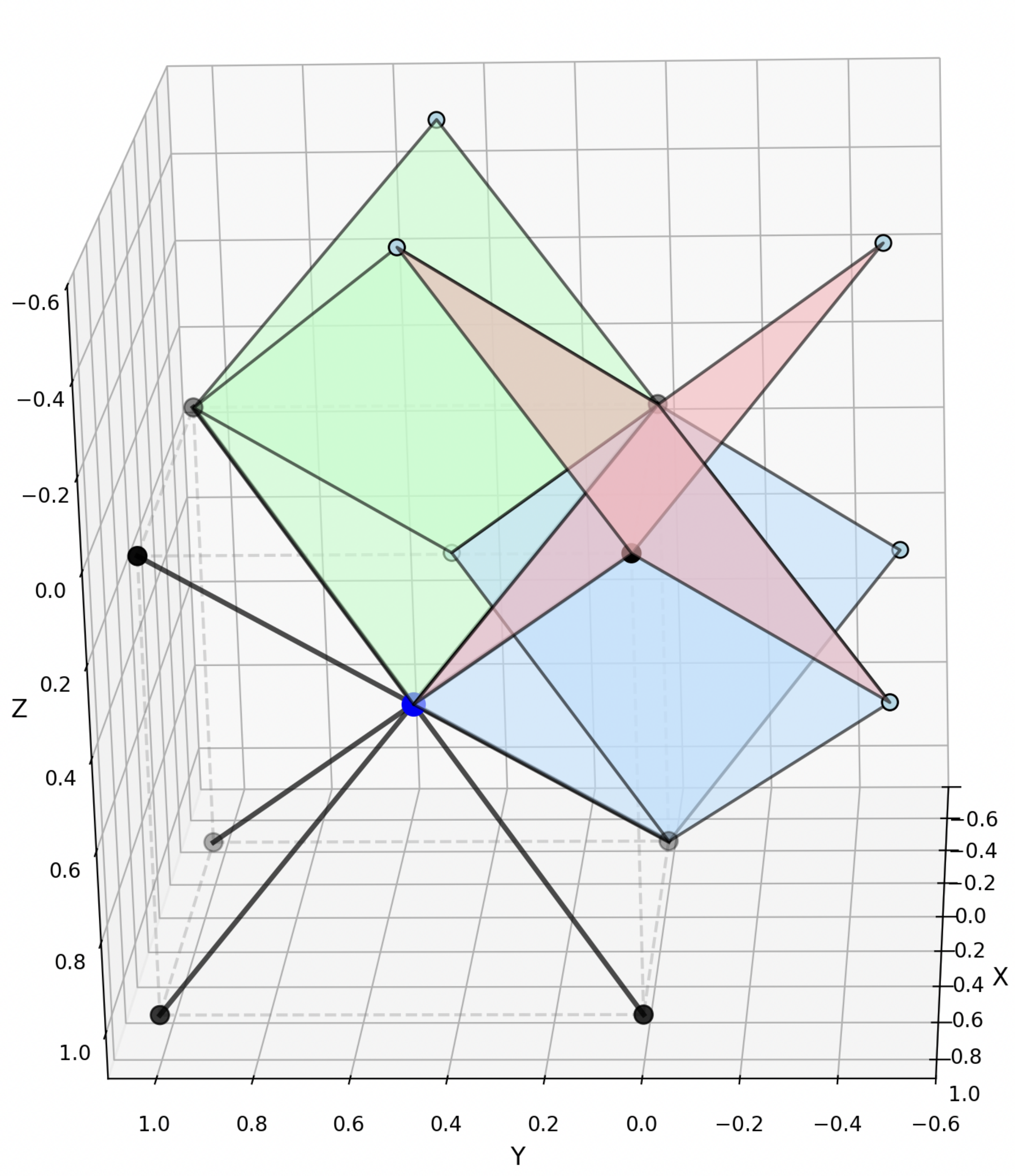}
        \caption{$6$ planar rhombic $4$-cycles}
    \end{subfigure}
    \hfill
    \begin{subfigure}[b]{0.42\linewidth}
        \centering
        \includegraphics[width=\linewidth]{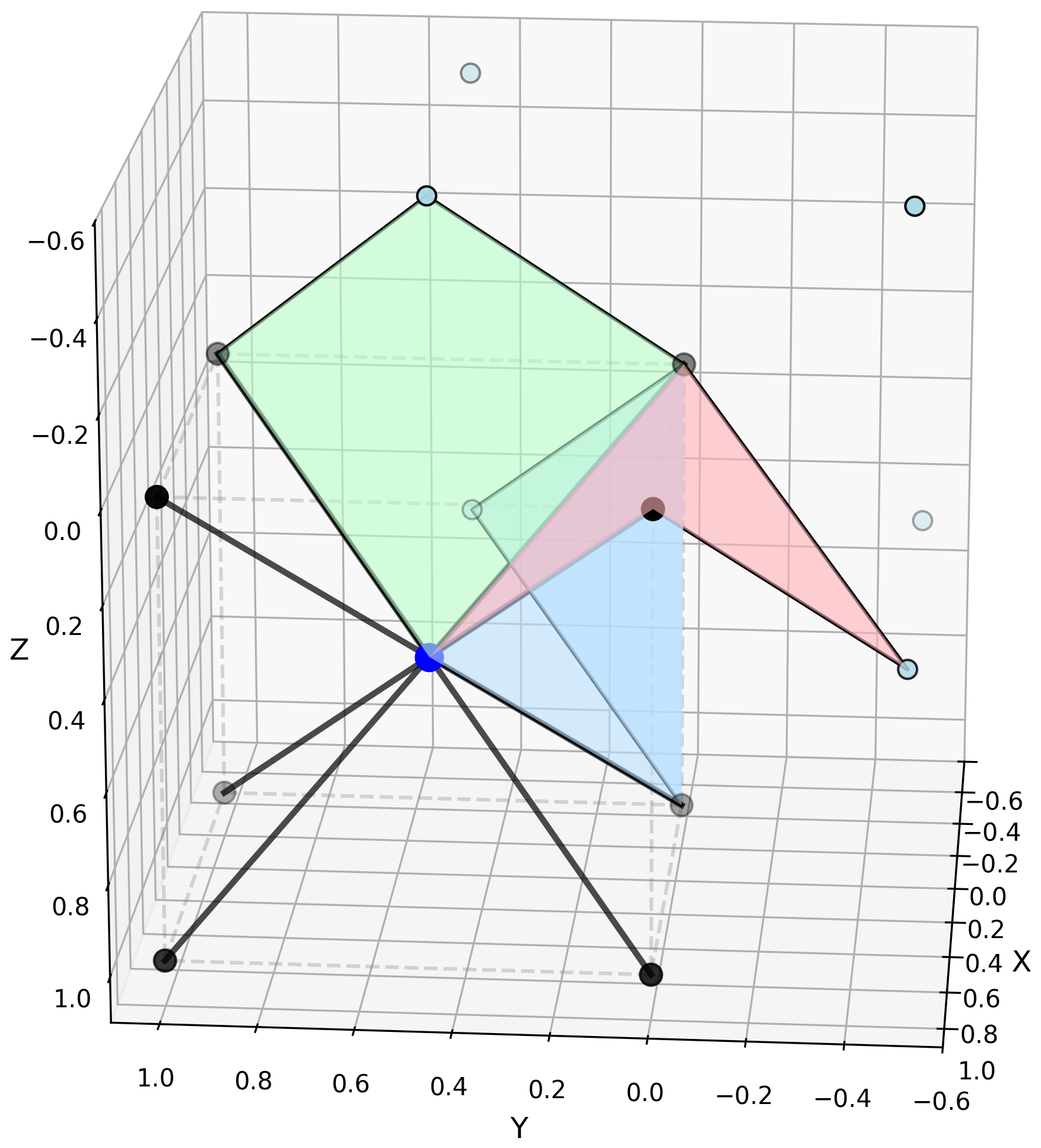}
        \caption{$6$ skew $4$-cycles ($1$ of each pair)}
    \end{subfigure}
    \caption{The $12$ distinct irreducible $4$-cycle classes per primitive unit cell of the BCC nearest-neighbour graph, shown with the conventional cubic cell: (a) the $6$ planar rhombic $4$-cycles and (b) one representative from each of the three pairs of skew non-planar $4$-cycles.}
    \label{fig:BCCFaces}
\end{figure}

\begin{figure}[p]
    \centering
    \includegraphics[width=0.90\linewidth]{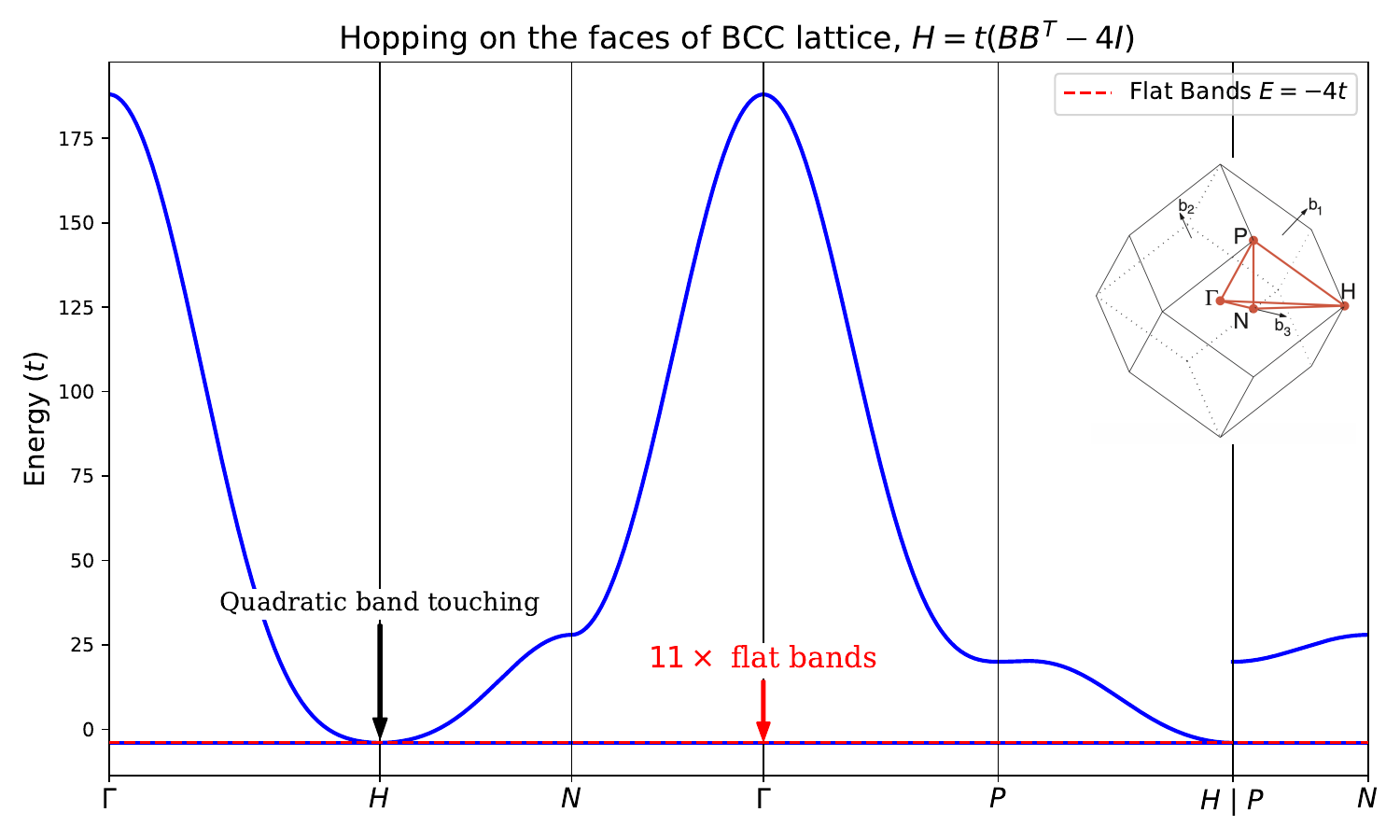}
    \caption{Band structure of the BCC face-graph hopping Hamiltonian $H = t(BB^T - 4I)$ on all $12$ irreducible $4$-cycles per primitive cell along the path in the inset, showing $11$ exact flat bands at $E = -4t$ touching the dispersive band at the $H$ point.}
    \label{fig:BCCBS}
\end{figure}

\section{Tight binding model of the faces of a simple hexagonal lattice}\label{sec:hex}

\begin{figure}[p]
    \centering
    \includegraphics[width=0.28\linewidth]{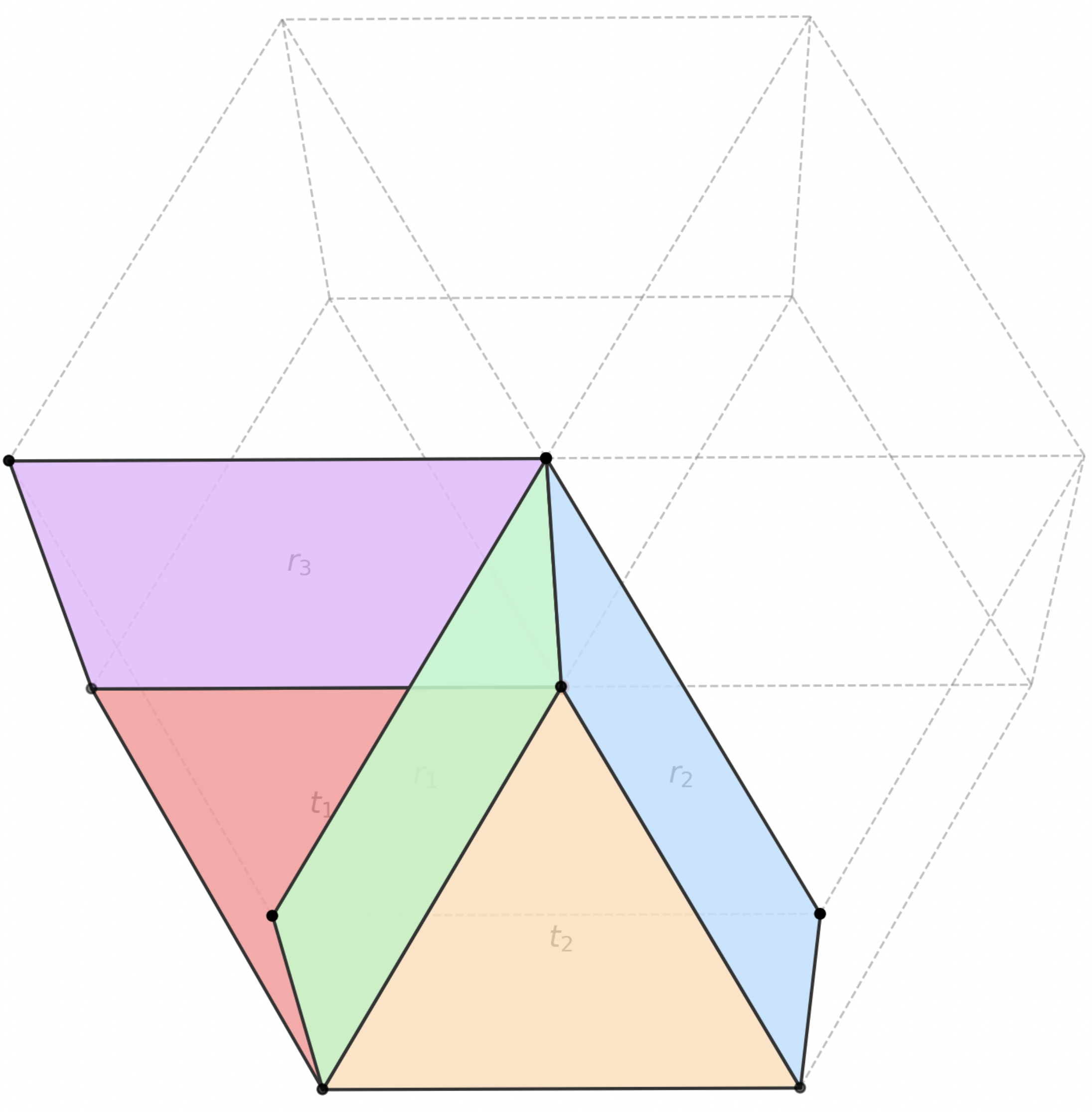}
    \caption{5 distinct faces of a simple hexagonal lattice, shown in the conventional unit cell.}
    \label{fig:HexFaces}
\end{figure}

We consider a three-dimensional simple hexagonal lattice defined by primitive vectors $\mathbf{a}_1 = (1, 0, 0)$, $\mathbf{a}_2 = (1/2, \sqrt{3}/2, 0)$, and $\mathbf{a}_3 = (0, 0, c)$. A single primitive unit cell contains 1 vertex centred at the origin, $\mathbf{r}_v = (0,0,0)$, and 5 elementary faces (see Fig.~\ref{fig:HexFaces}): 2 triangles on the basal planes ($|t_1\rangle, |t_2\rangle$) centred at $\mathbf{r}_{t1} = -\frac{1}{3}(\mathbf{a}_1+\mathbf{a}_2)$ and $\mathbf{r}_{t2} = \frac{1}{3}(\mathbf{a}_1+\mathbf{a}_2)$, and 3 rectangles forming the vertical prism walls ($|r_1\rangle, |r_2\rangle, |r_3\rangle$) centred at $\mathbf{r}_{r1} = \frac{1}{2}(\mathbf{a}_1+\mathbf{a}_3)$, $\mathbf{r}_{r2} = \frac{1}{2}(\mathbf{a}_2+\mathbf{a}_3)$, and $\mathbf{r}_{r3} = \frac{1}{2}(\mathbf{a}_1-\mathbf{a}_2+\mathbf{a}_3)$. The presence of a heterogeneous face set - 3-cycles and 4-cycles - makes this case fundamentally different from the previous cases where all the faces had the same degree.

Consider the real-space Hamiltonian $H = t B B^T$, where $B$ is the full face-vertex incidence matrix. By construction, $H$ has a nullspace of dimension at least $|F|-|V|=4N$ (throughout, we write extensive mode and rank counts to leading order, using $N$ as a proxy for $N + \mathcal{O}(1)$). This Hamiltonian can be separated into a diagonal $D$ equal to $t$ times the number of vertices incident on each face and an off-diagonal part representing hopping on the corner-sharing and edge-sharing face graphs as before:
\begin{equation}
H = D + t(A_{\text{corner}} + 2A_{\text{edge}})\label{eq:mixedH}.
\end{equation}
Because the face set has mixed degree, the diagonal matrix $D$ is not proportional to the identity, which means that the off-diagonal part does not necessarily inherit the full nullspace of $B B^T$. The hopping Hamiltonian can be explicitly partitioned into a homogeneous triangle-to-triangle sector $H_T$, a rectangle-to-rectangle sector $H_R$, and a bipartite cross-term sector $H_{TR}$. 
\begin{align}
H &= D + H_T+H_R+H_{TR}\\
H_T &= t(A^T_{\text{corner}} + 2A^T_{\text{edge}})\\
H_R &= t(A^R_{\text{corner}} + 2A^R_{\text{edge}})\\
H_{TR} &= t(A^{TR}_{\text{corner}} + 2A^{TR}_{\text{edge}}).
\end{align}
The homogeneous sub-graph Hamiltonians $H_T$ and $H_R$ can be identified as having the familiar structure 
\begin{align}
H_T &= t(B_T B_T^T - 3I)\\
H_R &= t(B_R B_R^T - 4I)
\end{align}
where $B_T$ and $B_R$ are the face-vertex incidence matrices of the homogeneous sub-sets of faces with dimensions $2N\times N$ and $3N\times N$ respectively. These incidence matrices mathematically guarantee $N$- and $2N$-fold degenerate macroscopic flat bands at $E=-3t$ and $E=-4t$, respectively. This accounts for 3 of the 4 flat bands of the full Hamiltonian $H = t B B^T$; the remaining flat band has support on both face-sets and originates from destructive interference between them in the full $H$, as analysed in Sec.~\ref{subsec:hetero_nullspace}.

In real space, the $4N$-dimensional flat-band manifold (up to non-contractible states winding around the 3-torus) is generated by lattice translations of the four elementary compact localised states (CLSs) shown in Fig.~\ref{fig:HexCLS}, matching this $1 + 2 + 1$ decomposition. The triangular flat band $\ker B_T^T$ is spanned by the alternating $\pm 1$ six-triangle basal wheel centred on a vertex (Fig.~\ref{fig:HexCLS}(a)). The two rectangular flat bands $\ker B_R^T$ are spanned by two linearly independent alternating $\pm 1$ four-rectangle belts on adjacent rhombic prisms (Fig.~\ref{fig:HexCLS}(b,c)); the third symmetry-related belt is their difference, just like the three cubic belts of Fig.~\ref{fig:cubicCLS}. Finally, the remaining $N$ mixed states are generated by translation of the $24$-face hexagonal-prism CLS centred on a vertical axis $(\mathbf{0}, \mathbf{a}_3)$ (Fig.~\ref{fig:HexCLS}(d)),
\begin{equation}
|\psi_{\text{CLS}}^{\text{mix}}\rangle = 2 \sum_{t \in T_{0} \cup T_{c}} |t\rangle - 2 \sum_{r \in R_{\text{rad}}} |r\rangle - \sum_{r \in R_{\text{per}}} |r\rangle ,
\label{eq:hex_mixed_cls}
\end{equation}
where $T_0$ and $T_c$ are the sets of $6$ basal triangles meeting at $\mathbf{0}$ and $\mathbf{a}_3$, $R_{\text{rad}}$ are the $6$ internal radial rectangles sharing $(\mathbf{0}, \mathbf{a}_3)$, and $R_{\text{per}}$ are the $6$ outer perimeter rectangles of the prism. Neither face class cancels on its own ($B_T^T \psi_T = -B_R^T \psi_R \neq 0$), yet their incident amplitudes cancel identically at all vertices. Note that $|\psi_{\text{CLS}}^{\text{mix}}\rangle$ has a non-zero overlap with rectangle-only zero modes in $\ker B_R^T$.

\begin{figure}[p]
    \centering
    \definecolor{clsplus}{RGB}{255,198,204}
    \definecolor{clsminus}{RGB}{198,228,252}
    \definecolor{clsminusdeep}{RGB}{148,194,240}
    \definecolor{clscancel}{RGB}{20,130,60}
    \tikzset{
        hexproj/.style={x={(0.966cm,-0.122cm)}, y={(0.259cm,0.453cm)}, z={(0cm,0.883cm)}},
        canceled/.style={circle, fill=clscancel, draw=black, line width=0.5pt, inner sep=0pt, minimum size=4.6pt},
        wire/.style={gray!55, dashed, line width=0.5pt}
    }
    \begin{subfigure}[b]{0.31\textwidth}
        \centering
        \begin{tikzpicture}[hexproj, scale=1.32]
            \def\h{1.15}
            \coordinate (O0) at (0,0,0);
            \coordinate (O1) at (0,0,\h);
            \coordinate (b0) at (1,0,0);
            \coordinate (b1) at (0.5,0.866,0);
            \coordinate (b2) at (-0.5,0.866,0);
            \coordinate (b3) at (-1,0,0);
            \coordinate (b4) at (-0.5,-0.866,0);
            \coordinate (b5) at (0.5,-0.866,0);
            \coordinate (t0) at (1,0,\h);
            \coordinate (t1) at (0.5,0.866,\h);
            \coordinate (t2) at (-0.5,0.866,\h);
            \coordinate (t3) at (-1,0,\h);
            \coordinate (t4) at (-0.5,-0.866,\h);
            \coordinate (t5) at (0.5,-0.866,\h);

            \fill[clsplus, fill opacity=0.82] (O0) -- (b0) -- (b1) -- cycle;
            \fill[clsminus, fill opacity=0.82] (O0) -- (b1) -- (b2) -- cycle;
            \fill[clsplus, fill opacity=0.82] (O0) -- (b2) -- (b3) -- cycle;
            \fill[clsminus, fill opacity=0.82] (O0) -- (b3) -- (b4) -- cycle;
            \fill[clsplus, fill opacity=0.82] (O0) -- (b4) -- (b5) -- cycle;
            \fill[clsminus, fill opacity=0.82] (O0) -- (b5) -- (b0) -- cycle;

            \draw[wire] (t0) -- (t1) -- (t2) -- (t3) -- (t4) -- (t5) -- cycle;
            \foreach \k in {0,...,5}{
                \draw[wire] (O1) -- (t\k);
                \draw[wire] (b\k) -- (t\k);
            }
            \draw[wire] (O0) -- (O1);

            \draw[thick] (b0) -- (b1) -- (b2) -- (b3) -- (b4) -- (b5) -- cycle;
            \foreach \k in {0,...,5} \draw[thick] (O0) -- (b\k);

            \node[font=\tiny] at (0.52,0.30,0) {$+1$};
            \node[font=\tiny] at (0.00,0.58,0) {$-1$};
            \node[font=\tiny] at (-0.52,0.30,0) {$+1$};
            \node[font=\tiny] at (-0.52,-0.30,0) {$-1$};
            \node[font=\tiny] at (0.00,-0.58,0) {$+1$};
            \node[font=\tiny] at (0.52,-0.30,0) {$-1$};

            \foreach \p in {O0,b0,b1,b2,b3,b4,b5} \node[canceled] at (\p) {};

            \draw[-{Stealth[length=3.5pt]}, thick] (-1.65,-0.95,0) -- (-1.20,-0.95,0) node[right=-1pt, font=\tiny] {$\mathbf{a}_1$};
            \draw[-{Stealth[length=3.5pt]}, thick] (-1.65,-0.95,0) -- (-1.43,-0.56,0) node[above right=-2pt, font=\tiny] {$\mathbf{a}_2$};
            \draw[-{Stealth[length=3.5pt]}, thick] (-1.65,-0.95,0) -- (-1.65,-0.95,0.48) node[above, font=\tiny] {$\mathbf{a}_3$};
        \end{tikzpicture}
        \caption{CLS of triangles $\subset \ker B_T^T$}
        \label{fig:HexCLS_tri}
    \end{subfigure}
    \hfill
    \begin{subfigure}[b]{0.31\textwidth}
        \centering
        \begin{tikzpicture}[hexproj, scale=1.32]
            \def\h{1.15}
            \coordinate (O0) at (0,0,0);
            \coordinate (O1) at (0,0,\h);
            \coordinate (b0) at (1,0,0);
            \coordinate (b1) at (0.5,0.866,0);
            \coordinate (b2) at (-0.5,0.866,0);
            \coordinate (b3) at (-1,0,0);
            \coordinate (b4) at (-0.5,-0.866,0);
            \coordinate (b5) at (0.5,-0.866,0);
            \coordinate (t0) at (1,0,\h);
            \coordinate (t1) at (0.5,0.866,\h);
            \coordinate (t2) at (-0.5,0.866,\h);
            \coordinate (t3) at (-1,0,\h);
            \coordinate (t4) at (-0.5,-0.866,\h);
            \coordinate (t5) at (0.5,-0.866,\h);

            \draw[wire] (b0) -- (b1) -- (b2) -- (b3) -- (b4) -- (b5) -- cycle;
            \draw[wire] (t0) -- (t1) -- (t2) -- (t3) -- (t4) -- (t5) -- cycle;
            \foreach \k in {0,...,5}{
                \draw[wire] (O0) -- (b\k);
                \draw[wire] (O1) -- (t\k);
                \draw[wire] (b\k) -- (t\k);
            }
            \draw[wire] (O0) -- (O1);

            \fill[clsminus, fill opacity=0.72] (b2) -- (b1) -- (t1) -- (t2) -- cycle;
            \fill[clsplus, fill opacity=0.72] (b3) -- (b2) -- (t2) -- (t3) -- cycle;
            \fill[clsplus, fill opacity=0.72] (O0) -- (b1) -- (t1) -- (O1) -- cycle;
            \fill[clsminus, fill opacity=0.72] (b3) -- (O0) -- (O1) -- (t3) -- cycle;

            \draw[thick] (b2) -- (b1) -- (t1) -- (t2) -- cycle;
            \draw[thick] (b3) -- (b2) -- (t2) -- (t3) -- cycle;
            \draw[thick] (O0) -- (b1) -- (t1) -- (O1) -- cycle;
            \draw[thick] (b3) -- (O0) -- (O1) -- (t3) -- cycle;

            \node[font=\tiny] at (0.0,0.866,0.68) {$-1$};
            \node[font=\tiny] at (-0.75,0.433,0.58) {$+1$};
            \node[font=\tiny] at (0.26,0.45,0.52) {$+1$};
            \node[font=\tiny] at (-0.45,0.0,0.66) {$-1$};

            \foreach \p in {O0,b1,b2,b3,O1,t1,t2,t3} \node[canceled] at (\p) {};
        \end{tikzpicture}
        \caption{CLS 1 of rectangles $\subset \ker B_R^T$}
        \label{fig:HexCLS_rect1}
    \end{subfigure}
    \hfill
    \begin{subfigure}[b]{0.31\textwidth}
        \centering
        \begin{tikzpicture}[hexproj, scale=1.32]
            \def\h{1.15}
            \coordinate (O0) at (0,0,0);
            \coordinate (O1) at (0,0,\h);
            \coordinate (b0) at (1,0,0);
            \coordinate (b1) at (0.5,0.866,0);
            \coordinate (b2) at (-0.5,0.866,0);
            \coordinate (b3) at (-1,0,0);
            \coordinate (b4) at (-0.5,-0.866,0);
            \coordinate (b5) at (0.5,-0.866,0);
            \coordinate (t0) at (1,0,\h);
            \coordinate (t1) at (0.5,0.866,\h);
            \coordinate (t2) at (-0.5,0.866,\h);
            \coordinate (t3) at (-1,0,\h);
            \coordinate (t4) at (-0.5,-0.866,\h);
            \coordinate (t5) at (0.5,-0.866,\h);

            \draw[wire] (b0) -- (b1) -- (b2) -- (b3) -- (b4) -- (b5) -- cycle;
            \draw[wire] (t0) -- (t1) -- (t2) -- (t3) -- (t4) -- (t5) -- cycle;
            \foreach \k in {0,...,5}{
                \draw[wire] (O0) -- (b\k);
                \draw[wire] (O1) -- (t\k);
                \draw[wire] (b\k) -- (t\k);
            }
            \draw[wire] (O0) -- (O1);

            \fill[clsplus, fill opacity=0.72] (O0) -- (b1) -- (t1) -- (O1) -- cycle;
            \fill[clsminus, fill opacity=0.72] (b1) -- (b0) -- (t0) -- (t1) -- cycle;
            \fill[clsminus, fill opacity=0.72] (O0) -- (b5) -- (t5) -- (O1) -- cycle;
            \fill[clsplus, fill opacity=0.72] (b5) -- (b0) -- (t0) -- (t5) -- cycle;

            \draw[thick] (O0) -- (b1) -- (t1) -- (O1) -- cycle;
            \draw[thick] (b1) -- (b0) -- (t0) -- (t1) -- cycle;
            \draw[thick] (O0) -- (b5) -- (t5) -- (O1) -- cycle;
            \draw[thick] (b5) -- (b0) -- (t0) -- (t5) -- cycle;

            \node[font=\tiny] at (0.24,0.44,0.68) {$+1$};
            \node[font=\tiny] at (0.76,0.44,0.58) {$-1$};
            \node[font=\tiny] at (0.24,-0.44,0.46) {$-1$};
            \node[font=\tiny] at (0.76,-0.44,0.56) {$+1$};

            \foreach \p in {O0,b0,b1,b5,O1,t0,t1,t5} \node[canceled] at (\p) {};
        \end{tikzpicture}
        \caption{CLS 2 of rectangles $\subset \ker B_R^T$}
        \label{fig:HexCLS_rect2}
    \end{subfigure}

    \vspace{0.6em}
    \begin{subfigure}[b]{0.96\textwidth}
        \centering
        \begin{tikzpicture}
            \begin{scope}[hexproj, scale=1.42, shift={(0,0,0)}]
                \def\h{1.10}
                \def\dz{0.34}
                \coordinate (O0) at (0,0,0);
                \coordinate (O1) at (0,0,\h);
                \coordinate (Ob) at (0,0,-\dz);
                \coordinate (Ot) at (0,0,\h+\dz);

                \coordinate (b0) at (1,0,0);       \coordinate (b1) at (0.5,0.866,0);
                \coordinate (b2) at (-0.5,0.866,0);\coordinate (b3) at (-1,0,0);
                \coordinate (b4) at (-0.5,-0.866,0);\coordinate (b5) at (0.5,-0.866,0);

                \coordinate (t0) at (1,0,\h);       \coordinate (t1) at (0.5,0.866,\h);
                \coordinate (t2) at (-0.5,0.866,\h);\coordinate (t3) at (-1,0,\h);
                \coordinate (t4) at (-0.5,-0.866,\h);\coordinate (t5) at (0.5,-0.866,\h);

                \coordinate (cb0) at (1,0,-\dz);       \coordinate (cb1) at (0.5,0.866,-\dz);
                \coordinate (cb2) at (-0.5,0.866,-\dz);\coordinate (cb3) at (-1,0,-\dz);
                \coordinate (cb4) at (-0.5,-0.866,-\dz);\coordinate (cb5) at (0.5,-0.866,-\dz);

                \coordinate (ct0) at (1,0,\h+\dz);       \coordinate (ct1) at (0.5,0.866,\h+\dz);
                \coordinate (ct2) at (-0.5,0.866,\h+\dz);\coordinate (ct3) at (-1,0,\h+\dz);
                \coordinate (ct4) at (-0.5,-0.866,\h+\dz);\coordinate (ct5) at (0.5,-0.866,\h+\dz);

                \foreach \a/\b in {0/1,1/2,2/3,3/4,4/5,5/0}{
                    \fill[clsplus, fill opacity=0.78] (Ob) -- (cb\a) -- (cb\b) -- cycle;
                    \draw[thick] (Ob) -- (cb\a) -- (cb\b) -- cycle;
                }
                \foreach \k in {0,...,5} \draw[gray!50, densely dotted, line width=0.5pt] (cb\k) -- (b\k);
                \draw[gray!50, densely dotted, line width=0.5pt] (Ob) -- (O0);

                \fill[clsminus, fill opacity=0.52] (b2) -- (b1) -- (t1) -- (t2) -- cycle;
                \fill[clsminus, fill opacity=0.52] (b3) -- (b2) -- (t2) -- (t3) -- cycle;
                \fill[clsminus, fill opacity=0.52] (b1) -- (b0) -- (t0) -- (t1) -- cycle;
                \draw[thick] (b2) -- (b1) -- (t1) -- (t2) -- cycle;
                \draw[thick] (b3) -- (b2) -- (t2) -- (t3) -- cycle;
                \draw[thick] (b1) -- (b0) -- (t0) -- (t1) -- cycle;

                \foreach \k in {2,1,3,0,4,5}{
                    \fill[clsminusdeep, fill opacity=0.68] (O0) -- (b\k) -- (t\k) -- (O1) -- cycle;
                    \draw[thick] (O0) -- (b\k) -- (t\k) -- (O1) -- cycle;
                }

                \fill[clsminus, fill opacity=0.52] (b3) -- (b4) -- (t4) -- (t3) -- cycle;
                \fill[clsminus, fill opacity=0.52] (b5) -- (b0) -- (t0) -- (t5) -- cycle;
                \fill[clsminus, fill opacity=0.52] (b4) -- (b5) -- (t5) -- (t4) -- cycle;
                \draw[thick] (b3) -- (b4) -- (t4) -- (t3) -- cycle;
                \draw[thick] (b5) -- (b0) -- (t0) -- (t5) -- cycle;
                \draw[thick] (b4) -- (b5) -- (t5) -- (t4) -- cycle;

                \foreach \p in {O0,b0,b1,b2,b3,b4,b5,O1,t0,t1,t2,t3,t4,t5} \node[canceled] at (\p) {};

                \foreach \k in {0,...,5} \draw[gray!50, densely dotted, line width=0.5pt] (t\k) -- (ct\k);
                \draw[gray!50, densely dotted, line width=0.5pt] (O1) -- (Ot);
                \foreach \a/\b in {0/1,1/2,2/3,3/4,4/5,5/0}{
                    \fill[clsplus, fill opacity=0.78] (Ot) -- (ct\a) -- (ct\b) -- cycle;
                    \draw[thick] (Ot) -- (ct\a) -- (ct\b) -- cycle;
                }

                \node[font=\scriptsize, anchor=west] (lblTc) at (1.28, 0.35, \h+\dz+0.05) {$T_c\ (+2)$};
                \draw[-{Stealth[length=4pt]}, line width=0.6pt] (lblTc.west) -- (0.65, 0.25, \h+\dz);

                \node[font=\scriptsize, anchor=west] (lblRrad) at (1.28, 0.20, 0.62*\h) {$R_{\text{rad}}\ (-2)$};
                \draw[-{Stealth[length=4pt]}, line width=0.6pt] (lblRrad.west) -- (0.55, 0.0, 0.58*\h);

                \node[font=\scriptsize, anchor=west] (lblRper) at (1.28, -0.25, 0.22*\h) {$R_{\text{per}}\ (-1)$};
                \draw[-{Stealth[length=4pt]}, line width=0.6pt] (lblRper.west) -- (0.76, -0.42, 0.28*\h);

                \node[font=\scriptsize, anchor=west] (lblT0) at (1.28, -0.25, -\dz-0.02) {$T_0\ (+2)$};
                \draw[-{Stealth[length=4pt]}, line width=0.6pt] (lblT0.west) -- (0.65, -0.28, -\dz);
            \end{scope}

            \begin{scope}[shift={(6.4, 0.80)}, scale=1.32]
                \coordinate (O) at (0,0);
                \foreach \k in {0,...,5} \coordinate (R\k) at ({cos(60*\k)},{sin(60*\k)});

                \fill[clsplus, fill opacity=0.55] (R0) -- (R1) -- (R2) -- (R3) -- (R4) -- (R5) -- cycle;
                \foreach \k in {0,...,5}{
                    \pgfmathtruncatemacro{\n}{mod(\k+1,6)}
                    \draw[clsminus, line width=3.0pt] (R\k) -- (R\n);
                    \draw[clsminusdeep, line width=3.0pt] (O) -- (R\k);
                }
                \foreach \k in {0,...,5} \node[canceled] at (R\k) {};
                \node[canceled] at (O) {};

                \node[font=\scriptsize, text=red!55!black] at (0.48,0.30) {$+2$};
                \node[font=\scriptsize, text=blue!55!black, fill=clsplus!55!white, inner sep=0.8pt] at (0.52,0.00) {$-2$};
                \node[font=\scriptsize, text=blue!45!black, anchor=south west] at (0.74,0.44) {$-1$};

                \node[font=\scriptsize, align=left, anchor=north] at (0,-1.18) {
                    central axis: $6(+2) + 6(-2) = 0$\\
                    rim vertices: $2(+2) + 1(-2) + 2(-1) = 0$
                };
            \end{scope}
        \end{tikzpicture}
        \caption{Mixed-sector CLS $|\psi_{\text{CLS}}^{\text{mix}}\rangle$ ($24$ faces): separated 3D prism (left) and top-down projection along $\mathbf{a}_3$ (right)}
        \label{fig:HexCLS_mix}
    \end{subfigure}
    \caption{The four elementary compact localised states (CLSs) whose lattice translations generate the $4N$-dimensional flat-band nullspace of the simple hexagonal face-graph. Light red shading denotes positive amplitudes ($+1$ in (a)-(c), $+2$ in (d)); pale and medium pastel blue denote negative amplitudes ($-1$ and $-2$, respectively); green circles mark vertices where the incident amplitudes cancel to zero. (a) Six-triangle alternating basal wheel spanning $\ker B_T^T$. (b,c) Two linearly independent four-rectangle belts spanning $\ker B_R^T$. (d) Twenty-four-face mixed hexagonal-prism state $|\psi_{\text{CLS}}^{\text{mix}}\rangle$ of Eq.~\eqref{eq:hex_mixed_cls}, where $12$ basal triangles ($+2$) cancel against $6$ internal radial rectangles ($-2$) and $6$ outer perimeter rectangles ($-1$).}
    \label{fig:HexCLS}
\end{figure}

\subsection{Explicit momentum-space Hamiltonian and flat bands}\label{subsec:hexTB_momentum}

The momentum-space basis functions are explicitly defined as the Bloch wave functions of the orbitals supported on each of the five elementary faces: $(|t_1\rangle, |t_2\rangle, |r_1\rangle, |r_2\rangle, |r_3\rangle)$. In this basis, we construct the momentum-space face-vertex incidence vector $B(\mathbf{k})$ using a symmetric gauge, as in Sec.~\ref{sec:Bmat}. Defining $k_i = \mathbf{k} \cdot \mathbf{a}_i$, the symmetric Fourier sums over the vertices of each face relative to its centre yield:
\begin{align}
\tilde{\gamma}_{t1}(\mathbf{k}) &\equiv \Delta(\mathbf{k}) = e^{i\frac{k_1+k_2}{3}} + e^{i\frac{-2k_1+k_2}{3}} + e^{i\frac{k_1-2k_2}{3}} \\
\tilde{\gamma}_{t2}(\mathbf{k}) &= \Delta^*(\mathbf{k}) \\
\tilde{\gamma}_{r1}(\mathbf{k}) &= 4 \cos\left(\frac{k_1}{2}\right) \cos\left(\frac{k_3}{2}\right) \\
\tilde{\gamma}_{r2}(\mathbf{k}) &= 4 \cos\left(\frac{k_2}{2}\right) \cos\left(\frac{k_3}{2}\right) \\
\tilde{\gamma}_{r3}(\mathbf{k}) &= 4 \cos\left(\frac{k_1 - k_2}{2}\right) \cos\left(\frac{k_3}{2}\right)
\end{align}

We partition the momentum-space incidence vector into two sub-vectors, $B_T(\mathbf{k})$ and $B_R(\mathbf{k})$, corresponding to the triangular ($p=3$) and rectangular ($q=4$) faces respectively:
\begin{equation}
B(\mathbf{k}) = \begin{bmatrix} B_T(\mathbf{k}) \\ B_R(\mathbf{k}) \end{bmatrix} = \left[\begin{array}{cc|ccc} \Delta & \Delta^* & \tilde{\gamma}_{r1} & \tilde{\gamma}_{r2} & \tilde{\gamma}_{r3} \end{array}\right]^T.
\end{equation}

\begin{figure}[!htbp]
    \centering
    \includegraphics[width=0.92\linewidth]{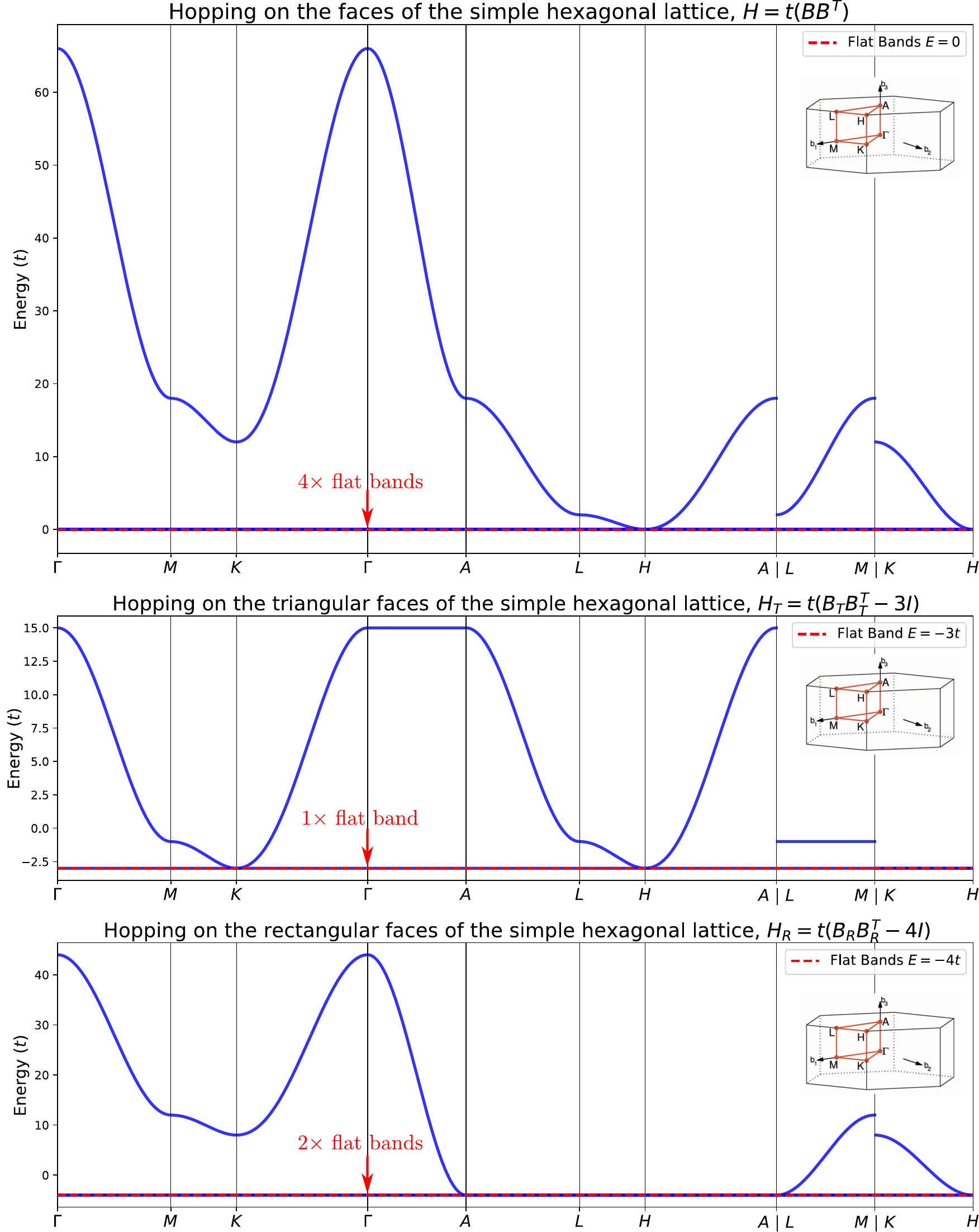}
    \caption{Band structure of the simple hexagonal face-graph Hamiltonian $H = tBB^T$ (together with its homogeneous triangular and rectangular sectors) along $\Gamma$-$M$-$K$-$\Gamma$-$A$-$L$-$H$-$A\mid L$-$M\mid K$-$H$, showing four exact flat bands at $E = 0$ touching the dispersive band at $H$ (and $H'$).}
    \label{fig:HexBS}
\end{figure}

The momentum-space hopping Hamiltonian is constructed as the outer product $H(\mathbf{k}) = t B(\mathbf{k}) B^\dagger(\mathbf{k})$. Utilising the partitioned incidence vectors $B_T(\mathbf{k})$ and $B_R(\mathbf{k})$, the Hamiltonian naturally inherits the block structure established in real space:
\begin{equation}
H(\mathbf{k}) = t \begin{pmatrix} B_T(\mathbf{k}) B_T^\dagger(\mathbf{k}) & B_T(\mathbf{k})B_R^\dagger(\mathbf{k})  \\ B_R(\mathbf{k}) B_T^\dagger(\mathbf{k}) & B_R(\mathbf{k}) B_R^\dagger(\mathbf{k}) \end{pmatrix}.
\end{equation}
This block matrix explicitly demonstrates the Fourier-space counterparts of the homogeneous and bipartite sectors. The diagonal blocks correspond to the homogeneous face-graph Hamiltonians shifted by their respective degrees:
\begin{align}
H_T(\mathbf{k}) &= t(B_T(\mathbf{k}) B_T^\dagger(\mathbf{k}) - 3I_2) = t(A^T_{\text{corner}}(\mathbf{k}) + 2A^T_{\text{edge}}(\mathbf{k})), \\
H_R(\mathbf{k}) &= t(B_R(\mathbf{k}) B_R^\dagger(\mathbf{k}) - 4I_3) = t(A^R_{\text{corner}}(\mathbf{k}) + 2A^R_{\text{edge}}(\mathbf{k})),
\end{align}
while the off-diagonal blocks $t B_T(\mathbf{k}) B_R^\dagger(\mathbf{k})$ and $t B_R(\mathbf{k}) B_T^\dagger(\mathbf{k})$ directly correspond to the bipartite cross-terms $H_{TR}(\mathbf{k})$.

Expanding $H(\mathbf{k})$ completely yields a $5 \times 5$ Hermitian matrix:
\begin{equation}
H(\mathbf{k}) = t
\begin{pmatrix}
|\Delta|^2 & \Delta^2 & \Delta \tilde{\gamma}_{r1} & \Delta \tilde{\gamma}_{r2} & \Delta \tilde{\gamma}_{r3} \\
(\Delta^*)^2 & |\Delta|^2 & \Delta^* \tilde{\gamma}_{r1} & \Delta^* \tilde{\gamma}_{r2} & \Delta^* \tilde{\gamma}_{r3} \\
\Delta^* \tilde{\gamma}_{r1} & \Delta \tilde{\gamma}_{r1} & \tilde{\gamma}_{r1}^2 & \tilde{\gamma}_{r1} \tilde{\gamma}_{r2} & \tilde{\gamma}_{r1} \tilde{\gamma}_{r3} \\
\Delta^* \tilde{\gamma}_{r2} & \Delta \tilde{\gamma}_{r2} & \tilde{\gamma}_{r1} \tilde{\gamma}_{r2} & \tilde{\gamma}_{r2}^2 & \tilde{\gamma}_{r2} \tilde{\gamma}_{r3} \\
\Delta^* \tilde{\gamma}_{r3} & \Delta \tilde{\gamma}_{r3} & \tilde{\gamma}_{r1} \tilde{\gamma}_{r3} & \tilde{\gamma}_{r2} \tilde{\gamma}_{r3} & \tilde{\gamma}_{r3}^2
\end{pmatrix}
\end{equation}
Because $H(\mathbf{k})$ is the outer product of a single vector $B(\mathbf{k})$, it has rank 1 at generic $\mathbf{k}$. By the rank-nullity theorem, at generic momentum $\mathbf{k}$, the matrix possesses $5 - 1 = 4$ zero eigenvalues and one dispersive band given by the trace,
\[
E(\mathbf{k}) = \text{Tr}[H(\mathbf{k})] = 2t \left[ 9 + 4 C_{\mathbf{k}} + 2 \left( 3 + C_{\mathbf{k}} \right) \cos(k_3) \right],
\]
where $C_{\mathbf{k}} = \cos(k_1) + \cos(k_2) + \cos(k_1 - k_2)$.
This guarantees 4 flat bands at $E=0$, touching the dispersive band at the $H$ and $H'$ points of the Brillouin zone where $B(\mathbf{k}) = 0$~\cite{mizoguchi2019} (see Fig.~\ref{fig:HexBS}).

As established in the real-space discussion, the homogeneous blocks $H_T(\mathbf{k})$ and $H_R(\mathbf{k})$ (whose zero modes span $\ker B_T^T$ and $\ker B_R^T$) account for $1$ and $2$ of the $4$ flat bands of $H(\mathbf{k})$, respectively. The remaining fourth flat band belongs to the mixed nullspace of the full Hamiltonian $H$: its eigenstates $\Psi = (\psi_T, \psi_R)^T$ satisfy $B_T^T \psi_T = -B_R^T \psi_R \neq 0$, originating from destructive interference between non-vanishing triangle and rectangle incidences at every vertex, as derived generally in Sec.~\ref{subsec:hetero_nullspace}.

\section{Generalisation to arbitrary heterogeneous face-graphs}\label{sec:hetero}

The construction demonstrated for the simple hexagonal lattice in Sec.~\ref{sec:hex} applies to any cell complex containing faces of mixed degrees. Consider an arbitrary graph whose heterogeneous face set $F$ has two disjoint homogeneous subsets according to cycle length: $p$-cycle faces ($F_p$) and $q$-cycle faces ($F_q$), with $p \neq q$ and $|F| = |F_p| + |F_q|$. Ordering the face basis into these two sectors splits the $|F| \times |V|$ face-vertex incidence matrix $B$ into vertical blocks,
\begin{equation}
B = \begin{pmatrix} B_p \\ B_q \end{pmatrix},
\end{equation}
where $B_p$ ($|F_p| \times |V|$) and $B_q$ ($|F_q| \times |V|$) are the incidence matrices of $F_p$ and $F_q$, respectively. For $t \neq 0$, the Hamiltonian $H = t B B^T$ decomposes into a block-diagonal intra-sector operator $H_{\text{diag}}$ and an off-diagonal bipartite inter-sector coupling $H_{pq}$:
\begin{equation}
H = t B B^T = t
\begin{pmatrix}
B_p B_p^T & B_p B_q^T \\
B_q B_p^T & B_q B_q^T
\end{pmatrix}
= H_{\text{diag}} + H_{pq},
\label{eq:hetero_blockH}
\end{equation}
where
\begin{equation}
H_{\text{diag}} = t \begin{pmatrix} B_p B_p^T & 0 \\ 0 & B_q B_q^T \end{pmatrix}, \qquad
H_{pq} = t \begin{pmatrix} 0 & B_p B_q^T \\ B_q B_p^T & 0 \end{pmatrix}.
\end{equation}
The diagonal entries of $H$ form a heterogeneous degree matrix
\begin{equation}
D = t \begin{pmatrix} p I_{|F_p|} & 0 \\ 0 & q I_{|F_q|} \end{pmatrix}.
\end{equation}
Subtracting $D$ leaves the pure off-diagonal hopping operator $H - D = t \sum_{k \ge 1} k A^{(k)}$ (Section~\ref{sec:summary}),\footnote{This reduces to $H - D = t(A_{\text{corner}} + 2 A_{\text{edge}})$ whenever any two distinct faces share at most one vertex or one edge.} whose diagonal blocks are the homogeneous face-graph hopping Hamiltonians $H_p = t(B_p B_p^T - p I_{|F_p|})$ and $H_q = t(B_q B_q^T - q I_{|F_q|})$, while $H_{pq}$ encodes the hopping between $F_p$ and $F_q$. Because $D$ is not proportional to $I_{|F|}$ when $p \neq q$, the zero modes of $H$ do not in general remain eigenstates of $H - D$; nevertheless, as shown in Section~\ref{sec:summary}, $H$ retains a nullspace of dimension at least $|F| - |V|$, forming exact flat bands at $E = 0$ on a periodic lattice.

\subsection{Nullspace decomposition for arbitrary face combinations}\label{subsec:hetero_nullspace}

Writing the face Hilbert space as $\mathbb{C}^{|F|} = \mathbb{C}^{|F_p|} \oplus \mathbb{C}^{|F_q|}$ and a general state as $\Psi = (\psi_p, \psi_q)^T$ with $\psi_p \in \mathbb{C}^{|F_p|}$ and $\psi_q \in \mathbb{C}^{|F_q|}$, the expectation value of $H$ with $B^T = \begin{bmatrix} B_p^T & B_q^T \end{bmatrix}$ is
\begin{equation}
\Psi^\dagger H \Psi = t \|B^T \Psi\|^2 = t \|B_p^T \psi_p + B_q^T \psi_q\|^2.
\label{eq:hetero_norm}
\end{equation}
As in Eq.~\eqref{eq:summary_interference}, for any $t \neq 0$ a state satisfies $H\Psi = 0$ if and only if $B^T \Psi = 0$, so $\ker(H) = \ker(B^T)$ has dimension
\begin{equation}
\mathcal{N}_{\text{total}} \equiv \dim\ker(B^T) = |F| - \text{rank}(B^T).
\end{equation}

To classify all zero modes of $H$, we first isolate the intra-sector nullspace $\mathcal{K}_0 \equiv \ker(B_p^T) \oplus \ker(B_q^T)$, consisting of states $(\psi_p, \psi_q)^T$ satisfying $B_p^T \psi_p = 0$ and $B_q^T \psi_q = 0$. At any vertex $v \in V$, the component $(B_p^T \psi_p)_v = \sum_{f \in F_p,\, v \in f} \psi_{p,f}$ sums the wavefunction amplitudes $\psi_{p,f}$ over all $p$-faces incident on $v$; the condition $B_p^T \psi_p = 0$ therefore expresses exact destructive interference at every vertex within $F_p$ alone, and likewise for $B_q^T \psi_q = 0$ within $F_q$. Acting with $H_{\text{diag}}$ and $H_{pq}$ separately on any $\Psi \in \mathcal{K}_0$ yields $H_{\text{diag}}\Psi = 0$ and $H_{pq}\Psi = 0$, so $\mathcal{K}_0 \subseteq \ker(B^T)$. Every zero mode of the isolated homogeneous blocks $B_p B_p^T$ (corresponding to flat bands of $H_p$ at $E = -pt$) and $B_q B_q^T$ (corresponding to flat bands of $H_q$ at $E = -qt$) therefore survives unchanged in $H$, supplying $\dim(\mathcal{K}_0) = \mathcal{N}_p + \mathcal{N}_q$ linearly independent intra-sector zero modes, where
\begin{equation}
\mathcal{N}_p \equiv \dim\ker(B_p^T) = |F_p| - \text{rank}(B_p^T), \qquad
\mathcal{N}_q \equiv \dim\ker(B_q^T) = |F_q| - \text{rank}(B_q^T).
\end{equation}

The remaining zero modes span the orthogonal complement $\mathcal{K}_{\text{mix}} \equiv \ker(B^T) \cap \mathcal{K}_0^\perp$, where $\psi_p \in (\ker B_p^T)^\perp$ and $\psi_q \in (\ker B_q^T)^\perp$. By Eq.~\eqref{eq:hetero_norm}, any non-zero state $\Psi \in \mathcal{K}_{\text{mix}}$ satisfies $B_p^T \psi_p = -B_q^T \psi_q \neq 0$: neither $H_{\text{diag}}\Psi$ nor $H_{pq}\Psi$ vanishes on its own, yet at every vertex the non-zero incidence from $F_p$ cancels identically against the incidence from $F_q$ so that $(H_{\text{diag}} + H_{pq})\Psi = 0$. Because $B_p^T$ and $B_q^T$ are injective on $(\ker B_p^T)^\perp$ and $(\ker B_q^T)^\perp$, each shared vertex-incidence vector $w = B_p^T \psi_p = -B_q^T \psi_q \in \text{col}(B_p^T) \cap \text{col}(B_q^T)$ uniquely determines $(\psi_p, \psi_q)$, supplying an inter-sector count of $\dim\big(\text{col}(B_p^T) \cap \text{col}(B_q^T)\big)$ mixed zero modes.

We can now verify that the three contributions sum to the total nullity of $H$. Since $B^T = \begin{bmatrix} B_p^T & B_q^T \end{bmatrix}$, its column space is the subspace sum $\text{col}(B^T) = \text{col}(B_p^T) + \text{col}(B_q^T)$, whose dimension satisfies
\begin{equation}
\text{rank}(B^T) = \text{rank}(B_p^T) + \text{rank}(B_q^T) - \dim\big(\text{col}(B_p^T) \cap \text{col}(B_q^T)\big).
\end{equation}
Adding the two intra-sector counts and the inter-sector count gives
\begin{align}
\mathcal{N}_{\text{total}} &= \mathcal{N}_p + \mathcal{N}_q + \dim\big(\text{col}(B_p^T) \cap \text{col}(B_q^T)\big) \nonumber \\
&= \big(|F_p| - \text{rank}(B_p^T)\big) + \big(|F_q| - \text{rank}(B_q^T)\big) + \dim\big(\text{col}(B_p^T) \cap \text{col}(B_q^T)\big) \nonumber \\
&= |F_p| + |F_q| - \text{rank}(B^T) = |F| - \text{rank}(B^T) \ge |F| - |V|,
\label{eq:nullity_sum}
\end{align}
recovering $\mathcal{N}_{\text{total}} = N + 2N + N = 4N$ for the simple hexagonal lattice of Sec.~\ref{sec:hex} ($\text{rank}(B_T^T) = \text{rank}(B_R^T) = \text{rank}(B^T) = N$).

The same argument extends directly to an arbitrary heterogeneous face set partitioned into $M$ degree classes $\{F_{p_1}, F_{p_2}, \ldots, F_{p_M}\}$ with $B^T = \begin{bmatrix} B_1^T & B_2^T & \cdots & B_M^T \end{bmatrix}$. The zero-mode condition becomes $\sum_{i=1}^M B_i^T \psi_i = 0$, the homogeneous sectors contribute $\sum_{i=1}^M \mathcal{N}_i = \sum_{i=1}^M (|F_{p_i}| - \text{rank}(B_i^T))$ intra-sector modes, and the inter-sector cancellation supplies $\sum_{i=1}^M \text{rank}(B_i^T) - \text{rank}(B^T)$ additional mixed modes, which again sum to $|F| - \text{rank}(B^T) \ge |F| - |V|$. Together, the homogeneous and heterogeneous constructions enumerate a countable infinity of exact flat-band Hamiltonians arising from destructive interference of hopping on the faces of any cell complex, with or without discrete translation invariance.

\section{Topological protection and disorder robustness of zero modes}\label{sec:topo}

In Sections~\ref{sec:faces}-\ref{sec:hetero}, we focused on face-graphs ($X = F$) to demonstrate explicit flat bands and compact localised states in the simplest extension of the familiar line-graphs. We now return to the general setting of an arbitrary hypergraph $G(V,X)$ introduced in Section~\ref{sec:summary}, where the cells $X$ may represent edges ($X = E$), faces ($X = F$), or higher-order hyperedges on any set of vertices $V$, with or without translation symmetry. The goal of this section is to show how connecting the dimension of the degenerate eigenspace to the Euler characteristic imbues it with a topological protection against certain kinds of disorder. We establish that the number of zero-modes and CLS is rigorously determined by the topology of graph connectivity~\cite{sutherland1986}. 

\subsection{Dual discrete curvature and the weighted incidence index}\label{subsec:topo_index}

Recall from Section~\ref{subsec:dual_curvature} that for any finite hypergraph $G(V,X)$ (with $d(v) \ge 1$ cells covering each vertex $v \in V$), the dual hypergraph $\widetilde{G}(X,V)$ interchanges cells and vertices, assigning to each cell $x \in X$ the dual discrete curvature of Eq.~\eqref{eq:summary_dual_curvature},
\begin{equation}
K_x = 1 - \sum_{v \in x} \frac{1}{d(v)}, \qquad \chi(\widetilde{G}) = \sum_{x \in X} K_x = |X| - |V|.
\label{eq:summary_curvature}
\end{equation}
This gives $\chi(\widetilde{G})$, the Euler characteristic of the dual hypergraph, and we now show that this topological index is sensitive only to the connectivity of the hypergraph, not to the weights of those connections, which may well be disordered. 
Holding the incidence structure of $G(V,X)$ fixed, we promote the $|X| \times |V|$ cell-vertex incidence matrix $B : \mathbb{C}^{|V|} \to \mathbb{C}^{|X|}$ from unit entries to arbitrary non-zero complex weights $w_{xv} \in \mathbb{C} \setminus \{0\}$:
\begin{equation}
B_{xv} = \begin{cases}
w_{xv}, & \text{if } v \in x, \\
0, & \text{if } v \notin x,
\end{cases}
\label{eq:topo_weighted_B}
\end{equation}
with Hermitian adjoint $(B^\dagger)_{vx} = w_{xv}^*$ for $v \in x$ (reducing to $B^T$ when $w_{xv} \in \mathbb{R}$). The tight-binding Hamiltonian $H = t B B^T$ of Section~\ref{sec:summary} then generalises to the weighted dual signless Laplacian Hamiltonian ($t \neq 0$)
\begin{equation}
H = t B B^\dagger, \qquad H_{xx'} = t \sum_{v \in x \cap x'} w_{xv} w_{x'v}^*.
\label{eq:topo_weighted_H}
\end{equation}
Allowing general weights $w_{xv}$ changes the diagonal shifts $H_{xx} = t \sum_{v \in x} |w_{xv}|^2$ from cell to cell and spoils the decomposition of $B B^T$ into integer multiples $\sum_k k A^{(k)}$ of $k$-vertex-sharing adjacency matrices. However, as long as cells are coupled solely through shared vertices, $H$ remains in the factorised semi-definite form $t B B^\dagger$, whose nullity is lower-bounded by the count $|X|-|V|$. The following two crucial properties also remain independently of the weights:
\begin{enumerate}
\item \textit{Zero-energy modes have vanishing incidence on every vertex.} For any cell state $|\psi\rangle \in \mathbb{C}^{|X|}$, \\$\langle \psi | B B^\dagger | \psi \rangle = \|B^\dagger \psi\|^2 = \sum_{v \in V} |\sum_{x \ni v} w_{xv}^* \psi_x|^2 \ge 0$, so $\ker H = \ker(B B^\dagger) = \ker B^\dagger$, and Eq.~\eqref{eq:summary_interference} becomes
\begin{equation}
(B^\dagger \psi)_v = \sum_{x \ni v} w_{xv}^* \psi_x = 0 \qquad \text{for every } v \in V.
\label{eq:topo_interference}
\end{equation}
Similarly, on the vertices $\langle \phi | B^\dagger B | \phi \rangle = \|B\phi\|^2$ gives $\ker(B^\dagger B) = \ker B$. The weights $w_{xv}^*$ rotate and rescale the zero-mode wavefunctions $\psi_x$, but Eq.~\eqref{eq:topo_interference} still imposes $|V|$ homogeneous linear constraints on $|X|$ amplitudes.

\item \textit{Rank-nullity and topological index.} Applying rank-nullity to $B^\dagger : \mathbb{C}^{|X|} \to \mathbb{C}^{|V|}$ and $B : \mathbb{C}^{|V|} \to \mathbb{C}^{|X|}$ gives $\dim\ker(B B^\dagger) = |X| - \operatorname{rank}(B^\dagger)$ and $\dim\ker(B^\dagger B) = |V| - \operatorname{rank}(B)$. Since $\operatorname{rank}(B^\dagger) = \operatorname{rank}(B)$, subtracting the two cancels $\operatorname{rank}(B)$ - the sole quantity sensitive to $\{w_{xv}\}$ - and combines with Eq.~\eqref{eq:summary_curvature} to yield
\begin{equation}
\dim\ker(B B^\dagger) - \dim\ker(B^\dagger B) = |X| - |V| = \chi(\widetilde{G}) = \sum_{x \in X} K_x.
\label{eq:topo_main_index}
\end{equation}
Neither the curvature $K_x$ nor the Euler characteristic $\chi(\widetilde{G})$ of the dual hypergraph depends on the weights $w_{xv}$, as $d(v)$ counts only the connectivity of $B$. Whenever $B$ is full-rank, the zero-mode count $\dim\ker(B B^\dagger) = \chi(\widetilde{G}) = |X| - |V|$ is independent of the weights.

\end{enumerate}

Applying the same arguments to the restricted submatrix $B_S$ on any finite subset $S \subset X$ shows that the lower bound $\dim\ker B_S^\dagger \ge |S| - |V(S)| \ge \bar{K}|S| - |\partial V(S)|$ of Eqs.~\eqref{eq:summary_subgraph} and~\eqref{eq:bst} does not depend on the weights, guaranteeing exact CLSs on amenable graphs once $|S| > |V(S)|$.

A direct corollary of Eq.~\eqref{eq:topo_main_index} is the discrete Atiyah-Singer index theorem~\cite{levitt1992,knill2012,knill2013}, proved in Appendix~\ref{app:atiyah_singer}, for a Dirac operator on the graded space $\mathcal{H} = \mathcal{H}_X \oplus \mathcal{H}_V$ with $\Gamma = \operatorname{diag}(I_{|X|}, -I_{|V|})$, 
\begin{equation}
D = \begin{pmatrix} 0 & B \\ B^\dagger & 0 \end{pmatrix}, \qquad D^2 = \begin{pmatrix} B B^\dagger & 0 \\ 0 & B^\dagger B \end{pmatrix}.
\label{eq:topo_dirac_corollary}
\end{equation}
The analytical index of this chiral Dirac operator is the left hand side of Eq.~\eqref{eq:topo_main_index} and it is equal to the topological index of the dual hypergraph $\chi(\widetilde{G}) = \sum_{x \in X} K_x$ for arbitrary non-zero weights $w_{xv}$ (Appendix~\ref{app:atiyah_singer}).

\subsection{Robustness to weight and flux disorder}\label{subsec:disorder_robustness}

We can now state concisely what physical perturbations this topological protection is robust against. In the crystalline face-graphs of Sections~\ref{sec:cubic}-\ref{sec:hetero}, the hopping Hamiltonians had specific integer ratios among edge-sharing and corner-sharing face-hopping amplitudes. In the microscopic vertex-mediated mechanism of Section~\ref{sec:motivation} (Eq.~\eqref{eq:mot_bipartite}), suppose instead that the orbitals supported on the cells couple to off-resonant vertex modes with site-dependent detunings $|\Delta_v| \gg |g_{xv}|$ of uniform sign via disordered complex hybridisations $g_{xv}$. Second-order perturbation theory yields $H_{\text{eff}} = -\operatorname{sgn}(\Delta_v) B B^\dagger$ with $w_{xv} = g_{xv}/\sqrt{|\Delta_v|}$. Consequently, the extensive degeneracy and the presence of CLSs on amenable graphs is robust against bond-strength $g_{xv}$ disorder, vertex-mode detuning disorder $\Delta_v$ (cf.\ Ref.~\cite{calugaru2022} for intra-sublattice perturbations in bipartite flat-band models), and external or synthetic magnetic fields entering $g_{xv}$ as arbitrary Peierls phases, extending to general hypergraphs the persistence of macroscopic zero modes under random $U(1)$ phases in molecular-orbital models~\cite{mizoguchi2023}. However, the CLSs may be different from the CLSs of the uniform weight hyper-graph, only their count is protected. Generic disorder merely gaps out accidental degeneracies (as demonstrated for random $U(1)$ phases in Ref.~\cite{mizoguchi2023}) so that the lower bound on the size of the degenerate nullspace $\dim\ker H_{\text{eff}} \ge \chi(\widetilde{G})$ holds as long as the connectivity - and hence the constraint count~\cite{roychowdhury2024} - of the graph itself remains.

Crucially, this protection holds only as long as cells are connected solely through the vertices $V$, with no direct cell-to-cell connection. Adding direct inter-cell hopping $H_{\text{direct}}$ (such as direct $\mathrm{O\text{-}O}$ overlap in the perovskite realisation of Section~\ref{sec:motivation}) yields $H = t B B^\dagger + H_{\text{direct}}$, which broadens the $E = 0$ manifold to order $\|H_{\text{direct}}\|$. Bare on-site disorder on the cells also lifts the degeneracy, but it is important to note that amenable graphs enjoy an additional protection against local disorder, since disorder at a cell $x$ cannot affect the energy of the extensive number of CLS that have \emph{no} support at $x$.

\section{Discussion}\label{sec:disc}

In computer science and spectral graph theory \cite{godsil2001,cvetkovic2009towards,cvetkovic2010,cvetkovic2010a}, the nullspaces of such signless Laplacians ($B B^T$) are known to govern network dynamics in strongly constrained environments. 
In particular, high-dimensional cell complexes are increasingly used in machine learning to represent higher-order relational information in complex datasets\cite{hajij2023}.
Random walks on these generalised graph structures provide an overarching algorithmic framework for tasks such as representation learning and semi-supervised learning\cite{mukherjee2016,yang2022}.
Unlike the standard graph Laplacian, where the null space simply counts disconnected components, the macroscopic null space of the signless Laplacian operator $B B^T$ describes localised \emph{quantum} states on a fully connected graph\cite{leykam2018} with compact support. 
Because these compact localised states are exact eigenmodes, they represent quantum wavefunctions that fail to ergodically explore the network due to perfect destructive interference\cite{bergman2008}. 
This has consequences not only for quantum machine learning algorithms\cite{biamonte2017}, but also quite generally for quantum search algorithms\cite{childs2004} that rely on the physics of a quantum walk on a generalised graph, where compact localised eigenmodes break the implicit assumption that the perturbed walk is ergodic and always able to reach the target node.

Consequently, identifying and manipulating these extended degeneracies is critical for resolving arrested dynamics and information trapping in complex discrete quantum networks, whether they are built of quantum simulators in arbitrary graphs\cite{kollarLineGraphLatticesEuclidean2020}, Josephson junction arrays\cite{chandra1995,bonamassa2023} in the quantum regime\cite{dutta2021,nambisan2026} or the connectivity of many-body states in Hilbert space\cite{tan2025}. 

In particular, the geometry of incidence matrices can be readily applied to the many-body transition graphs of kinetically constrained systems such as dimer models, where extensive degeneracies in the many-body spectrum have been reported and in some cases, connected explicitly to compact localised states on the transition graph. Just as we defined faces as irreducible cycles on the graph and the adjacency matrix of the face-graphs in terms of the face-vertex incidence matrix, we can define higher-dimensional analogues of faces as irreducible cycles of the line-graphs, the face-graphs and so on, to identify higher-order compact localised states defined on these higher-dimensional cell-complexes\footnote{Formally, this can be done by defining a poset with ground set $P = V \cup E \cup \cdots \cup F_n$, whose covering relations encode the incidences among vertices, edges, and generalised faces.}. The prescription demonstrated here for face-graphs can be used to systematically enumerate such many-body Aharonov-Bohm cages\cite{vidal1998} and the resulting extensive degeneracies in the many-body eigenspectrum that have been recently discussed as a new form of eigenstate order called quantum many-body caged spin-glass\cite{ben-amiManybodyCagesDisorderfree2025}, wherein strong correlations in the form of kinetic constraints cause arrested dynamics in the many-body Hilbert space.

The robustness of this macroscopic degeneracy also extends to disordered networks where a fraction of the edges violate the perfect local frustration-free connectivity implied by the positive-semidefinite operator $B B^T$. Destructive interference itself does not require local graph automorphisms - it can arise in asymmetric networks from balanced couplings or latent symmetries~\cite{roentgen2018,morfonios2021}, and in our framework requires no symmetry at all so long as hopping remains purely vertex-mediated (Section~\ref{sec:topo}). Rather, structural defects that introduce direct cell-to-cell links or uncompensated pathways locally violate the positive-semidefinite $B B^\dagger$ factorisation, partially lifting the exact degeneracy. However, the extensive nature of the degeneracy is expected to survive below a critical threshold, beyond which disorder-delocalised puddles may percolate to form a giant connected cluster. This persistence in disordered environments is facilitated by the spatial compactness of the localised states, which remain shielded from distant structural defects. These natural corollaries of the prescription for extensive degeneracies of the kind that we discuss are ripe for demonstration in simulations and experiments. 

The extensive degeneracies of adjacency matrices of face-graphs and line-graphs obtained from this prescription of incidence geometries represent a novel failure mode of quantum networks with no analogue in classical networks. A signal that is initiated with the particular quantum superposition of a compact localised state fails to propagate through the network even when it is fully connected. In stark contrast to classical networks, local defects in the connectivity or classical failure of nodes or edges of these graphs serve to heal these defect modes when they open additional non-frustrated hopping pathways, so that beyond a critical threshold of defects there are no localised eigenmodes that prevent quantum signals from propagating across a fully connected network. In this sense, this provides a complementary perspective to the standard lore of connectivity and dependency in interdependent networks pioneered by Havlin and collaborators\cite{buldyrev2010,bonamassa2023} where dependencies between nodes in two networks lead to cascades that cause both networks to fail together. In quantum networks on the line-graphs and face-graphs that are dependent on the networks on the root graph, failures on the latter make the former robust against arrested dynamics of quantum information. Conversely, controlled local switching of couplings provides a mechanism to store and coherently release or route quantum states trapped in compact localised modes, as proposed for symmetry-induced CLSs in switchable quantum networks~\cite{roentgen2019}.

As noted in Section~\ref{sec:motivation}, in quantum condensed matter physics and chemistry the face-graph construction may be relevant for systems whose orbitals have maximum probability amplitude away from the geometric centre, supported on three or more spatial locations. Prominent examples are the so-called `fidget-spinner' orbitals of magic-angle twisted bilayer graphene~\cite{koshinoMaximallyLocalizedWannier2018,po-prx-2018,kangSymmetryMaximallyLocalized2018}, the molecular orbitals of star-shaped organic polymers (e.g., starphenes~\cite{clar1968}), and open-shell graphene nanoflakes~\cite{clar1953,melle-franco2017,pavlicek2017}.

Networks built out of such orbitals allow novel possibilities in the realm of spin-qubit-based quantum computing. Open-shell graphene nanoflakes are non-Kekul\'e structures (also known as unbalanced bipartite lattices in physics) with high-spin ground states and a degenerate set of zero-energy non-bonding orbitals because of Ovchinnikov’s rule\cite{ovchinnikov1978} (Lieb's theorem\cite{lieb1989} in physics). The resulting $N$-level system of $N$-triangulene is therefore recognised as a potential quantum memory and as a generalised qubit for quantum computation\cite{mishra2021,wei2012,sato2009,mishra2020,valenta2022}, where the quantum information of the spin-state is stored in the Singly Occupied Molecular Orbitals that are delocalised over the flake boundaries\footnote{In the thermodynamic limit, these are the familiar topologically protected edge-modes at the zigzag edges of graphene\cite{fujita1996,ryu2002,castroneto2009}.}. Constructing arbitrary lattices of such flakes allows the possibility of compact localised states on the face-graphs and line-graphs, that have no overlap with any decoherence sources beyond a small finite number of flakes. The quantum information in the generalised qubit is then topologically protected in a way that is fundamentally distinct from the conventional routes to topological quantum computation which depend on the elusive Majorana bound states or Majorana quasiparticles\cite{nayak-rmp-2008,aliceaNewDirectionsPursuit2012}. 

Having established the hypergraph and face-graph framework, we can now compare it directly with prior flat-band constructions. Tasaki's cell construction, introduced to prove flat-band ferromagnetism in repulsive Hubbard models\cite{tasakiFerromagnetismHubbardModels1992,mielkeTasaki1993,tasaki1998,tasakiPhysicsMathematicsQuantum2020}, writes the hopping Hamiltonian as a sum of local positive operators $t \sum_v b_v^\dagger b_v$. Our Hamiltonian $t B B^T$ has this exact form: each vertex $v \in V$ acts as a cell, and $b_v$ sums the orbitals on all hyperedges $x \ni v$ meeting at $v$. In Tasaki's original lattices, each cell also contains a private internal site that keeps the operators $b_v$ linearly independent and gaps the flat band. Here every hyperedge covers $|x| \ge 2$ shared vertices, so no vertex carries a private orbital, placing our models in the shared-site regime of Tanaka's extended cell construction\cite{tanaka2020}. Consequently, the operators $b_v$ can become linearly dependent at high-symmetry momenta, which is why dispersive bands touch the flat manifold. The same algebra underlies the molecular-orbital representation of Hatsugai and collaborators\cite{hatsugai2011,mizoguchi2019,mizoguchi2023} and related Gram-matrix models\cite{xu2020}, where the columns of $B$ serve as non-orthogonal molecular orbitals. This guarantees at least $|X| - \operatorname{rank}(B)$ zero modes, with band touchings occurring wherever the Bloch molecular orbitals become linearly dependent\cite{mizoguchi2019}.

Viewed through the parent bipartite cell-vertex Hamiltonian of Section~\ref{sec:motivation}, these zero modes also connect to sublattice-imbalance flat bands\cite{sutherland1986,lieb1989}. In crystalline settings, C{\u a}lug{\u a}ru \textit{et al.}\cite{calugaru2022} showed that such flat bands survive arbitrary intra-vertex perturbations and classified their symmetry eigenvalues across space groups. Likewise, Roychowdhury \textit{et al.}\cite{roychowdhury2024} analysed the partner Hamiltonians $B B^\dagger$ and $B^\dagger B$ via lattice supersymmetry, identifying $|X| - |V|$ with the Witten index. Section~\ref{sec:topo} and Appendix~\ref{app:atiyah_singer} re-derive this partner-spectrum index via a discrete Atiyah-Singer theorem in order to decompose $|X| - |V|$ into a local curvature sum. What is new here builds on this shared count in three directions. First, we choose $B$ canonically from graph connectivity alone as the face-vertex (or $k$-cell-vertex) incidence matrix with intrinsically defined faces, and obtain explicit band structures on three-dimensional Bravais lattices. Second, restricting the local curvature sum to finite subgraphs yields a bulk-boundary amenability criterion that guarantees compact localised states without translation symmetry. Finally, for heterogeneous graphs with faces of mixed degree, we derive the exact nullspace decomposition and identify hybrid cages supported jointly on distinct face types.

Complementary to the recent interest in the Riemannian structure~\cite{provostRiemannianStructureManifolds1980,berryFiveYearsLater,chengQuantumGeometricTensor2010,marzariMaximallyLocalizedGeneralized1997} of flat-band Bloch wavefunctions in momentum space - widely known as quantum geometry~\cite{ahn2022,verma2025} - we provide an orthogonal perspective on the quantum mechanical implications of the discrete geometry of the real-space graphs on which these flat bands are defined. Positive dual discrete curvature topologically guarantees an extensive degeneracy on the edges, faces, or higher-dimensional cells of an arbitrary hypergraph, even without translation symmetry when bands are not well-defined. When the network is amenable, this bulk curvature inevitably overtakes the boundary penalty on finite subgraphs, yielding compact localised states, which are localised without disorder by destructive interference - a quantum mechanical effect enabled by the geometry of graphs.

\section{Acknowledgements}
I am grateful to Nick Sander and Lars Franke for pointing me to Ref.\cite{kane2014}, to Peter Schmelcher for pointing me to Refs.~\cite{roentgen2018,roentgen2019,roentgen2021,morfonios2021,roentgen2024}, to J\"org Schmalian and Piers Coleman for encouraging feedback and to Daniel Schultz for a well-posed question that prompted Appendix~\ref{app:face_definition_physics}.

\begin{appendix}
\numberwithin{equation}{section}

\section{Physical Motivation for the Chosen Definition of Faces}\label{app:face_definition_physics}
In the main text, we defined the faces of an arbitrary graph as the set of its relevant (or irreducible) cycles. This definition emerges naturally from the physical requirements of defining gauge fields and fluxes on discrete networks.

Consider a graph $G(V,E)$ as a network of wires or bonds hosting a vector field, such as a vector potential $\mathbf{A}$. By the discrete Hodge-Helmholtz theorem, any edge-defined vector field on a graph can be uniquely decomposed into three orthogonal components:
\begin{equation}
    \mathbf{A} = \nabla \phi + \nabla \times \mathbf{\Phi} + \mathbf{h}
\end{equation}
comprising an irrotational gradient field ($\nabla \phi$), a solenoidal rotational field ($\nabla \times \mathbf{\Phi}$), and a global harmonic component ($\mathbf{h}$). The harmonic component corresponds to an Aharonov-Bohm flux threaded through the holes of the periodic system, in practice implemented as a large gauge transformation. The physical observables are the magnetic fluxes, defined as the circulation of the vector potential around closed loops.

By the discrete Stokes' theorem \cite{grady2010}, the total circulation around any cycle in the graph is given by the oriented sum of the fluxes through a set of plaquettes spanning the enclosed surface. Therefore, the rotational component of the vector field is uniquely determined by its fluxes through the elementary cycles that tile the graph in the sense that a modulo-2 sum of these cycles can describe any closed cycle of the graph. This requirement naturally leads to the graph-theoretic notion of the cycle space \cite{biggs1993}. 
The closest analogue to a measurable magnetic field is the discrete curl on the smallest possible cycles on the graph.
This motivates the use of a Minimum Weight Cycle Basis (MWCB)\footnote{Consider the edge space $C_1(G; \mathbb{F}_2)$ and the vertex space $C_0(G; \mathbb{F}_2)$ over the finite field of two elements. The boundary operator $\partial_1: C_1 \to C_0$ maps edges to their incident vertices. Formally, the cycle space of the graph is the kernel of the boundary operator: $\mathcal{Z} = \ker(\partial_1)$. The dimension of this space is given by the cyclomatic number $\mu = |E| - |V| + c$, where $c$ is the number of connected components. The minimum weight cycle basis is defined by selecting a basis of $\mathcal{Z}$ that minimises the total length (or weight) of its constituent cycles.} - a basis whose total cycle length is minimal. 
However, choosing a single MWCB generally breaks some spatial symmetries that may exist on regular graphs, as there are often multiple equivalent shortest loops. 
By taking the union of all MWCBs, which define the \textit{relevant cycles}, we construct an overcomplete basis for the cycle space. 
This definition uniquely identifies faces of a graph without any reference to spatial embedding. 
For example, while the FCC equatorial square is reducible (as discussed in the main text), all $12$ length-$4$ cycles per primitive cell in a bipartite lattice like the Body-Centred Cubic (BCC) lattice (the $6$ planar rhombi and $6$ skew $4$-cycles shown in Fig.~\ref{fig:BCCFaces}) are irreducible because the lattice contains no length-$3$ triangles to tile them.

\section{Discrete Atiyah-Singer Index Theorem on Hypergraphs}\label{app:atiyah_singer}

In this appendix, we present the discrete Atiyah-Singer index theorem~\cite{levitt1992,knill2012,knill2013} on a finite hypergraph $G(V,X)$ (with vertex set $V$, cell set $X$, and vertex degree $d(v) \ge 1$ for all $v \in V$) and its dual hypergraph $\widetilde{G}(X,V)$, establishing via McKean-Singer spectral symmetry~\footnote{In the language of supersymmetric quantum mechanics and lattice supersymmetry~\cite{witten1982constraints,rana1993,roychowdhury2024}, $D$ is the Hermitian supercharge pairing the isospectral partner Hamiltonians $L_X = B B^\dagger$ and $L_V = B^\dagger B$ away from zero energy, and $\operatorname{ind}_a(D) = |X| - |V|$ is the Witten index; formulating this index via the Levitt-Knill Gauss-Bonnet identity~\cite{levitt1992,knill2012,knill2013} decomposes it into the local curvature sum $\sum_{x \in X} K_x$, whose restriction to finite subgraphs yields the CLS criterion of Eq.~\eqref{eq:bst}.} that the analytical index of the discrete Dirac operator equals the dual Euler characteristic $\chi(\widetilde{G}) = \sum_{x \in X} K_x$.

\subsection{Graded Hilbert space and discrete Dirac and Laplace-Beltrami operators}

Let $\mathcal{H}$ be the Hilbert space of states supported on the cells and vertices, decomposed into orthogonal even and odd sectors by the grading operator $\Gamma = \operatorname{diag}(I_{|X|}, -I_{|V|})$:
\begin{equation}
\mathcal{H} = \mathcal{H}_X \oplus \mathcal{H}_V,
\end{equation}
where $\mathcal{H}_X \cong \mathbb{C}^{|X|}$ ($\Gamma = +1$) is spanned by states localised on the cells $X$ (the $0$-cells of $\widetilde{G}$, such as the faces $F$ of Section~\ref{sec:faces}) and $\mathcal{H}_V \cong \mathbb{C}^{|V|}$ ($\Gamma = -1$) is spanned by states localised on the vertices $V$ (the $1$-cells of $\widetilde{G}$).

For any $|X| \times |V|$ weighted cell-vertex incidence matrix $B$ as in Eq.~\eqref{eq:topo_weighted_B}, the Hermitian discrete Dirac operator $D : \mathcal{H} \to \mathcal{H}$ is defined as
\begin{equation}
D = \begin{pmatrix} 0 & B \\ B^\dagger & 0 \end{pmatrix},
\end{equation}
which is odd under the grading, $\{D, \Gamma\} = 0$. Its square defines the discrete Laplace-Beltrami operator $L = D^2$, which is block-diagonal and preserves the grading, $[L, \Gamma] = 0$:
\begin{equation}
L = D^2 = \begin{pmatrix} B B^\dagger & 0 \\ 0 & B^\dagger B \end{pmatrix} = \begin{pmatrix} L_X & 0 \\ 0 & L_V \end{pmatrix}.
\end{equation}
Here $L_X = B B^\dagger$ is the dual signless Laplacian on the cells and $L_V = B^\dagger B$ is the signless Laplacian on the vertices.

\subsection{McKean-Singer spectral symmetry and the analytical index}

Following Knill~\cite{knill2013}, the analytical index of $D$ is defined as the supertrace of the heat kernel $e^{-\beta L}$ for $\beta \ge 0$:
\begin{equation}
\operatorname{ind}_a(D) \equiv \operatorname{str}(e^{-\beta L}) = \operatorname{Tr}(\Gamma e^{-\beta L}) = \operatorname{Tr}(e^{-\beta L_X}) - \operatorname{Tr}(e^{-\beta L_V}).
\label{eq:app_analytical_index}
\end{equation}

\textbf{Theorem (McKean-Singer spectral symmetry).} \textit{Every positive eigenvalue $\lambda > 0$ of $L_X = B B^\dagger$ is also an eigenvalue of $L_V = B^\dagger B$ with identical multiplicity.}

\textit{Proof.} The operators $B^\dagger$ and $B$ intertwine $L_X$ and $L_V$: $L_V B^\dagger = B^\dagger(B B^\dagger) = B^\dagger L_X$ and $L_X B = (B B^\dagger)B = B L_V$. Let $\psi_X \in \mathcal{H}_X$ be a normalised eigenvector of $L_X$ with eigenvalue $\lambda > 0$, so $B B^\dagger \psi_X = \lambda \psi_X$, and define $\psi_V = \lambda^{-1/2} B^\dagger \psi_X \in \mathcal{H}_V$. Then $\|\psi_V\|^2 = \lambda^{-1} \langle \psi_X | B B^\dagger | \psi_X \rangle = 1$ (so $\psi_V \neq 0$) and
\begin{equation}
L_V \psi_V = \lambda^{-1/2} (B^\dagger B)(B^\dagger \psi_X) = \lambda^{-1/2} B^\dagger (B B^\dagger \psi_X) = \lambda \psi_V.
\end{equation}
Conversely, $\lambda^{-1/2} B \psi_V = \lambda^{-1} B B^\dagger \psi_X = \psi_X$, so $\lambda^{-1/2} B^\dagger$ and $\lambda^{-1/2} B$ are mutually inverse unitary maps between the $\lambda$-eigenspaces of $L_X$ and $L_V$.

Because all $\lambda > 0$ eigenspaces cancel pairwise between $\operatorname{Tr}(e^{-\beta L_X})$ and $\operatorname{Tr}(e^{-\beta L_V})$, $\operatorname{str}(e^{-\beta L})$ is independent of $\beta \ge 0$. Taking $\beta \to \infty$ projects onto the $\lambda = 0$ kernels and yields
\begin{equation}
\operatorname{ind}_a(D) = \dim\ker L_X - \dim\ker L_V = \dim\ker(B B^\dagger) - \dim\ker(B^\dagger B).
\label{eq:app_index_nullity}
\end{equation}

\subsection{Dual discrete curvature, Gauss-Bonnet, and the index theorem}

On the dual hypergraph $\widetilde{G}(X,V)$, each cell $x \in X$ is a $0$-cell carrying even parity weight $\omega(x) = +1$, while each vertex $v \in V$ is a dual hyperedge covering $d(v)$ cells and carrying odd parity weight $\omega(v) = -1$. Applying the Levitt-Knill local Euler curvature~\cite{levitt1992,knill2012} to each $0$-cell $x \in X$ of $\widetilde{G}(X,V)$ gives the dual discrete curvature of Eq.~\eqref{eq:summary_dual_curvature},
\begin{equation}
K_x = \frac{\omega(x)}{1} + \sum_{v \in x} \frac{\omega(v)}{d(v)} = 1 - \sum_{v \in x} \frac{1}{d(v)}.
\end{equation}
Summing $K_x$ over $x \in X$ gives the discrete Gauss-Bonnet identity on $\widetilde{G}(X,V)$:
\begin{equation}
\sum_{x \in X} K_x = \sum_{x \in X} \left( 1 - \sum_{v \in x} \frac{1}{d(v)} \right) = |X| - \sum_{v \in V} \sum_{x \ni v} \frac{1}{d(v)} = |X| - |V| = \chi(\widetilde{G}).
\label{eq:app_gauss_bonnet}
\end{equation}
(Equivalently, distributing each cell's $+1$ weight equally across its $|x|$ vertices expresses the same topological index as a sum over $V$, $\chi(\widetilde{G}) = \sum_{v \in V}\big(-1 + \sum_{x \ni v} 1/|x|\big)$.)

Finally, evaluating the $\beta$-independent supertrace in Eq.~\eqref{eq:app_analytical_index} at $\beta = 0$ gives \\
$\operatorname{str}(I_{\mathcal{H}}) = \operatorname{Tr}(I_{|X|}) - \operatorname{Tr}(I_{|V|}) = |X| - |V|$. Equating this $\beta = 0$ evaluation with Eq.~\eqref{eq:app_index_nullity} and Eq.~\eqref{eq:app_gauss_bonnet} completes the proof of the discrete Atiyah-Singer index theorem:
\begin{equation}
\operatorname{ind}_a(D) = \dim\ker(B B^\dagger) - \dim\ker(B^\dagger B) = \sum_{x \in X} K_x = \chi(\widetilde{G}),
\end{equation}
independent of the non-zero coupling weights in $B$.

\end{appendix}
\bibliography{FaceGraph}

@article{mielkeFerromagnetismHubbardmodel1991,
  title = {Ferromagnetism in the {{Hubbard}} Model on Line Graphs and Further Considerations},
  author = {Mielke, A.},
  year = 1991,
  month = jul,
  journal = {Journal of Physics A: Mathematical and General},
  volume = {24},
  number = {14},
  pages = {3311},
  issn = {0305-4470},
  doi = {10.1088/0305-4470/24/14/018},
  url = {https://dx.doi.org/10.1088/0305-4470/24/14/018},
  urldate = {2024-08-05},
  langid = {english}
}

@article{maimaiti2017,
  title = {Compact Localized States and Flat-Band Generators in One Dimension},
  author = {Maimaiti, Wulayimu and Andreanov, Alexei and Park, Hee Chul and Gendelman, Oleg and Flach, Sergej},
  year = 2017,
  month = mar,
  journal = {Physical Review B},
  volume = {95},
  number = {11},
  pages = {115135},
  publisher = {American Physical Society},
  doi = {10.1103/PhysRevB.95.115135},
  url = {https://link.aps.org/doi/10.1103/PhysRevB.95.115135},
  urldate = {2024-07-18}
}

@article{maimaiti2021,
  title = {Flat-Band Generator in Two Dimensions},
  author = {Maimaiti, Wulayimu and Andreanov, Alexei and Flach, Sergej},
  year = 2021,
  month = apr,
  journal = {Physical Review B},
  volume = {103},
  number = {16},
  pages = {165116},
  publisher = {American Physical Society},
  doi = {10.1103/PhysRevB.103.165116},
  url = {https://link.aps.org/doi/10.1103/PhysRevB.103.165116},
  urldate = {2024-07-18}
}

@article{bergman2008,
  title = {Band Touching from Real-Space Topology in Frustrated Hopping Models},
  author = {Bergman, Doron L. and Wu, Congjun and Balents, Leon},
  year = 2008,
  month = sep,
  journal = {Physical Review B},
  volume = {78},
  number = {12},
  pages = {125104},
  publisher = {American Physical Society},
  doi = {10.1103/PhysRevB.78.125104},
  url = {https://link.aps.org/doi/10.1103/PhysRevB.78.125104},
  urldate = {2024-07-13}
}

@article{kollarLineGraphLatticesEuclidean2020,
  title = {Line-{{Graph Lattices}}: {{Euclidean}} and {{Non-Euclidean Flat Bands}}, and {{Implementations}} in {{Circuit Quantum Electrodynamics}}},
  shorttitle = {Line-{{Graph Lattices}}},
  author = {Koll{\'a}r, Alicia J. and Fitzpatrick, Mattias and Sarnak, Peter and Houck, Andrew A.},
  year = 2020,
  month = jun,
  journal = {Communications in Mathematical Physics},
  volume = {376},
  number = {3},
  pages = {1909--1956},
  issn = {1432-0916},
  doi = {10.1007/s00220-019-03645-8},
  url = {https://doi.org/10.1007/s00220-019-03645-8},
  urldate = {2025-09-03},
  langid = {english}
}

@article{lieb1989,
  title = {Two Theorems on the {{Hubbard}} Model},
  author = {Lieb, Elliott H.},
  year = 1989,
  month = mar,
  journal = {Physical Review Letters},
  volume = {62},
  number = {10},
  pages = {1201--1204},
  publisher = {American Physical Society},
  doi = {10.1103/PhysRevLett.62.1201},
  url = {https://link.aps.org/doi/10.1103/PhysRevLett.62.1201},
  urldate = {2024-08-05}
}

@book{tasakiPhysicsMathematicsQuantum2020,
  title = {Physics and Mathematics of Quantum Many-Body Systems},
  author = {Tasaki, Hal},
  year = {2020},
  publisher = {Springer},
  location = {Cham},
  isbn = {978-3-030-41265-4},
  doi = {10.1007/978-3-030-41265-4},
  langid = {english}
}

@article{tasakiFerromagnetismHubbardModels1992,
  title = {Ferromagnetism in the {{Hubbard}} Models with Degenerate Single-Electron Ground States},
  author = {Tasaki, Hal},
  year = 1992,
  month = sep,
  journal = {Physical Review Letters},
  volume = {69},
  number = {10},
  pages = {1608--1611},
  issn = {0031-9007},
  doi = {10.1103/PhysRevLett.69.1608},
  url = {https://link.aps.org/doi/10.1103/PhysRevLett.69.1608},
  urldate = {2024-08-05},
  copyright = {http://link.aps.org/licenses/aps-default-license},
  langid = {english}
}

@article{landau1930,
  title = {Diamagnetismus Der {{Metalle}}},
  author = {Landau, L.},
  year = 1930,
  month = sep,
  journal = {Zeitschrift f\"ur Physik},
  volume = {64},
  number = {9},
  pages = {629--637},
  issn = {0044-3328},
  doi = {10.1007/BF01397213},
  url = {https://doi.org/10.1007/BF01397213}
}

@article{caoCorrelatedInsulatorBehaviour2018,
  title = {Correlated Insulator Behaviour at Half-Filling in Magic-Angle Graphene Superlattices},
  author = {Cao, Yuan and Fatemi, Valla and Demir, Ahmet and Fang, Shiang and Tomarken, Spencer L. and Luo, Jason Y. and {Sanchez-Yamagishi}, Javier D. and Watanabe, Kenji and Taniguchi, Takashi and Kaxiras, Efthimios and Ashoori, Ray C. and {Jarillo-Herrero}, Pablo},
  year = 2018,
  month = mar,
  journal = {Nature},
  volume = {556},
  number = {7699},
  pages = {80--84},
  issn = {0028-0836, 1476-4687},
  doi = {10.1038/nature26154},
  url = {http://www.nature.com/doifinder/10.1038/nature26154},
  urldate = {2018-09-17}
}

@article{caoUnconventionalSuperconductivityMagicangle2018,
  title = {Unconventional Superconductivity in Magic-Angle Graphene Superlattices},
  author = {Cao, Yuan and Fatemi, Valla and Fang, Shiang and Watanabe, Kenji and Taniguchi, Takashi and Kaxiras, Efthimios and {Jarillo-Herrero}, Pablo},
  year = 2018,
  month = mar,
  journal = {Nature},
  volume = {556},
  number = {7699},
  pages = {43--50},
  issn = {0028-0836, 1476-4687},
  doi = {10.1038/nature26160},
  urldate = {2018-09-14}
}

@article{li2010,
  title = {Observation of {{Van Hove}} Singularities in Twisted Graphene Layers},
  author = {Li, Guohong and Luican, A. and {Lopes dos Santos}, J. M. B. and Castro Neto, A. H. and Reina, A. and Kong, J. and Andrei, E. Y.},
  year = 2010,
  month = feb,
  journal = {Nature Physics},
  volume = {6},
  number = {2},
  pages = {109--113},
  publisher = {Nature Publishing Group},
  issn = {1745-2481},
  doi = {10.1038/nphys1463},
  url = {https://www.nature.com/articles/nphys1463},
  urldate = {2023-05-19},
  copyright = {2009 Springer Nature Limited},
  langid = {english}
}

@article{andrei2021,
  title = {The Marvels of Moir\'e Materials},
  author = {Andrei, Eva Y. and Efetov, Dmitri K. and {Jarillo-Herrero}, Pablo and MacDonald, Allan H. and Mak, Kin Fai and Senthil, T. and Tutuc, Emanuel and Yazdani, Ali and Young, Andrea F.},
  year = 2021,
  month = mar,
  journal = {Nature Reviews Materials},
  volume = {6},
  number = {3},
  pages = {201--206},
  publisher = {Nature Publishing Group},
  issn = {2058-8437},
  doi = {10.1038/s41578-021-00284-1},
  url = {https://www.nature.com/articles/s41578-021-00284-1},
  urldate = {2023-05-20},
  copyright = {2021 Springer Nature Limited},
  langid = {english}
}

@article{vonklitzing1986,
  title = {The Quantized {{Hall}} Effect},
  author = {Von Klitzing, Klaus},
  year = 1986,
  month = jul,
  journal = {Reviews of Modern Physics},
  volume = {58},
  number = {3},
  pages = {519--531},
  issn = {0034-6861},
  doi = {10.1103/RevModPhys.58.519},
  url = {https://link.aps.org/doi/10.1103/RevModPhys.58.519},
  urldate = {2026-07-18},
  copyright = {http://link.aps.org/licenses/aps-default-license},
  langid = {english}
}

@article{zeng2023,
  title = {Thermodynamic Evidence of Fractional {{Chern}} Insulator in Moir\'e {MoTe$_2$}},
  author = {Zeng, Yihang and Xia, Zhengchao and Kang, Kaifei and Zhu, Jiacheng and Kn{\"u}ppel, Patrick and Vaswani, Chirag and Watanabe, Kenji and Taniguchi, Takashi and Mak, Kin Fai and Shan, Jie},
  year = 2023,
  month = oct,
  journal = {Nature},
  volume = {622},
  number = {7981},
  pages = {69--73},
  publisher = {Nature Publishing Group},
  issn = {1476-4687},
  doi = {10.1038/s41586-023-06452-3},
  url = {https://www.nature.com/articles/s41586-023-06452-3},
  urldate = {2026-07-18},
  copyright = {2023 The Author(s), under exclusive licence to Springer Nature Limited},
  langid = {english}
}

@article{cai2023,
  title = {Signatures of Fractional Quantum Anomalous {{Hall}} States in Twisted {MoTe$_2$}},
  author = {Cai, Jiaqi and Anderson, Eric and Wang, Chong and Zhang, Xiaowei and Liu, Xiaoyu and Holtzmann, William and Zhang, Yinong and Fan, Fengren and Taniguchi, Takashi and Watanabe, Kenji and Ran, Ying and Cao, Ting and Fu, Liang and Xiao, Di and Yao, Wang and Xu, Xiaodong},
  year = 2023,
  month = oct,
  journal = {Nature},
  volume = {622},
  number = {7981},
  pages = {63--68},
  publisher = {Nature Publishing Group},
  issn = {1476-4687},
  doi = {10.1038/s41586-023-06289-w},
  url = {https://www.nature.com/articles/s41586-023-06289-w},
  urldate = {2026-07-18},
  copyright = {2023 The Author(s), under exclusive licence to Springer Nature Limited},
  langid = {english}
}

@article{lu2024a,
  title = {Fractional Quantum Anomalous {{Hall}} Effect in Multilayer Graphene},
  author = {Lu, Zhengguang and Han, Tonghang and Yao, Yuxuan and Reddy, Aidan P. and Yang, Jixiang and Seo, Junseok and Watanabe, Kenji and Taniguchi, Takashi and Fu, Liang and Ju, Long},
  year = 2024,
  month = feb,
  journal = {Nature},
  volume = {626},
  number = {8000},
  pages = {759--764},
  publisher = {Nature Publishing Group},
  issn = {1476-4687},
  doi = {10.1038/s41586-023-07010-7},
  url = {https://www.nature.com/articles/s41586-023-07010-7},
  urldate = {2026-07-18},
  copyright = {2024 The Author(s), under exclusive licence to Springer Nature Limited},
  langid = {english}
}

@article{xu2023,
  title = {Observation of {{Integer}} and {{Fractional Quantum Anomalous Hall Effects}} in {{Twisted Bilayer}} {${\mathrm{MoTe}}_{2}$}},
  author = {Xu, Fan and Sun, Zheng and Jia, Tongtong and Liu, Chang and Xu, Cheng and Li, Chushan and Gu, Yu and Watanabe, Kenji and Taniguchi, Takashi and Tong, Bingbing and Jia, Jinfeng and Shi, Zhiwen and Jiang, Shengwei and Zhang, Yang and Liu, Xiaoxue and Li, Tingxin},
  year = 2023,
  month = sep,
  journal = {Physical Review X},
  volume = {13},
  number = {3},
  pages = {031037},
  publisher = {American Physical Society},
  doi = {10.1103/PhysRevX.13.031037},
  url = {https://link.aps.org/doi/10.1103/PhysRevX.13.031037},
  urldate = {2026-07-18}
}

@article{nuckolls2024a,
  title = {A Microscopic Perspective on Moir\'e Materials},
  author = {Nuckolls, Kevin P. and Yazdani, Ali},
  year = 2024,
  month = jul,
  journal = {Nature Reviews Materials},
  volume = {9},
  number = {7},
  pages = {460--480},
  publisher = {Nature Publishing Group},
  issn = {2058-8437},
  doi = {10.1038/s41578-024-00682-1},
  url = {https://www.nature.com/articles/s41578-024-00682-1},
  urldate = {2026-07-18},
  copyright = {2024 Springer Nature Limited},
  langid = {english}
}

@article{andrei2020,
  title = {Graphene Bilayers with a Twist},
  author = {Andrei, Eva Y. and MacDonald, Allan H.},
  year = 2020,
  month = dec,
  journal = {Nature Materials},
  volume = {19},
  number = {12},
  pages = {1265--1275},
  publisher = {Nature Publishing Group},
  issn = {1476-4660},
  doi = {10.1038/s41563-020-00840-0},
  url = {https://www.nature.com/articles/s41563-020-00840-0},
  urldate = {2022-09-07},
  copyright = {2020 Springer Nature Limited},
  langid = {english}
}

@article{mak2022a,
  title = {Semiconductor Moir\'e Materials},
  author = {Mak, Kin Fai and Shan, Jie},
  year = 2022,
  month = jul,
  journal = {Nature Nanotechnology},
  volume = {17},
  number = {7},
  pages = {686--695},
  publisher = {Nature Publishing Group},
  issn = {1748-3395},
  doi = {10.1038/s41565-022-01165-6},
  url = {https://www.nature.com/articles/s41565-022-01165-6},
  urldate = {2026-07-18},
  copyright = {2022 Springer Nature Limited},
  langid = {english}
}

@article{mellado2025,
  title = {Magnetic Moir\'e Systems: A Review},
  shorttitle = {Magnetic Moir\'e Systems},
  author = {Mellado, Paula},
  year = 2025,
  month = aug,
  journal = {Journal of Physics: Condensed Matter},
  volume = {37},
  number = {32},
  pages = {323001},
  publisher = {IOP Publishing},
  issn = {0953-8984},
  doi = {10.1088/1361-648X/adf483},
  url = {https://doi.org/10.1088/1361-648X/adf483},
  urldate = {2026-07-18},
  langid = {english}
}

@article{kane2014,
  title = {Topological Boundary Modes in Isostatic Lattices},
  author = {Kane, C. L. and Lubensky, T. C.},
  year = 2014,
  month = jan,
  journal = {Nature Physics},
  volume = {10},
  number = {1},
  pages = {39--45},
  publisher = {Nature Publishing Group},
  issn = {1745-2481},
  doi = {10.1038/nphys2835},
  url = {https://www.nature.com/articles/nphys2835},
  urldate = {2026-06-06},
  copyright = {2013 Springer Nature Limited},
  langid = {english}
}

@article{rana1993,
  title = {Soluble Supersymmetric Quantum {{XY}} Model},
  author = {Rana, A. E. and Girvin, S. M.},
  year = 1993,
  month = jul,
  journal = {Physical Review B},
  volume = {48},
  number = {1},
  pages = {360--364},
  publisher = {American Physical Society},
  doi = {10.1103/PhysRevB.48.360},
  url = {https://link.aps.org/doi/10.1103/PhysRevB.48.360},
  urldate = {2025-01-02}
}

@article{roychowdhury2024,
  title = {Supersymmetry on the Lattice: {{Geometry}}, Topology, and Flat Bands},
  shorttitle = {Supersymmetry on the Lattice},
  author = {Roychowdhury, Krishanu and Attig, Jan and Trebst, Simon and Lawler, Michael J.},
  year = 2024,
  month = dec,
  journal = {Physical Review Research},
  volume = {6},
  number = {4},
  pages = {043273},
  issn = {2643-1564},
  doi = {10.1103/PhysRevResearch.6.043273},
  url = {https://link.aps.org/doi/10.1103/PhysRevResearch.6.043273},
  urldate = {2026-06-29},
  langid = {english}
}

@article{fendley2003,
  title = {Lattice {{Models}} with {$\mathcal{N}=2$} {{Supersymmetry}}},
  author = {Fendley, Paul and Schoutens, Kareljan and {de Boer}, Jan},
  year = 2003,
  month = mar,
  journal = {Physical Review Letters},
  volume = {90},
  number = {12},
  pages = {120402},
  publisher = {American Physical Society},
  doi = {10.1103/PhysRevLett.90.120402},
  url = {https://link.aps.org/doi/10.1103/PhysRevLett.90.120402},
  urldate = {2026-07-18}
}

@article{sutherland1986,
  title = {Localization of Electronic Wave Functions Due to Local Topology},
  author = {Sutherland, Bill},
  year = 1986,
  month = oct,
  journal = {Physical Review B},
  volume = {34},
  number = {8},
  pages = {5208--5211},
  issn = {0163-1829},
  doi = {10.1103/PhysRevB.34.5208},
  url = {https://link.aps.org/doi/10.1103/PhysRevB.34.5208},
  urldate = {2025-10-05},
  copyright = {http://link.aps.org/licenses/aps-default-license},
  langid = {english}
}

@article{witten1982constraints,
  title={Constraints on supersymmetry breaking},
  author={Witten, Edward},
  journal={Nuclear Physics B},
  volume={202},
  number={2},
  pages={253--316},
  year={1982},
  doi={10.1016/0550-3213(82)90071-2}
}

@incollection{arovas1992,
  title = {Exact {{Questions}} to {{Some Interesting Answers}} in {{Many Body Physics}}},
  booktitle = {Recent {{Progress}} in {{Many-Body Theories}}: {{Volume}} 3},
  author = {Arovas, D. P. and Girvin, S. M.},
  editor = {Ainsworth, T. L. and Campbell, C. E. and Clements, B. E. and Krotscheck, E.},
  year = 1992,
  pages = {315--344},
  publisher = {Springer US},
  address = {Boston, MA},
  doi = {10.1007/978-1-4615-3466-2_21},
  url = {https://doi.org/10.1007/978-1-4615-3466-2_21},
  urldate = {2025-01-12},
  isbn = {978-1-4615-3466-2},
  langid = {english}
}

@article{ardonneTopologicalOrderConformal2004,
  title = {Topological Order and Conformal Quantum Critical Points},
  author = {Ardonne, Eddy and Fendley, Paul and Fradkin, Eduardo},
  year = 2004,
  month = apr,
  journal = {Annals of Physics},
  volume = {310},
  number = {2},
  pages = {493--551},
  issn = {0003-4916},
  doi = {10.1016/j.aop.2004.01.004},
  url = {https://www.sciencedirect.com/science/article/pii/S0003491604000247},
  urldate = {2025-01-02}
}

@misc{tan2025,
  title = {Interference-Caged Quantum Many-Body Scars: Local {{ETH}}-Violation Witnesses and Deformation Stability},
  shorttitle = {Interference-Caged Quantum Many-Body Scars},
  author = {Tan, Tao-Lin and Huang, Yi-Ping},
  year = 2025,
  month = apr,
  number = {arXiv:2504.07780},
  primaryclass = {cond-mat},
  publisher = {arXiv},
  doi = {10.48550/arXiv.2504.07780},
  url = {http://arxiv.org/abs/2504.07780},
  urldate = {2026-11-25},
  archiveprefix = {arXiv}
}

@article{ben-amiManybodyCagesDisorderfree2025,
  title = {Many-Body Cages: Disorder-Free Glassiness from Flat Bands in {{Fock}} Space, and Many-Body {{Rabi}} Oscillations},
  shorttitle = {Many-Body Cages},
  author = {{Ben-Ami}, Tom and Heyl, Markus and Moessner, Roderich},
  year = 2026,
  month = sep,
  journal = {Physical Review Letters},
  volume = {137},
  number = {10},
  pages = {100401},
  publisher = {American Physical Society},
  doi = {10.1103/gmw6-x94c},
  url = {https://link.aps.org/doi/10.1103/gmw6-x94c},
  urldate = {2026-11-25}
}

@article{jonay2026,
  title = {Localized {{Fock}} Space Cages in Kinetically Constrained Models},
  author = {Jonay, Cheryne and Pollmann, Frank},
  year = 2026,
  month = apr,
  journal = {Physical Review B},
  volume = {113},
  number = {13},
  pages = {134313},
  publisher = {American Physical Society},
  doi = {10.1103/wz33-vczt},
  url = {https://link.aps.org/doi/10.1103/wz33-vczt},
  urldate = {2026-06-28}
}

@article{nicolau2026,
  title = {Fragmentation, {{Zero Modes}}, and {{Collective Bound States}} in {{Constrained Models}}},
  author = {Nicolau, Eloi and Ljubotina, Marko and Serbyn, Maksym},
  year = 2026,
  month = mar,
  journal = {PRX Quantum},
  volume = {7},
  number = {1},
  pages = {010352},
  issn = {2691-3399},
  doi = {10.1103/sl79-1xgb},
  url = {https://link.aps.org/doi/10.1103/sl79-1xgb},
  urldate = {2026-04-21},
  langid = {english}
}

@article{mohapatra2026,
  title = {Additional Quantum Many-Body Scars of the {spin-$1$ $XY$} Model with {{Fock-space}} Cages and Commutant Algebras},
  author = {Mohapatra, Sashikanta and Moudgalya, Sanjay and Balram, Ajit C.},
  year = 2026,
  month = feb,
  journal = {Physical Review B},
  volume = {113},
  number = {5},
  primaryclass = {cond-mat},
  pages = {054310},
  issn = {2469-9950, 2469-9969},
  doi = {10.1103/4tv9-q7g7},
  url = {https://link.aps.org/doi/10.1103/4tv9-q7g7},
  urldate = {2026-04-14},
  archiveprefix = {arXiv}
}

@article{hazra2026,
  title = {Obstructed {{Cooper}} Pairs in Flat Band Systems -- Weakly-Coherent Superfluids and Exact Spin Liquids},
  author = {Hazra, Tamaghna and Verma, Nishchhal and Schmalian, J{\"o}rg},
  year = 2026,
  month = nov,
  journal = {SciPost Physics},
  volume = {21},
  number = {5},
  pages = {126},
  issn = {2542-4653},
  doi = {10.21468/SciPostPhys.21.5.126},
  url = {https://scipost.org/SciPostPhys.21.5.126},
  urldate = {2026-11-25},
  langid = {english}
}

@article{cvetkovic2009towards,
  title={Towards a spectral theory of graphs based on the {Signless Laplacian, I}},
  author={Cvetkovi{\'c}, Drago{\v{s}} and Simi{\'c}, Slobodan K},
  journal={Publications de l'Institut Mathematique},
  volume={85},
  number={105},
  pages={19--33},
  year={2009},
  doi={10.2298/pim0999019c},
  publisher={Matemati{\v{c}}ki institut SANU}
}

@article{cvetkovic2010,
  title = {Towards a Spectral Theory of Graphs Based on the Signless {{Laplacian}}, {{II}}},
  author = {Cvetkovi{\'c}, Drago{\v s} and Simi{\'c}, Slobodan K.},
  year = 2010,
  month = apr,
  journal = {Linear Algebra and its Applications},
  volume = {432},
  number = {9},
  pages = {2257--2272},
  issn = {00243795},
  doi = {10.1016/j.laa.2009.05.020},
  url = {https://linkinghub.elsevier.com/retrieve/pii/S0024379509002808},
  urldate = {2026-07-21},
  copyright = {https://www.elsevier.com/tdm/userlicense/1.0/},
  langid = {english}
}

@article{cvetkovic2010a,
  title = {Towards a {{Spectral Theory}} of {{Graphs Based}} on the {{Signless Laplacian}}, {{III}}},
  author = {Cvetkovi{\'c}, Drago{\v s} and Simi{\'c}, Slobodan K.},
  year = 2010,
  journal = {Applicable Analysis and Discrete Mathematics},
  volume = {4},
  number = {1},
  pages = {156--166},
  publisher = {University of Belgrade, Serbia},
  issn = {1452-8630},
  doi = {10.2298/aadm1000001c},
  url = {https://www.jstor.org/stable/43671298},
  urldate = {2026-07-21}
}

@book{godsil2001,
  title = {Algebraic {{Graph Theory}}},
  author = {Godsil, Chris and Royle, Gordon},
  year = 2001,
  month = jan,
  series = {Graduate Texts in Mathematics},
  volume = {207},
  publisher = {Springer},
  doi = {10.1007/978-1-4613-0163-9},
  isbn = {978-0-387-95220-8}
}

@article{childs2004,
  title = {Spatial Search by Quantum Walk},
  author = {Childs, Andrew M. and Goldstone, Jeffrey},
  year = 2004,
  month = aug,
  journal = {Physical Review A},
  volume = {70},
  number = {2},
  pages = {022314},
  publisher = {American Physical Society},
  doi = {10.1103/PhysRevA.70.022314},
  url = {https://link.aps.org/doi/10.1103/PhysRevA.70.022314},
  urldate = {2026-07-21}
}

@article{leykam2018,
  title = {Artificial Flat Band Systems: From Lattice Models to Experiments},
  shorttitle = {Artificial Flat Band Systems},
  author = {Leykam, Daniel and Andreanov, Alexei and Flach, Sergej},
  year = 2018,
  month = jan,
  journal = {Advances in Physics: X},
  volume = {3},
  number = {1},
  pages = {1473052},
  publisher = {Taylor \& Francis},
  issn = {null},
  doi = {10.1080/23746149.2018.1473052},
  url = {https://doi.org/10.1080/23746149.2018.1473052},
  urldate = {2026-07-21}
}

@article{mukherjee2016,
  title = {Random Walks on Simplicial Complexes and Harmonics},
  author = {Mukherjee, Sayan and Steenbergen, John},
  year = 2016,
  journal = {Random Structures \& Algorithms},
  volume = {49},
  number = {2},
  pages = {379--405},
  issn = {1098-2418},
  doi = {10.1002/rsa.20645},
  url = {https://onlinelibrary.wiley.com/doi/abs/10.1002/rsa.20645},
  urldate = {2026-07-21},
  copyright = {\copyright{} 2016 The Authors Random Structures \& Algorithms Published by Wiley Periodicals, Inc.},
  langid = {english}
}

@misc{hajij2023,
  title = {Topological {{Deep Learning}}: {{Going Beyond Graph Data}}},
  shorttitle = {Topological {{Deep Learning}}},
  author = {Hajij, Mustafa and Zamzmi, Ghada and Papamarkou, Theodore and Miolane, Nina and {Guzm{\'a}n-S{\'a}enz}, Aldo and Ramamurthy, Karthikeyan Natesan and Birdal, Tolga and Dey, Tamal K. and Mukherjee, Soham and Samaga, Shreyas N. and Livesay, Neal and Walters, Robin and Rosen, Paul and Schaub, Michael T.},
  year = 2023,
  month = may,
  number = {arXiv:2206.00606},
  primaryclass = {cs.LG},
  publisher = {arXiv},
  doi = {10.48550/arXiv.2206.00606},
  url = {http://arxiv.org/abs/2206.00606},
  urldate = {2026-07-21},
  archiveprefix = {arXiv}
}

@inproceedings{yang2022,
  title = {Efficient {{Representation Learning}} for {{Higher-Order Data With Simplicial Complexes}}},
  booktitle = {Proceedings of the {{First Learning}} on {{Graphs Conference}}},
  author = {Yang, Ruochen and Sala, Frederic and Bogdan, Paul},
  year = 2022,
  month = dec,
  pages = {13:1-13:21},
  publisher = {PMLR},
  issn = {2640-3498},
  url = {https://proceedings.mlr.press/v198/yang22a.html},
  urldate = {2026-07-21},
  langid = {english}
}

@article{biamonte2017,
  title = {Quantum Machine Learning},
  author = {Biamonte, Jacob and Wittek, Peter and Pancotti, Nicola and Rebentrost, Patrick and Wiebe, Nathan and Lloyd, Seth},
  year = 2017,
  month = sep,
  journal = {Nature},
  volume = {549},
  number = {7671},
  pages = {195--202},
  publisher = {Nature Publishing Group},
  issn = {1476-4687},
  doi = {10.1038/nature23474},
  url = {https://www.nature.com/articles/nature23474},
  urldate = {2026-07-21},
  copyright = {2017 Macmillan Publishers Limited, part of Springer Nature. All rights reserved.},
  langid = {english}
}

@article{chandra1995,
  title = {Possible {{Glassiness}} in a {{Periodic Long-Range Josephson Array}}},
  author = {Chandra, P. and Ioffe, L. B. and Sherrington, D.},
  year = 1995,
  month = jul,
  journal = {Physical Review Letters},
  volume = {75},
  number = {4},
  pages = {713--716},
  issn = {0031-9007, 1079-7114},
  doi = {10.1103/PhysRevLett.75.713},
  url = {https://link.aps.org/doi/10.1103/PhysRevLett.75.713},
  urldate = {2026-05-14},
  copyright = {http://link.aps.org/licenses/aps-default-license},
  langid = {english}
}

@article{bonamassa2023,
  title = {Interdependent Superconducting Networks},
  author = {Bonamassa, I. and Gross, B. and Laav, M. and Volotsenko, I. and Frydman, A. and Havlin, S.},
  year = 2023,
  month = aug,
  journal = {Nature Physics},
  volume = {19},
  number = {8},
  pages = {1163--1170},
  publisher = {Nature Publishing Group},
  issn = {1745-2481},
  doi = {10.1038/s41567-023-02029-z},
  url = {https://www.nature.com/articles/s41567-023-02029-z},
  urldate = {2026-05-14},
  copyright = {2023 The Author(s), under exclusive licence to Springer Nature Limited},
  langid = {english}
}

@article{nambisan2026,
  title = {Quantum Coherent Manipulation and Readout of Superconducting Vortex States},
  author = {Nambisan, Ameya and G{\"u}nzler, Simon and Rieger, Dennis and Gosling, Nicolas and Geisert, Simon and Carpentier, Victor and Zapata, Nicolas and Field, Mitchell and Milo{\v s}evi{\'c}, Milorad V. and Lopez, Carlos A. Diaz and Padurariu, Ciprian and Kubala, Bj{\"o}rn and Ankerhold, Joachim and Wernsdorfer, Wolfgang and Spiecker, Martin and Pop, Ioan M.},
  year = 2026,
  month = may,
  journal = {Nature},
  volume = {653},
  number = {8113},
  pages = {63--67},
  publisher = {Nature Publishing Group},
  issn = {1476-4687},
  doi = {10.1038/s41586-026-10441-7},
  url = {https://www.nature.com/articles/s41586-026-10441-7},
  urldate = {2026-07-21},
  copyright = {2026 The Author(s)},
  langid = {english}
}

@article{dutta2021,
  title = {Evidence of Zero-Point Fluctuation of Vortices in a Very Weakly Pinned {$a$}-{{MoGe}} Thin Film},
  author = {Dutta, Surajit and Roy, Indranil and Jesudasan, John and Sachdev, Subir and Raychaudhuri, Pratap},
  year = 2021,
  month = jun,
  journal = {Physical Review B},
  volume = {103},
  number = {21},
  pages = {214512},
  publisher = {American Physical Society},
  doi = {10.1103/PhysRevB.103.214512},
  url = {https://link.aps.org/doi/10.1103/PhysRevB.103.214512},
  urldate = {2026-07-21}
}

@misc{knill2012,
  title = {The Theorems of {{Green-Stokes}}, {{Gauss-Bonnet}} and {{Poincare-Hopf}} in {{Graph Theory}}},
  author = {Knill, Oliver},
  year = 2012,
  month = jan,
  number = {arXiv:1201.6049},
  primaryclass = {math.DG},
  publisher = {arXiv},
  doi = {10.48550/arXiv.1201.6049},
  url = {http://arxiv.org/abs/1201.6049},
  urldate = {2026-07-23},
  archiveprefix = {arXiv}
}

@misc{knill2013,
  title = {The {{McKean-Singer Formula}} in {{Graph Theory}}},
  author = {Knill, Oliver},
  year = 2013,
  month = jan,
  number = {arXiv:1301.1408},
  primaryclass = {math.CO},
  publisher = {arXiv},
  doi = {10.48550/arXiv.1301.1408},
  url = {http://arxiv.org/abs/1301.1408},
  urldate = {2026-07-23},
  archiveprefix = {arXiv}
}

@book{grady2010,
  title = {Discrete {{Calculus}}},
  author = {Grady, Leo J. and Polimeni, Jonathan R.},
  year = 2010,
  publisher = {Springer},
  address = {London},
  doi = {10.1007/978-1-84996-290-2},
  url = {http://link.springer.com/10.1007/978-1-84996-290-2},
  urldate = {2026-07-23},
  copyright = {http://www.springer.com/tdm},
  isbn = {978-1-84996-289-6 978-1-84996-290-2},
  langid = {english}
}

@article{levitt1992,
  title = {The {Euler} Characteristic Is the Unique Locally Determined Numerical Homotopy Invariant of Finite Complexes},
  author = {Levitt, Norman},
  year = 1992,
  month = jan,
  journal = {Discrete \& Computational Geometry},
  volume = {7},
  number = {1},
  pages = {59--67},
  issn = {1432-0444},
  doi = {10.1007/BF02187824},
  url = {https://doi.org/10.1007/BF02187824},
  urldate = {2026-07-23},
  langid = {english}
}

@article{horton1987,
  title = {A {{Polynomial-Time Algorithm}} to {{Find}} the {{Shortest Cycle Basis}} of a {{Graph}}},
  author = {Horton, J. D.},
  year = 1987,
  month = apr,
  journal = {SIAM Journal on Computing},
  volume = {16},
  number = {2},
  pages = {358--366},
  publisher = {{Society for Industrial and Applied Mathematics}},
  issn = {0097-5397},
  doi = {10.1137/0216026},
  url = {https://epubs.siam.org/doi/10.1137/0216026},
  urldate = {2026-07-24}
}

@article{mishra2021,
  title = {Observation of Fractional Edge Excitations in Nanographene Spin Chains},
  author = {Mishra, Shantanu and Catarina, Gon{\c c}alo and Wu, Fupeng and Ortiz, Ricardo and Jacob, David and Eimre, Kristjan and Ma, Ji and Pignedoli, Carlo A. and Feng, Xinliang and Ruffieux, Pascal and {Fern{\'a}ndez-Rossier}, Joaqu{\'i}n and Fasel, Roman},
  year = 2021,
  month = oct,
  journal = {Nature},
  volume = {598},
  number = {7880},
  pages = {287--292},
  publisher = {Nature Publishing Group},
  issn = {1476-4687},
  doi = {10.1038/s41586-021-03842-3},
  url = {https://www.nature.com/articles/s41586-021-03842-3},
  urldate = {2026-07-24},
  copyright = {2021 The Author(s), under exclusive licence to Springer Nature Limited},
  langid = {english}
}

@article{wei2012,
  title = {Two-Dimensional {{Affleck-Kennedy-Lieb-Tasaki}} State on the Honeycomb Lattice Is a Universal Resource for Quantum Computation},
  author = {Wei, Tzu-Chieh and Affleck, Ian and Raussendorf, Robert},
  year = 2012,
  month = sep,
  journal = {Physical Review A},
  volume = {86},
  number = {3},
  pages = {032328},
  publisher = {American Physical Society},
  doi = {10.1103/PhysRevA.86.032328},
  url = {https://link.aps.org/doi/10.1103/PhysRevA.86.032328},
  urldate = {2026-07-24}
}

@article{sato2009,
  title = {Molecular Electron-Spin Quantum Computers and Quantum Information Processing: Pulse-Based Electron Magnetic Resonance Spin Technology Applied to Matter Spin-Qubits},
  shorttitle = {Molecular Electron-Spin Quantum Computers and Quantum Information Processing},
  author = {Sato, Kazunobu and Nakazawa, Shigeaki and Rahimi, Robabeh and Ise, Tomoaki and Nishida, Shinsuke and Yoshino, Tomohiro and Mori, Nobuyuki and Toyota, Kazuo and Shiomi, Daisuke and Yakiyama, Yumi and Morita, Yasushi and Kitagawa, Masahiro and Nakasuji, Kazuhiro and Nakahara, Mikio and Hara, Hideyuki and Carl, Patrick and H{\"o}fer, Peter and Takui, Takeji},
  year = 2009,
  month = jun,
  journal = {Journal of Materials Chemistry},
  volume = {19},
  number = {22},
  pages = {3739--3754},
  issn = {0959-9428},
  doi = {10.1039/b819556k},
  url = {https://doi.org/10.1039/b819556k},
  urldate = {2026-07-24}
}

@article{mishra2020,
  title = {Topological Frustration Induces Unconventional Magnetism in a Nanographene},
  author = {Mishra, Shantanu and Beyer, Doreen and Eimre, Kristjan and Kezilebieke, Shawulienu and Berger, Reinhard and Gr{\"o}ning, Oliver and Pignedoli, Carlo A. and M{\"u}llen, Klaus and Liljeroth, Peter and Ruffieux, Pascal and Feng, Xinliang and Fasel, Roman},
  year = 2020,
  month = jan,
  journal = {Nature Nanotechnology},
  volume = {15},
  number = {1},
  pages = {22--28},
  issn = {1748-3395},
  doi = {10.1038/s41565-019-0577-9},
  langid = {english},
  pmid = {31819244}
}

@article{valenta2022,
  title = {The Taming of {{Clar}}'s Hydrocarbon},
  author = {Valenta, Leo{\v s} and Jur{\'i}{\v c}ek, Michal},
  year = 2022,
  journal = {Chemical Communications},
  volume = {58},
  number = {78},
  pages = {10896--10906},
  issn = {1359-7345, 1364-548X},
  doi = {10.1039/D2CC03720C},
  url = {https://pubs.rsc.org/cc/article/58/78/10896-10906/758052},
  urldate = {2026-07-17},
  langid = {english}
}

@article{ovchinnikov1978,
  title = {Multiplicity of the Ground State of Large Alternant Organic Molecules with Conjugated Bonds},
  author = {Ovchinnikov, Alexandr A.},
  year = 1978,
  month = dec,
  journal = {Theoretica chimica acta},
  volume = {47},
  number = {4},
  pages = {297--304},
  issn = {1432-2234},
  doi = {10.1007/BF00549259},
  url = {https://doi.org/10.1007/BF00549259},
  urldate = {2026-07-17},
  langid = {english}
}

@article{clar1953,
  title = {Aromatic {{Hydrocarbons}}. {{LXV}}. {{Triangulene Derivatives}}},
  author = {Clar, E. and Stewart, D. G.},
  year = 1953,
  month = jun,
  journal = {Journal of the American Chemical Society},
  volume = {75},
  number = {11},
  pages = {2667--2672},
  publisher = {American Chemical Society},
  issn = {0002-7863},
  doi = {10.1021/ja01107a035},
  url = {https://doi.org/10.1021/ja01107a035},
  urldate = {2026-07-17}
}

@article{clar1968,
  title = {The Non-Existence of a Threefold Aromatic Conjugation in Linear Benzologues of Triphenylene (Starphenes)},
  author = {Clar, E. and Mullen, A.},
  year = 1968,
  month = jan,
  journal = {Tetrahedron},
  volume = {24},
  number = {23},
  pages = {6719--6724},
  issn = {0040-4020},
  doi = {10.1016/S0040-4020(01)96845-0},
  url = {https://www.sciencedirect.com/science/article/pii/S0040402001968450},
  urldate = {2026-07-17}
}

@article{melle-franco2017,
  title = {When 1 + 1 Is Odd},
  author = {{Melle-Franco}, Manuel},
  year = 2017,
  month = apr,
  journal = {Nature Nanotechnology},
  volume = {12},
  number = {4},
  pages = {292--293},
  publisher = {Nature Publishing Group},
  issn = {1748-3395},
  doi = {10.1038/nnano.2017.9},
  url = {https://www.nature.com/articles/nnano.2017.9},
  urldate = {2026-07-17},
  copyright = {2017 Springer Nature Limited},
  langid = {english}
}

@article{pavlicek2017,
  title = {Synthesis and Characterization of Triangulene},
  author = {Pavli{\v c}ek, Niko and Mistry, Anish and Majzik, Zsolt and Moll, Nikolaj and Meyer, Gerhard and Fox, David J. and Gross, Leo},
  year = 2017,
  month = apr,
  journal = {Nature Nanotechnology},
  volume = {12},
  number = {4},
  pages = {308--311},
  publisher = {Nature Publishing Group},
  issn = {1748-3395},
  doi = {10.1038/nnano.2016.305},
  url = {https://www.nature.com/articles/nnano.2016.305},
  urldate = {2026-07-17},
  copyright = {2017 Springer Nature Limited},
  langid = {english}
}

@article{koshinoMaximallyLocalizedWannier2018,
  title = {Maximally {{Localized Wannier Orbitals}} and the {{Extended Hubbard Model}} for {{Twisted Bilayer Graphene}}},
  author = {Koshino, Mikito and Yuan, Noah F. Q. and Koretsune, Takashi and Ochi, Masayuki and Kuroki, Kazuhiko and Fu, Liang},
  year = 2018,
  month = sep,
  journal = {Physical Review X},
  volume = {8},
  number = {3},
  pages = {031087},
  doi = {10.1103/PhysRevX.8.031087},
  url = {https://link.aps.org/doi/10.1103/PhysRevX.8.031087},
  urldate = {2018-11-08}
}

@article{po-prx-2018,
  title = {Origin of {{Mott}} Insulating Behavior and Superconductivity in Twisted Bilayer Graphene},
  author = {Po, Hoi Chun and Zou, Liujun and Vishwanath, Ashvin and Senthil, T.},
  year = 2018,
  month = sep,
  journal = {Physical Review X},
  volume = {8},
  number = {3},
  pages = {031089},
  publisher = {American Physical Society},
  doi = {10.1103/PhysRevX.8.031089},
  url = {https://link.aps.org/doi/10.1103/PhysRevX.8.031089}
}

@article{kangSymmetryMaximallyLocalized2018,
  title = {Symmetry, {{Maximally Localized Wannier States}}, and a {{Low-Energy Model}} for {{Twisted Bilayer Graphene Narrow Bands}}},
  author = {Kang, Jian and Vafek, Oskar},
  year = 2018,
  month = sep,
  journal = {Physical Review X},
  volume = {8},
  number = {3},
  pages = {031088},
  doi = {10.1103/PhysRevX.8.031088},
  url = {https://link.aps.org/doi/10.1103/PhysRevX.8.031088},
  urldate = {2019-04-27}
}

@article{aliceaNewDirectionsPursuit2012,
  ids = {alicea-rpp-2012},
  title = {New Directions in the Pursuit of {{Majorana}} Fermions in Solid State Systems},
  author = {Alicea, Jason},
  year = 2012,
  journal = {Reports on Progress in Physics},
  volume = {75},
  number = {7},
  pages = {076501},
  issn = {0034-4885},
  doi = {10.1088/0034-4885/75/7/076501},
  url = {http://stacks.iop.org/0034-4885/75/i=7/a=076501},
  urldate = {2018-02-06},
  langid = {english}
}

@article{vidal1998,
  title = {Aharonov-{{Bohm Cages}} in {{Two-Dimensional Structures}}},
  author = {Vidal, Julien and Mosseri, R{\'e}my and Dou{\c c}ot, Benoit},
  year = 1998,
  month = dec,
  journal = {Physical Review Letters},
  volume = {81},
  number = {26},
  pages = {5888--5891},
  publisher = {American Physical Society},
  doi = {10.1103/PhysRevLett.81.5888},
  url = {https://link.aps.org/doi/10.1103/PhysRevLett.81.5888},
  urldate = {2026-07-24}
}

@article{castroneto2009,
  title = {The Electronic Properties of Graphene},
  author = {Castro Neto, A. H. and Guinea, F. and Peres, N. M. R. and Novoselov, K. S. and Geim, A. K.},
  year = 2009,
  month = jan,
  journal = {Reviews of Modern Physics},
  volume = {81},
  number = {1},
  pages = {109--162},
  publisher = {American Physical Society},
  doi = {10.1103/RevModPhys.81.109},
  url = {https://link.aps.org/doi/10.1103/RevModPhys.81.109},
  urldate = {2023-05-16}
}

@article{fujita1996,
  title = {Peculiar {{Localized State}} at {{Zigzag Graphite Edge}}},
  author = {Fujita, Mitsutaka and Wakabayashi, Katsunori and Nakada, Kyoko and Kusakabe, Koichi},
  year = 1996,
  month = jul,
  journal = {Journal of the Physical Society of Japan},
  volume = {65},
  number = {7},
  pages = {1920--1923},
  publisher = {The Physical Society of Japan},
  issn = {0031-9015},
  doi = {10.1143/JPSJ.65.1920},
  url = {https://journals.jps.jp/doi/10.1143/JPSJ.65.1920},
  urldate = {2026-07-24}
}

@article{ryu2002,
  title = {Topological {{Origin}} of {{Zero-Energy Edge States}} in {{Particle-Hole Symmetric Systems}}},
  author = {Ryu, Shinsei and Hatsugai, Yasuhiro},
  year = 2002,
  month = jul,
  journal = {Physical Review Letters},
  volume = {89},
  number = {7},
  pages = {077002},
  publisher = {American Physical Society},
  doi = {10.1103/PhysRevLett.89.077002},
  url = {https://link.aps.org/doi/10.1103/PhysRevLett.89.077002},
  urldate = {2026-07-24}
}

@article{verma2025,
  title = {Quantum {{Geometry}} and the {{Hidden Scales}} in {{Materials}}},
  author = {Verma, Nishchhal and Moll, Philip J. W. and Holder, Tobias and Queiroz, Raquel},
  year = 2026,
  month = apr,
  journal = {Nature Reviews Physics},
  volume = {8},
  number = {4},
  pages = {226--239},
  doi = {10.1038/s42254-026-00923-y},
  url = {https://doi.org/10.1038/s42254-026-00923-y},
  urldate = {2026-07-24},
  langid = {english}
}

@article{ahn2022,
  title = {Riemannian Geometry of Resonant Optical Responses},
  author = {Ahn, Junyeong and Guo, Guang-Yu and Nagaosa, Naoto and Vishwanath, Ashvin},
  year = 2022,
  month = mar,
  journal = {Nature Physics},
  volume = {18},
  number = {3},
  pages = {290--295},
  publisher = {Nature Publishing Group},
  issn = {1745-2481},
  doi = {10.1038/s41567-021-01465-z},
  url = {https://www.nature.com/articles/s41567-021-01465-z},
  urldate = {2026-07-24},
  copyright = {2021 The Author(s), under exclusive licence to Springer Nature Limited},
  langid = {english}
}

@article{provostRiemannianStructureManifolds1980,
  title = {Riemannian Structure on Manifolds of Quantum States},
  author = {Provost, J. P. and Vallee, G.},
  year = 1980,
  month = sep,
  journal = {Communications in Mathematical Physics},
  volume = {76},
  number = {3},
  pages = {289--301},
  issn = {1432-0916},
  doi = {10.1007/BF02193559},
  url = {https://doi.org/10.1007/BF02193559},
  urldate = {2024-01-05},
  langid = {english}
}

@incollection{berryFiveYearsLater,
  title = {The Quantum Phase, Five Years After},
  booktitle = {Geometric Phases in Physics},
  author = {Berry, M. V.},
  year = {1989},
  pages = {7--28},
  publisher = {World Scientific},
  doi = {10.1142/9789812798381_0001},
  url = {https://www.worldscientific.com/doi/abs/10.1142/9789812798381_0001}
}

@article{chengQuantumGeometricTensor2010,
  title = {Quantum {{Geometric Tensor}} ({{Fubini-Study Metric}}) in {{Simple Quantum System}}: {{A}} Pedagogical {{Introduction}}},
  shorttitle = {Quantum {{Geometric Tensor}} ({{Fubini-Study Metric}}) in {{Simple Quantum System}}},
  author = {Cheng, Ran},
  year = 2010,
  month = dec,
  journal = {arXiv:1012.1337 [math-ph, physics:quant-ph]},
  primaryclass = {math-ph, physics:quant-ph},
  doi = {10.48550/arXiv.1012.1337},
  url = {http://arxiv.org/abs/1012.1337},
  urldate = {2019-07-23},
  archiveprefix = {arXiv}
}

@article{marzariMaximallyLocalizedGeneralized1997,
  title = {Maximally Localized Generalized {{Wannier}} Functions for Composite Energy Bands},
  author = {Marzari, Nicola and Vanderbilt, David},
  year = 1997,
  month = nov,
  journal = {Physical Review B},
  volume = {56},
  number = {20},
  pages = {12847--12865},
  doi = {10.1103/PhysRevB.56.12847},
  url = {https://link.aps.org/doi/10.1103/PhysRevB.56.12847},
  urldate = {2019-10-24}
}

@article{vismara1997,
  title = {Union of All the {{Minimum Cycle Bases}} of a {{Graph}}},
  author = {Vismara, Philippe},
  year = 1997,
  month = jan,
  journal = {The Electronic Journal of Combinatorics},
  volume = {4},
  number = {1},
  pages = {R9},
  issn = {1077-8926},
  doi = {10.37236/1294},
  url = {https://www.combinatorics.org/ojs/index.php/eljc/article/view/v4i1r9},
  urldate = {2026-07-24}
}

@article{kavitha2009,
  title = {Cycle Bases in Graphs Characterization, Algorithms, Complexity, and Applications},
  author = {Kavitha, Telikepalli and Liebchen, Christian and Mehlhorn, Kurt and Michail, Dimitrios and Rizzi, Romeo and Ueckerdt, Torsten and Zweig, Katharina A.},
  year = 2009,
  month = nov,
  journal = {Computer Science Review},
  volume = {3},
  number = {4},
  pages = {199--243},
  issn = {15740137},
  doi = {10.1016/j.cosrev.2009.08.001},
  url = {https://linkinghub.elsevier.com/retrieve/pii/S1574013709000483},
  urldate = {2026-07-24},
  copyright = {https://www.elsevier.com/tdm/userlicense/1.0/},
  langid = {english}
}

@article{buldyrev2010,
  title = {Catastrophic Cascade of Failures in Interdependent Networks},
  author = {Buldyrev, Sergey V. and Parshani, Roni and Paul, Gerald and Stanley, H. Eugene and Havlin, Shlomo},
  year = 2010,
  month = apr,
  journal = {Nature},
  volume = {464},
  number = {7291},
  pages = {1025--1028},
  publisher = {Nature Publishing Group},
  issn = {1476-4687},
  doi = {10.1038/nature08932},
  url = {https://www.nature.com/articles/nature08932},
  urldate = {2026-07-17},
  copyright = {2010 Macmillan Publishers Limited. All rights reserved},
  langid = {english}
}

@book{biggs1993,
  title = {Algebraic {{Graph Theory}}},
  author = {Biggs, Norman},
  year = 1993,
  publisher = {Cambridge University Press},
  googlebooks = {6TasRmIFOxQC},
  isbn = {978-0-521-45897-9},
  langid = {english}
}

@article{biswas2022,
  title = {Scars from Protected Zero Modes and beyond in {$U(1)$} Quantum Link and Quantum Dimer Models},
  author = {Biswas, Saptarshi and Banerjee, Debasish and Sen, Arnab},
  year = 2022,
  month = may,
  journal = {SciPost Physics},
  volume = {12},
  number = {5},
  pages = {148},
  issn = {2542-4653},
  doi = {10.21468/SciPostPhys.12.5.148},
  url = {https://scipost.org/SciPostPhys.12.5.148},
  urldate = {2026-10-31},
  langid = {english}
}

@book{cvetkovic2004spectral,
  author    = {Cvetkovi{\'c}, Drago{\v s} and Rowlinson, Peter and Simi{\'c}, Slobodan},
  title     = {Spectral Generalizations of Line Graphs: On Graphs with Least Eigenvalue $-2$},
  series    = {London Mathematical Society Lecture Note Series},
  volume    = {314},
  publisher = {Cambridge University Press},
  address   = {Cambridge},
  year      = {2004},
  doi       = {10.1017/CBO9780511751752}
}

@article{nayak-rmp-2008,
  title = {Non-{{Abelian}} Anyons and Topological Quantum Computation},
  author = {Nayak, Chetan and Simon, Steven H. and Stern, Ady and Freedman, Michael and Das Sarma, Sankar},
  year = 2008,
  journal = {Reviews of Modern Physics},
  volume = {80},
  number = {3},
  pages = {1083},
  publisher = {American Physical Society},
  doi = {10.1103/RevModPhys.80.1083}
}

@article{xu2020,
  title = {Building Flat-Band Lattice Models from {{Gram}} Matrices},
  author = {Xu, Youjiang and Pu, Han},
  year = 2020,
  month = nov,
  journal = {Physical Review A},
  volume = {102},
  number = {5},
  pages = {053305},
  publisher = {American Physical Society},
  doi = {10.1103/PhysRevA.102.053305}
}

@article{roentgen2018,
  title = {Compact Localized States and Flat Bands from Local Symmetry Partitioning},
  author = {R{\"o}ntgen, M. and Morfonios, C. V. and Schmelcher, P.},
  year = 2018,
  month = jan,
  journal = {Physical Review B},
  volume = {97},
  number = {3},
  pages = {035161},
  publisher = {American Physical Society},
  doi = {10.1103/PhysRevB.97.035161}
}

@article{roentgen2021,
  title = {Latent Symmetry Induced Degeneracies},
  author = {R{\"o}ntgen, M. and Pyzh, M. and Morfonios, C. V. and Palaiodimopoulos, N. E. and Diakonos, F. K. and Schmelcher, P.},
  year = 2021,
  month = may,
  journal = {Physical Review Letters},
  volume = {126},
  number = {18},
  pages = {180601},
  publisher = {American Physical Society},
  doi = {10.1103/PhysRevLett.126.180601}
}

@article{morfonios2021,
  title = {Flat Bands by Latent Symmetry},
  author = {Morfonios, C. V. and R{\"o}ntgen, M. and Pyzh, M. and Schmelcher, P.},
  year = 2021,
  month = jul,
  journal = {Physical Review B},
  volume = {104},
  number = {3},
  pages = {035105},
  publisher = {American Physical Society},
  doi = {10.1103/PhysRevB.104.035105}
}

@article{roentgen2019,
  title = {Quantum Network Transfer and Storage with Compact Localized States Induced by Local Symmetries},
  author = {R{\"o}ntgen, M. and Morfonios, C. V. and Brouzos, I. and Diakonos, F. K. and Schmelcher, P.},
  year = 2019,
  month = aug,
  journal = {Physical Review Letters},
  volume = {123},
  number = {8},
  pages = {080504},
  publisher = {American Physical Society},
  doi = {10.1103/PhysRevLett.123.080504}
}

@article{roentgen2024,
  title = {Topological States Protected by Hidden Symmetry},
  author = {R{\"o}ntgen, Malte and Chen, Xuelong and Gao, Wenlong and Pyzh, Maxim and Schmelcher, Peter and Pagneux, Vincent and Achilleos, Vassos and Coutant, Antonin},
  year = 2024,
  month = jul,
  journal = {Physical Review B},
  volume = {110},
  number = {3},
  pages = {035106},
  publisher = {American Physical Society},
  doi = {10.1103/PhysRevB.110.035106}
}

@article{mielkeTasaki1993,
  title = {Ferromagnetism in the {{Hubbard}} Model: {{Examples}} from Models with Degenerate Single-Electron Ground States},
  author = {Mielke, Andreas and Tasaki, Hal},
  year = {1993},
  journal = {Communications in Mathematical Physics},
  volume = {158},
  number = {2},
  pages = {341--371},
  doi = {10.1007/BF02108079}
}

@article{tasaki1998,
  title = {From {{Nagaoka}}'s Ferromagnetism to Flat-Band Ferromagnetism and Beyond: {{An}} Introduction to Ferromagnetism in the {{Hubbard}} Model},
  author = {Tasaki, Hal},
  year = {1998},
  journal = {Progress of Theoretical Physics},
  volume = {99},
  number = {4},
  pages = {489--548},
  doi = {10.1143/PTP.99.489}
}

@article{tanaka2020,
  title = {An Extension of the Cell-Construction Method for the Flat-Band Ferromagnetism},
  author = {Tanaka, Akinori},
  year = {2020},
  journal = {Journal of Statistical Physics},
  volume = {181},
  number = {3},
  pages = {897--916},
  doi = {10.1007/s10955-020-02610-3}
}

@article{hatsugai2011,
  title = {{$Z_Q$} Topological Invariants for Polyacetylene, Kagome and Pyrochlore Lattices},
  author = {Hatsugai, Yasuhiro and Maruyama, Isao},
  year = {2011},
  journal = {Europhysics Letters},
  volume = {95},
  number = {2},
  pages = {20003},
  doi = {10.1209/0295-5075/95/20003}
}

@article{mizoguchi2019,
  title = {Molecular-Orbital Representation of Generic Flat-Band Models},
  author = {Mizoguchi, Tomonari and Hatsugai, Yasuhiro},
  year = {2019},
  journal = {Europhysics Letters},
  volume = {127},
  number = {4},
  pages = {47001},
  doi = {10.1209/0295-5075/127/47001}
}

@article{mizoguchi2023,
  title = {Molecular-Orbital Representation with Random {$U(1)$} Variables},
  author = {Mizoguchi, Tomonari and Hatsugai, Yasuhiro},
  year = {2023},
  journal = {Physical Review B},
  volume = {107},
  number = {9},
  pages = {094201},
  doi = {10.1103/PhysRevB.107.094201}
}

@article{calugaru2022,
  title = {General Construction and Topological Classification of Crystalline Flat Bands},
  author = {C{\u a}lug{\u a}ru, Dumitru and Chew, Aaron and Elcoro, Luis and Xu, Yuanfeng and Regnault, Nicolas and Song, Zhi-Da and Bernevig, B. Andrei},
  year = {2022},
  journal = {Nature Physics},
  volume = {18},
  number = {2},
  pages = {185--189},
  doi = {10.1038/s41567-021-01445-3}
}
\end{document}